\documentclass[twocolumn,english,aps,pra,longbibliography,superscriptaddress, floatfix,10pt]{revtex4-2}

\usepackage[utf8]{inputenc}
\usepackage{xcolor}
\usepackage{amsthm, amssymb}
\usepackage{amsmath}
\usepackage{braket}
\usepackage{txfonts,stmaryrd}
\usepackage{graphicx}
\usepackage{enumerate}
\usepackage[colorlinks=true,urlcolor=blue,citecolor=blue,linkcolor=blue]{hyperref}

\usepackage{bm}

\usepackage[normalem]{ulem}

\newcommand{\ZN}{\mathbb{Z}_N}
\newcommand{\G}{\mathcal{G}}
\newcommand{\Ztwo}{\mathbb{Z}_2}
\newcommand{\Dthree}{\mathrm{D}_3}

\begin{document}
\title{Quantum Resources in Non-Abelian Lattice Gauge Theories: Nonstabilizerness, Multipartite Entanglement, and Fermionic Non-Gaussianity}
\date{\today}

\author{Gopal Chandra Santra}
\email{gopalchandra.santra@unitn.it}
\affiliation{Pitaevskii BEC Center, CNR-INO and Department of Physics, University  of  Trento,  Via Sommarive 14, I-38123 Trento, Italy}
\affiliation{INFN-TIFPA, Trento Institute for Fundamental Physics and Applications, Via Sommarive 14, I-38123 Trento, Italy}
\affiliation{Kirchhoff-Institut f\"ur Physik, Universit\"at Heidelberg, Im Neuenheimer Feld 227, 69120 Heidelberg, Germany}

\author{Julius Mildenberger}
\email{mildenberger@lorentz.leidenuniv.nl}
\affiliation{Pitaevskii BEC Center, CNR-INO and Department of Physics, University  of  Trento,  Via Sommarive 14, I-38123 Trento, Italy}
\affiliation{INFN-TIFPA, Trento Institute for Fundamental Physics and Applications, Via Sommarive 14, I-38123 Trento, Italy}
\affiliation{Lorentz Institute for Theoretical Physics, Leiden University, The Netherlands}
\affiliation{$\langle aQa^L\rangle$ Applied Quantum Algorithms Leiden, The Netherlands}

\author{Edoardo Ballini}
\email{edoardo.ballini@unitn.it}
\affiliation{Pitaevskii BEC Center, CNR-INO and Department of Physics, University  of  Trento,  Via Sommarive 14, I-38123 Trento, Italy}
\affiliation{INFN-TIFPA, Trento Institute for Fundamental Physics and Applications, Via Sommarive 14, I-38123 Trento, Italy}

\author{Alberto Bottarelli}
\email{alberto.bottarelli@unitn.it}
\affiliation{Pitaevskii BEC Center, CNR-INO and Department of Physics, University  of  Trento,  Via Sommarive 14, I-38123 Trento, Italy}
\affiliation{INFN-TIFPA, Trento Institute for Fundamental Physics and Applications, Via Sommarive 14, I-38123 Trento, Italy}

\author{Matteo M. Wauters}
\email{matteo.wauters@unitn.it}
\affiliation{Pitaevskii BEC Center, CNR-INO and Department of Physics, University  of  Trento,  Via Sommarive 14, I-38123 Trento, Italy}
\affiliation{INFN-TIFPA, Trento Institute for Fundamental Physics and Applications, Via Sommarive 14, I-38123 Trento, Italy}

\author{Philipp Hauke}
\email{philipp.hauke@unitn.it}
\affiliation{Pitaevskii BEC Center, CNR-INO and Department of Physics, University  of  Trento,  Via Sommarive 14, I-38123 Trento, Italy}
\affiliation{INFN-TIFPA, Trento Institute for Fundamental Physics and Applications, Via Sommarive 14, I-38123 Trento, Italy}

\begin{abstract}
Lattice gauge theories (LGTs) represent one of the most ambitious goals of quantum simulation. 
From a practical implementation perspective, non-Abelian theories present significantly tougher challenges than Abelian LGTs. 
However, it is unknown whether this is also reflected in increased values of quantum resources relating to the complexity of simulating quantum many-body models. 
Here, we compare three paradigmatic measures of quantum resources---stabilizer Rényi entropy, generalized geometric measure of entanglement, and fermionic antiflatness---for pure-gauge theories on a ladder with Abelian $\ZN$ as well as non-Abelian $\Dthree$ and SU(2) gauge symmetries. 
We find that non-Abelian symmetries are not necessarily inherently harder to simulate than Abelian ones, but rather the required quantum resources depend nontrivially on the interplay between the group structure, superselection sector, and encoding of the gauge constraints. 
Our findings help indicate where quantum advantage could emerge in simulations of LGTs, both in NISQ and fault-tolerant eras.
\end{abstract}
\maketitle

\emph{Introduction}---Quantum simulation harnesses quantum-mechanical devices to address the challenges of solving complex many-body problems, which are often intractable using classical methods~\cite{Hauke_RepProgPhys2012,Georgescu_RMP2014,Daley2022}. Due to their interest spanning from particle to condensed matter physics, lattice gauge theories (LGTs) represent a particularly promising domain for quantum simulation~\cite{dalmonte2016,Preskill_arxiv2018,Aidelsburger_LGT2021,Zohar_PhilTransA2022,halimeh2025cold}. Among these, non-Abelian LGTs stand out as a critical step toward making quantitative predictions with relevance in high-energy physics~\cite{klco2020su2,Bauer_PRXQ2023}. They are also deeply connected to the rich physics of universal anyons and topological quantum computation~\cite{kitaev2002topological,KITAEV2003,Andersen_Nature2023,Iqbal2024}. However, the intricate structure of non-Abelian symmetries suggests that (quantum) computational complexity in these systems may unfold in fundamentally different ways compared to their Abelian counterparts.
In fact, non-Abelian LGTs remain exceptionally difficult to simulate, and existing quantum experiments have been limited to small systems~\cite{Mezzacapo_PRL2015,klco2020su2,Atas_NatComm2021,rahman2021su2,rahman2022realtime,rahman2022self,Fromm2023,Atas_PRR2023, Farrell_PRD2023, Turro_PRD2024,Than_2024, Ciavarella_PRL2024}. This raises a compelling question: if simulating such systems is challenging both for classical and quantum computers, do their Hamiltonians encode distinctive quantum-mechanical properties that must either be leveraged or overcome as resource bottlenecks in the simulation process?

In this work, we characterize the ground states of various LGTs in terms of three complementary resources: nonstabilizerness~\cite{liu2022many,leone2022stabilizer}, multipartite entanglement~\cite{walter2016multipartite,bengtsson2016brief}, and fermionic non-Gaussianity~\cite{froland2025entanglement,hebenstreit2019all,collura2024quantum,sierant2025fermionic}. 
A scalable quantum computational advantage requires multipartite entanglement to grow with system size~\cite{jozsa2003role}. 
However, even if a platform provides extensive entanglement, it does not mean it reaches beyond classical simulability, as Clifford circuits generate stabilizer states that can be highly entangled but are classically simulable due to the Gottesman--Knill theorem~\cite{gottesman1998heisenberg,aaronson2004improved}. 
Nonstabilizerness characterizes the deviation from such ``easy'' states. 
It bounds the minimum number of T-gates required to generate ``magic'' states that present an exponential hardness for classical stabilizer tableaus~\cite{nest2008classical} and are a bottleneck ingredient in fault-tolerant platforms built on, e.g., the surface code~\cite{bravyi2005universal,fowler2012surface}. 
Reference~\cite{Falcao_PRB2025} illustrates the subtle connection between nonstabilizerness and critical phenomena in the case of a (1+1)D U(1) LGT, but leaves the fundamental question open of how the group-structure of LGTs relates to their quantum and classical simulation complexity.
A complementary framework for efficient classical simulation is provided by matchgate circuits, encapsulating systems of free fermions~\cite{valiant2001quantum, terhal2002classical, jozsa2008matchgates,Brod_PRA2016}. 
In this framework, the relevant resource for realizing universal quantum computation and for achieving an exponential difficulty for classical simulation is given by fermionic non-Gaussianity~\cite{hebenstreit2019all}, 
whose study thus offers valuable complementary insights into the resources needed for quantum advantage. These quantum resources, previously employed to study many-body physics~\cite{roy2022genuine,lakkaraju2024predicting,sapui2025genuine,jasser2025stabilizer,santra2025complexity, Falcao_PRB2025,brokemeier2025quantum,cherynshev2025quantum,viscardi2025interplay,sierant2025fermionic} and quantum computation~\cite{fux2024entanglement,KUMAR2024129668,santra2025genuine,turkeshi2025magic,capecci2025role}, thus provide a natural framework for analyzing quantum resources in lattice gauge theory.

\begin{figure*}[t!]
    \centering
    \includegraphics[width=\textwidth, clip]{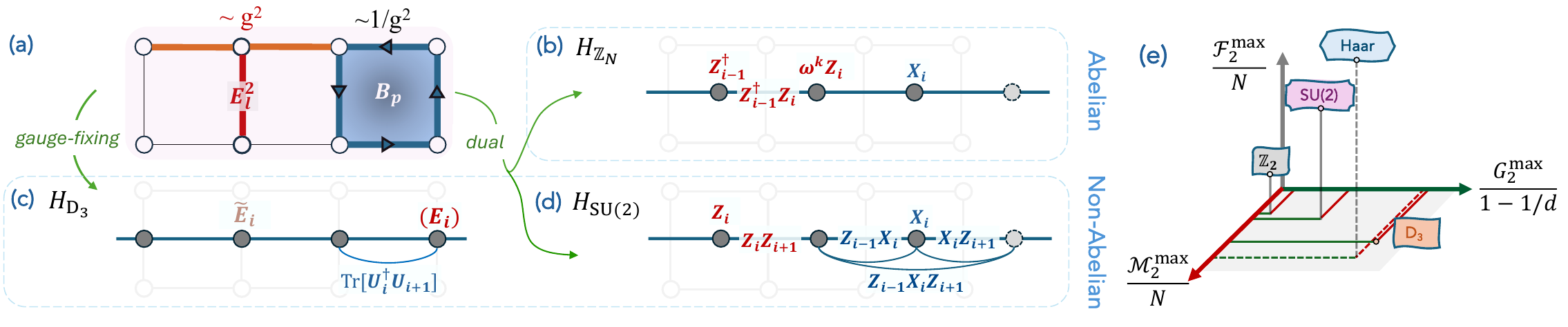}
    \caption{(a) Sketch of the (2+1)D LGT flux ladder and the three different mappings to one-dimensional chains, depending on the underlying gauge symmetry: (b) $\ZN$, (c) $\Dthree$, (d) SU(2). 
    (e) Sketch of the maximum values of the resources across the phase diagrams of the different models we consider: Multipartite entanglement $G_2$, stabilizer R\'enyi entropy $\mathcal{M}_2$, and fermionic antiflatness $\mathcal{F}_2$. All values except the maximal GGM for $\Dthree$ remain below those of Haar random states (fermionic antiflatness is not computed for $\Dthree$). 
    }
    \label{fig:figure1}
\end{figure*} 

We compare these three quantum resources on different pure-gauge models on a ladder geometry: the
Abelian gauge group $\mathbb{Z}_N$ and the non-Abelian groups $\Dthree$ and (truncated) SU(2). We find that non-Abelian gauge symmetries do not necessarily generate more ground-state complexity than Abelian ones. 
For instance, in the absence of background charges, only SU(2)---but not $\Dthree$ nor $\mathbb{Z}_N$---displays an extended regime with finite nonstabilizer density, hinting that Lie groups might be tougher to simulate than discrete ones.
As we illustrate with the example of $\ZN$ LGTs, the superselection sector, i.e., the background charge, can significantly alter the behavior of quantum resources.  
By studying the intricate relation between different local symmetries and simulation complexity, our results indicate that classically hard-to-simulate regimes exist for both Abelian and non-Abelian LGTs.
Our work paves the way for a deeper understanding of which and where quantum resources are necessary to reach a potential quantum advantage in simulating LGTs.

\emph{Models}---To probe the connection between resources and different gauge-symmetry groups, we focus on pure-gauge theories in the quasi-(2+1)D geometry of a plaquette (or flux) ladder, see Fig.~\ref{fig:figure1}(a).
This is the simplest scenario to observe the competition between the magnetic and electric fields, representing a key ingredient for mature quantum simulations of LGTs~\cite{Barbiero2019,Gonzalezcuadra_PRX2020,rahman2021su2,rahman2022self,grabowska2024fully,rahman2022realtime, Zohar_PhilTransA2022, yao2023su,Fromm2023,Ballini_2024,Wang_PRR2025, Fontana_PRX2025}.

For a generic group $\G$, we can write the Kogut--Susskind Hamiltonian~\cite{kogut1975hamiltonian} as 
\begin{equation}
    \label{eq:HKS_general}
    H = \frac{g^2}{2}\sum_{\rm links} \mathbf{E}_{l}^2 - \frac{2}{g^2} \sum_{\rm plaquettes} B_p \ .\\
\end{equation}
Here, $g$ is the adimensional coupling constant. $\mathbf{E}^2_l=\sum_J\alpha^J\hat{P}^J_l$ is the weighted sum of projectors $\hat{P}_J$ on the irreducible representations ({\em irreps}) $J$ of the group, where $\alpha^J$ is a function of the {\em irrep}.
This term represents the local electric field density on the link $l$ for Lie groups, for instance, $\alpha^J=J(J+1)$ if $\G=\mathrm{SU(2)}$.
The plaquette operator is $B_p = {\rm Tr}(U_{p_1} U_{p_2} U^\dagger_{p_3} U^\dagger_{p_4} + {\rm H.c.}) $, where $U_{p_i}$ is the parallel transporter, associated to a faithful {\em irrep}, acting on the ${p_i}-$th link (labeled anticlockwise) of the plaquette $p$. Physically, it corresponds to the action of a magnetic flux and the associated Aharonov--Bohm phase acquired by a charged particle on closed loops. The lattice spacing has been set to 1.

At each vertex ${\bf v}$ of the lattice $\Lambda$, we define the local gauge transformations associated to the group element $h\in\ G $ as $\Theta_{\bf v}(h)=\prod_i \theta^R_i(h) \prod_o \theta^L_o(h)$, where $i$ ($o$) labels the ingoing (outgoing) edges connected to the vertex ${\bf v}$. $\theta^{L/R}_l(h)$ is the left/right unitary transformation on the link $l$ associated to $h$. 
In Abelian groups, $\theta^{R}_l(h)=\theta^L_l(h^{-1})=\theta^{\dagger L}_l(h)$.
For a state to be physical, we require that $\Theta_{\bf v}(h)|\psi\rangle_{\rm phys}=|\psi \rangle_{\rm phys}\ \forall h \in \G, {\bf v} \in \Lambda$.
References~\cite{zohar2015formulation,Mariani_PRD2023, lamm2019prd} discuss in greater depth the formulations of LGTs for quantum simulations.

At strong coupling $g^2\gg 1$, the system is in a confined phase whose ground state is well approximated by a trivial product state with the lowest-energy {\em irrep}, typically the identity, on every link.
At weak coupling $g^2\ll 1$ 
\footnote{In extended (2+1)D geometries, the system can undergo a phase transition at a finite coupling, while in the ladder geometry it occurs only at $g_c^2=0$.},
the ground state minimizes the number of magnetic vortices by ``locking'' together the gauge field, in the group-element basis, on each link. This, combined with the gauge constraints, generates complex entangled states, whose specific structure depends on the symmetry group. 
In this regime, a variety of phenomena appear, such as deconfinement, topological order, spin-liquid phases, and anyonic excitations~\cite{fradkin1978,Nayak_RMP2008,Zohar2016,Munk_PRB2018,Gonzalezcuadra_PRX2020,Semeghini_Sci2021,Satzinger_Science2021,Lumia_PRXQ2022,Cataldi_PRR2024,Iqbal2024,Pradhan2024prd}.
 
We compare the ground-state resources of flux ladders with different gauge groups: SU(2), $\Dthree$, and $\mathbb{Z}_N$. SU(2)
represents an important intermediate step towards quantum simulations of QCD~\cite{Mezzacapo_PRL2015,Zohar2016,Cataldi_PRR2024,rahman2021su2,grabowska2024fully,Silvi2017finitedensityphase,Bauer_PRXQ2023,DiMeglio_PRXQ2024}.
Being a continuous Lie group, it requires some approximation to be represented on discrete variables. In this work, we will adopt the smallest nontrivial truncation of its {\em irreps} to $J=0$ and $J=1/2$, the hardcore-gluon approximation~\cite{Silvi2017finitedensityphase}.  
As for $\Dthree$, this is the smallest discrete non-Abelian group. This class of models gives the exciting prospect of exact implementation in state-of-the-art qudit-platforms~\cite{Gonzalezcuadra_PRL2022,Gustafson_PRD2022,Ballini_2024,Gaz_PRR2025}.
Moreover, dihedral groups are associated with non-Abelian topological order and universal anyons~\cite{kitaev2002topological,KITAEV2003,Iqbal2024}. To contrast these non-Abelian with Abelian symmetries, we further analyze $\mathbb{Z}_N$ models, well-studied workhorses that interpolate between $\mathbb{Z}_2$ (relevant, e.g., for the toric code~\cite{KITAEV2003}) and
U(1) (the gauge symmetry of QED)~\cite{Notarnicola_JPA2015,Magnifico_PRB2019,Magnifico_Quantum2020,meth2023simulating,calliari_2025}.

The local gauge symmetries allow for mapping these systems onto effective one-dimensional Hamiltonians. One way to achieve this is by identifying gauge-invariant operators and states defined on the dual lattice. We adopt this strategy for SU(2) and $\mathbb{Z}_N$, following the mapping from the flux ladder to a spin chain of Refs.~\cite{yao2023su,Pradhan2024prd}. 
For $\Dthree$, instead, because of the more complex fusion rules, we pursue a second strategy, which consists of fixing the gauge to remove redundant degrees of freedom on the ladder legs, leading to a one-dimensional chain with only dynamical degrees of freedom on the rungs~\cite{Munk_PRB2018}. 
Both mappings are sketched in Fig.~\ref{fig:figure1}(a-c), and we present further details in App.~\ref{app:models}.

\begin{figure*}[t!]
    \centering
    \includegraphics[width=\textwidth]{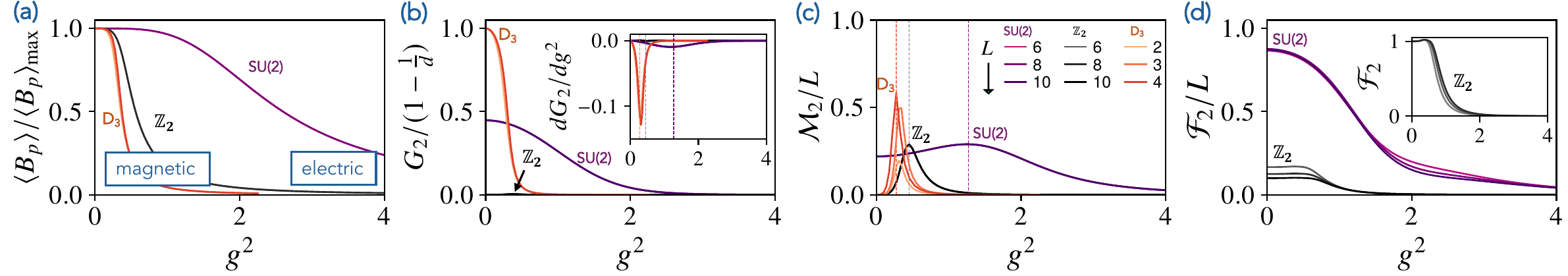}
    \caption{(a) Expectation value of the plaquette operator, (b) GGM, (c) SRE density, and (d) FAF density, computed for the ground states of SU(2), $\Ztwo$, and $\Dthree$ pure LGTs. 
    For all considered theories, results converge fast with increasing system size.  The derivative of $G_2$, see the inset of panel (b), peaks at the same position as $\mathcal{M}_2$.
    SU(2) displays sizable values of all quantum resources at small coupling, where quantum fluctuations proliferate. 
    In contrast, the studied LGTs with discrete groups display mixed regimes, which are easy in terms of one resource and hard for another.  FAF is only computed for SU(2) and $\Ztwo$, where the Jordan--Wigner transformation enables a clear mapping between Majorana and qubit operators. 
    }
    \label{fig:fig_two}
\end{figure*}

With these procedures, the effective Hamiltonians take the following forms:
\begin{align}
    H_{\rm SU(2)}  = & g^2 \sum_{i} \left[ -\frac{3}{2} Z_i Z_{i+1} + 3 Z_i\right] \nonumber \\ 
    & -\frac{1}{g^2} \sum_i (1-3Z_{i-1}) X_i (1-3Z_{i+1}) \label{eq: su2 ham},\\
    H_{\mathbb{Z}_N} = & - \frac{g^2}{2}\sum_i  \left[ Z_{i-1}^\dagger Z_i + (1 + \omega^k)Z_i + {\rm H.c} \right] \nonumber \\
    & + \frac{1}{2g^2} \sum_i \left( X_i + X^\dagger_i \right) \label{eq: zn ham}\ , \\
    H_{\Dthree} = & -g^2\sum_i\sum_J \alpha_J \left[ \hat{P}^J_i + 2 \prod_{i'<i} \hat{P}^J_{i'} \right] 
    \nonumber \\  
    & - \frac{1}{2g^2} \sum_i(\textrm{Tr}[U^\dagger_iU_{i+1}] + \textrm{H.c.}) \ .
    \label{eq: D3Ham}
\end{align}
For $H_{\mathrm{SU(2)}}$ and $H_{\mathbb{Z}_2}$, the Hamiltonians are represented by qubits using Pauli matrices, whereas, for $H_{\mathbb{Z}_N}$, the system is mapped to qudits with local dimension $d = N$, described by the $\mathbb{Z}_N$ clock operators $Z$ and $X$, which satisfy $ZX = \omega XZ$ with $\omega = e^{ 2\pi i/N}$, and index $k$ labels the superselection sector corresponding to a background field $\omega^k$. For the case of $ H_{\mathrm{D}_3}$, the operators are represented in qudits with $d = 6$.

\emph{Quantum resource measures}---The Generalized Geometric Measure (GGM) extends the geometric measure of entanglement to quantify genuine multipartite entanglement~\cite{shimony1995degree,barnum2001monotones,wei2003geometric, sende2010pra,biswas2014pra,das2016generalized}. Intuitively, it measures the minimal distance between a quantum state $\ket{\psi}$ and the set $\mathcal{S}_2$ of $2$-separable product states. 
The GGM is efficiently computable for pure states as 
\begin{equation} \label{eq: GGM_using_schmidt}
    G_2(\ket{\psi}) =1 - \max_{|\pi\rangle \in \mathcal{S}_2} |\langle \pi | \psi \rangle|^2= 1 - \max_{\mathcal{A}:\mathcal{B}} (\lambda^{\max}_{\mathcal{A}:\mathcal{B}})^2,
\end{equation}
where $\lambda^{\max}_{\mathcal{A}:\mathcal{B}}$ is the largest Schmidt coefficient across the bipartition $\mathcal{A}:\mathcal{B}$. The maximization runs over all nontrivial bipartitions of the system. The maximum possible value is $G_2^{\max} = 1 - 1/d$, where $d$ is the local Hilbert space dimension.

In classical stabilizer-tableau simulations as well as certain fault-tolerant quantum-computation schemes, entanglement is no longer a scarce resource; instead, magic states are. 
The Stabilizer Rényi Entropy (SRE) measures this resource by quantifying how a pure state $|\psi\rangle$ of $L$ qudits is distributed over the basis of Pauli strings, defined as $\mathcal{P}_L \equiv \{P_{\vec{v}^{(1)}} \otimes P_{\vec{v}^{(2)}} \otimes \cdots \otimes P_{\vec{v}^{(L)}}\}$. For qudits of local dimension $d$, each generalized Pauli operator takes the form $P_{\vec{v}} = X^{v_1} Z^{v_2}$, with $\vec{v} = (v_1, v_2) \in [0, d{-}1]^2$. 
The SRE is defined as~\cite{leone2022stabilizer}
\begin{equation}\label{eq: sre definition}
    \mathcal{M}_k(|\psi\rangle) = \frac{1}{1-k} \log \left[ \sum_{P \in \mathcal{P}_L} \frac{|\langle \psi | P | \psi \rangle|^{2k}}{d^L} \right].
\end{equation}
The SRE is non-negative, vanishes if and only if $|\psi\rangle$ is a stabilizer state~\cite{haug2023stabilizer, gross2021schurweylduality}, and is experimentally accessible~\cite{Niroula2024}.
Here, we focus on the second Rényi entropy, $\mathcal{M}_2$. 

Matchgate circuits, while classically simulatable \cite{jozsa2008matchgates}, become classically intractable when combined with fermionic non-Gaussian states, highlighting fermionic non-Gaussianity as a key resource \cite{hebenstreit2019all}.  It can be quantified by the fermionic anti-flatness (FAF)~\cite{sierant2025fermionic}, 
\begin{equation} \label{eq:faf}
    \mathcal{F}_k(\ket{\psi})= L-\frac{1}{2}{\rm tr}[(M^TM)^k],\  
    M_{mn}= -\frac{i}{2} \langle \psi|[\gamma_m, \gamma_n]|\psi\rangle\,,
\end{equation}
where $M$ is the covariance matrix of the $2L$ Majorana operators $\{\gamma_n\}$. 
The FAF vanishes if and only if $\psi$ is a fermionic Gaussian state, 
invariant under Gaussian unitaries, and (sub)additive for product states.
It is efficiently computable~\cite{collura2024quantum,sierant2025fermionic} and experimentally accessible~\cite{sierant2025fermionic}. 
Haar random states achieve nearly maximal values of all of the above resources~\cite {bera2020growth, turkeshi2025pauli, sierant2025fermionic}.

\emph{Resources and group structure}---All models considered display a crossover between a strong-coupling regime, where the energy of the electric field $\mathbf{E}$ is minimized, and a weak-coupling regime, where $\mathbf{E}$ has large fluctuations. 
This crossover can be identified via the expectation value of the plaquette operator $\langle B_p\rangle $, see Fig.~\ref{fig:fig_two}(a). 
Importantly, in the ladder geometry, we do not expect any appreciable finite-size dependence in either the order parameter or in the minimum energy gap (see App.~\ref{app:models_d3}), since the edges induce an effective longitudinal field that renders the two regimes adiabatically connected. The absence of a phase transition leads to a fast convergence of the considered observables, making the relevant physics accessible within numerically tractable system sizes~\footnote{In an extended (2+1)D system, the crossover becomes a continuous phase transition, at least for $\ZN$ LGTs that can be mapped to Potts models. For SU(2), there are indications of a phase transition~\cite{Cataldi_PRR2024}, but numerical limitations prevent a precise characterization of the critical behavior.}.

The crossover region between the electric and magnetic regimes is reflected in the behavior of all resource measures---GGM, SRE, and FAF---as shown
in Fig.~\ref{fig:fig_two}. 
In all three models, entanglement vanishes in the electric limit, see panel~(b). The reason is straightforward to see for $\Ztwo$ and SU(2), which both map to an Ising model with a longitudinal field, whose ground state is a product state. In the $\Dthree$ theory, the energy at strong coupling is minimized by setting all sites to the trivial {\em irrep}, resulting in a product state as well. 
In contrast, deep in the magnetic regime, entanglement vanishes only for the $\Ztwo$ theory, as all links become polarized by the local $X$ field~\footnote{This would not be the case in the full Hilbert space, where the ground-state at $g=0$ is an entangled quantum-spin liquid.}. 
In both non-Abelian theories, instead, the plaquette terms correlate different sites, leading to nonzero (in fact, their maximal) entanglement in the magnetic regime.

The picture changes when considering nonstabilizerness, shown in Fig.~\ref{fig:fig_two}(c), although a clear correlation with GGM remains. In discrete gauge groups, nonvanishing SRE appears only within the crossover regime where both terms compete: Deep in the magnetic region, the $\ZN$ LGT maps onto a simple $\ZN$ paramagnet, while the ground state of the $D_3$ theory takes the form of a generalized GHZ state in the group element basis. Both are stabilizer states. Deep in the electric region, the ground state is separable in the representation basis, also leading to vanishing SRE.
In the crossover between these two regimes, SRE assumes a maximum that qualitatively resembles the derivative of the GGM, shown in the inset of Fig.~\ref{fig:fig_two}(b).
In contrast, for SU(2), the SRE does peak in the crossover region, but it remains finite throughout the magnetic regime. This originates from the form of the SU(2) dual Hamiltonian: 
at $g\rightarrow 0$, it almost becomes a cluster Ising model~\cite{Smacchia_PRA2011}, which has a stabilizer ground state, but with further terms that destroy this structure. 
 
Figure~\ref{fig:fig_two}(d) shows the FAF for the $\Ztwo$ and SU(2) theories (both map to qubit systems where Eq.~\eqref{eq:faf} can be naturally applied using a Jordan--Wigner transformation). 
In both cases, the FAF shows qualitative similarity to $\langle B_p \rangle$, though with a stronger dependence on the system size. That is, it plateaus in the magnetic regime and quickly decreases once the system reaches the crossover region. 
A crucial difference emerges, though: In SU(2), we observe a fast convergence of the FAF density with system size at small $g$; in contrast, for $\Ztwo$ the plateau of the FAF density decreases as $1/L$ (i.e., the FAF itself is independent of $L$). 
Indeed, a Jordan--Wigner transformation maps the $\Ztwo$ Hamiltonian onto free fermions with only a small nonintegrable perturbation (the longitudinal field). The plaquette term of SU(2), instead, is not quadratic in fermionic operators, inducing a larger non-Gaussianity in the ground state.

\emph{Group order and superselection sectors}---To understand how sensitive the above findings are towards details of the theory (beyond its underlying gauge symmetry), we compare the above quantum resources for $\ZN$ LGTs with varying $N$ and superselection sector $k$, which labels the background charge that induces a uniform $\ZN$ electric field $\omega^k=e^{i2\pi k/N}$ on the boundaries of the ladder.
In the following, we show that the behavior of the ground state can be different, due to the degeneracies in the spectrum of Eq.~\eqref{eq: zn ham} appearing at specific combinations of $N$ and $k$.  

\begin{figure}
    \centering
    \includegraphics[width=\columnwidth]{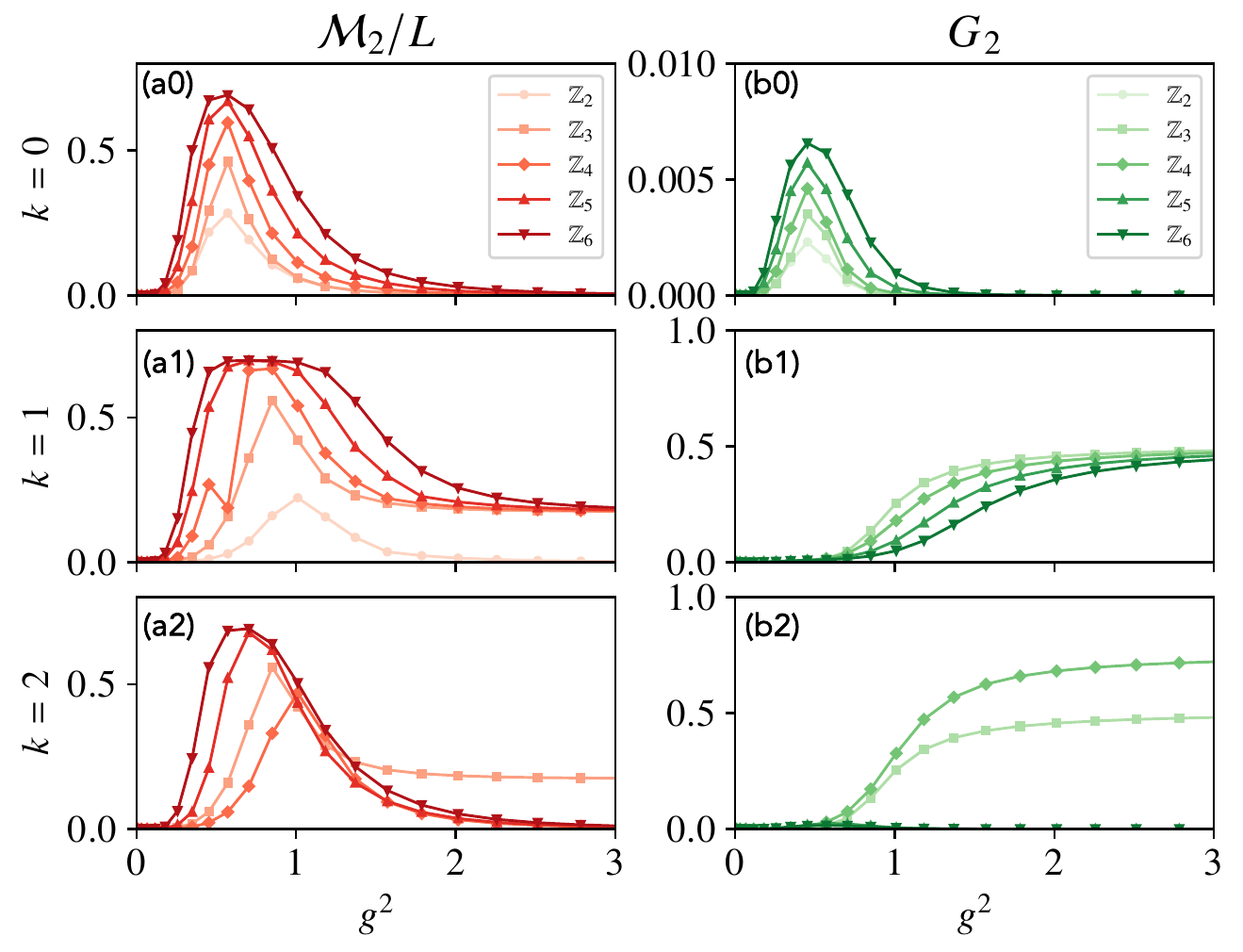}
    \caption{
    Impact of superselection sector $k$ and group order $N$ on quantum resources, illustrated for a $\ZN$ LGT, for $N = 2,3,4,5,6$ and $k=0,1,2$, with $L=4$ fixed. 
    (a) In all cases considered, SRE peaks when electric and magnetic terms compete. 
    (b) GGM has a significant peak in the crossover region only in the absence of background charges $k=0$. 
    Both quantities may reach a plateau at strong coupling, whose emergence and height critically depend on the combination of $k$ and $N$. 
    }
    \label{fig: resources ZN}
\end{figure}

In the $k=0$ sector (i.e., no background charges), the ground state of the $\ZN$ Hamiltonian 
is fully-polarized in both the strong- and weak-coupling limits, leading to vanishing nonstabilizerness and entanglement independently from $N$. 
Around the crossover, both quantities display a peak whose magnitude increases monotonically with $N$, see Fig.~\ref{fig: resources ZN}(a0,b0).

When adding background charges ($k>0$), the physics changes qualitatively. At weak coupling, the large electric fluctuations screen the background field, and the ground-state resource is independent of $k$. For large $g$, instead, the combination of $N$ and $k$ induces degeneracies in the spectrum, associated with regions of extensive SRE and GGM.

For $k=\frac{N}{2}$, $1+\omega^k=0$ and the longitudinal field vanishes~\cite{Pradhan2024prd}, restoring the global $\ZN$ symmetry. Its ground state at $g\to\infty$ is an entangled $\ZN$ GHZ state with $G_2=1-\frac{1}{N}$ but vanishing SRE. This is illustrated in Fig.~\ref{fig: resources ZN}(a1-b1) and (a2-b2) for a $\Ztwo(\mathbb{Z}_4)$ theory with $k=1(2)$. 
Besides this special value, the electric Hamiltonian has degenerate pairs of eigenvalues associated with angles $2\pi a/N$, $2\pi b/N$, where $a+b=N-k$ and $a,b=0,\dots, N-1$, see App.~\ref{app:zn}.
When the pair $\{\ket{a}, \ket{b}\}$ minimizes the electric Hamiltonian, the $g\to\infty$ ground state takes the form $\ket{\psi}=\frac{\ket{a}^{\otimes L}+\ket{b}^{\otimes L}}{\sqrt{2}}$. This is not a GHZ state except for $N=2$, leading to finite values of both entanglement,  $G_2=\frac{1}{2}$, and nonstabilizerness, $M_2=0.32$ (see App.~\ref{app: anal} for details). 
Figure~\ref{fig: resources ZN}(a1-b2) clearly showcases this phenomenology: At $k=1$, all $\ZN$ models except $\Ztwo$ display finite nonstabilizerness and entanglement at strong-coupling, consistent with the analytical prediction.
For $k=2$, only $\mathbb{Z}_3$ presents a two-fold-degenerate ground state with both nonzero GGM and SRE at large $g$, while  
$\mathbb{Z}_4$ falls into the case $k=\frac{N}{2}$ with finite entanglement but vanishing nonstabilizerness, and all other considered models have product ground states. 

\emph{Remarks on extended (2+1)D LGTs}--- 
We focused on ladder geometries, as extended (2+1)D systems pose a significant numerical challenge. 
Nonetheless, we can make some statements in limiting cases. 
At $g\to \infty$, the LGT Hamiltonian becomes $H = \frac{g^2}{2}\sum_{\rm links} \mathbf{E}_{l}^2$. Thus, without background charges and independently of $\G$, we expect a product state of polarized links, with vanishing resources.
In contrast, at $g\to 0$, the ground state for discrete groups can be derived from the associated surface code~\cite{KITAEV2003}. For the Abelian $\ZN$, the eigenvalues of the set of stabilizers (plaquette and vertex operators) uniquely define each eigenstate. Analogously to the $\Ztwo$ case, they have large entanglement (in the full Hilbert space) but vanishing SRE. Their dual representation typically modifies the entanglement structures, as plaquette operators become single-body terms, but not the nonstabilizerness, since the nonlocal transformations involve mapping between Pauli strings.  Non-Abelian topological order, instead, breaks this picture~\cite{Iqbal2024,TheChosenOne}, possibly leading to larger values of quantum resources independently of the chosen representation. 
On a ladder, this is already hinted at by ${\rm SU}(2)$, which maintains large $M_2$ at weak coupling even after the dual mapping. 
Plausibly, its complexity should increase in a full (2+1)D scenario, although it is less clear what to expect if we relax the truncation on its {\em irreps}.

\emph{Conclusions}---We have investigated the connection between the symmetry group in LGTs and ground-state complexity through quantum resources in (2+1)D pure-gauge theories.
An intricate picture emerges, where quantum resources depend on multiple factors, including the gauge group, background charges, and the strategy to project the Hamiltonian onto independent degrees of freedom.
Our results indicate that quantum simulations of LGTs display resource-demanding regimes for both NISQ and fault-tolerant scenarios, due to large entanglement (e.g., CNOT gate count) and nonstabilizerness (e.g., T-gate count), respectively. 
Perhaps counterintuitively, non-Abelian groups are not necessarily associated with larger nonstabilizerness, although they seem to induce larger entanglement than Abelian ones.

In $\ZN$ LGTs, the background charges in conjunction with the group order play a central role in determining the ground-state complexity, both for nonstabilizerness and entanglement. 
Our findings suggest that future quantum simulation experiments, both in terms of aiming for beyond-classical regimes and in estimating the required quantum resources, will have to carefully consider chosen superselection sectors, group orders, and coupling regimes. 

Our study raises a plethora of further questions, such as understanding how quantum resources depend on the presence of dynamical matter, the truncation of continuous groups, or the dimensionality of the lattice. 
It will also be interesting to delve deeper into the possible advantage offered by employing high-dimensional quantum-information carriers, qudits. Our data shows that both nonstabilizerness and multipartite entanglement increase in $\ZN$ LGTs with the group order. This effect may be exacerbated when breaking the theory down to qubits, instead of using qudit operators. On the flip side, current qudit hardware comes with increased control and accuracy cost and circuit implementation complexity. Comparing the respective resource requirements is, therefore, highly interesting. Such an analysis will be particularly relevant for large-dimensional groups or continuous groups with large truncation, where the encoding of the physical degrees of freedom in qubits or qudits drastically changes the implementation strategy.

\emph{Acknowledgements}---The authors are grateful to Michele Burrello and Ricardo Costa de Almeida for helpful discussions.
This project has received funding from the European Union’s Horizon Europe research and innovation programme under grant agreement No 101080086 NeQST, from the Italian Ministry of University and Research (MUR) through the FARE grant for the project DAVNE (Grant R20PEX7Y3A). This work was supported by the Dutch National Growth Fund (NGF), as part of the Quantum Delta NL Programme.
A.B.\ acknowledges funding from the Honda Research Institute Europe.
P.H. has further received funding from the Swiss State Secretariat for Education, Research and Innovation (SERI) under contract number UeMO19-5.1, from the QuantERA II Programme through the European Union’s Horizon 2020 research and innovation programme under Grant Agreement No 101017733, from the European Union under NextGenerationEU, PRIN 2022 Prot. n. 2022ATM8FY (CUP: E53D23002240006), from the European Union under NextGenerationEU via the ICSC – Centro Nazionale di Ricerca in HPC, Big Data and Quantum Computing.
Views and opinions expressed are however those of the author(s) only and do not necessarily reflect those of the European Union or the European Commission. Neither the European Union nor the granting authority can be held responsible for them.
This work was supported by Q@TN, the joint lab between the University of Trento, FBK—Fondazione Bruno Kessler, INFN—National Institute for Nuclear Physics, and CNR—National Research Council.

\appendix

\section{Details on the pure-gauge flux ladders}\label{app:models}
Kogut--Susskind LGT Hamiltonians on a (2+1)D lattice take the general form of Eq.~\eqref{eq:HKS_general}, which we rewrite for convenience
\begin{equation}
    \label{eq:HKSgeneral}
    H = \frac{g^2}{2}\sum_{\rm links} (E_{l})^2 - \frac{2}{g^2} \sum_{\rm plaquettes} B_p \ .
\end{equation}
The two competing terms, which in QED correspond to the electric- and magnetic-field contributions, have different structures and favor states with starkly opposite properties.
The ``electric'' contribution $E^2_l=\sum_J\alpha^J P^J_l$ is a single-body term that acts on individual gauge degrees of freedom and it is diagonal in the irreducible representations ({\em irreps}) of the group $J$. Typically, the energies $\alpha^J$ associated with each representation are derived from the group Laplacian, which is natural for Lie groups but less trivial for discrete ones~$\cite{Mariani_PRD2023}$.
With this construction, the lowest energy {\em irrep} is always the identity $J=e$, i.e., the one that transforms trivially under the group action. Being a one-dimensional {\em irrep}, it completely defines the ground state in the strong-coupling limit 
\begin{equation}
    \ket{GS(g\to\infty)}=\bigotimes_{\rm links}\ket{e}_l \ .
\end{equation}
Because $\theta^{L/R}(g)\ket{e}=\ket{e}$, this state is also gauge invariant. For discrete groups, it can be rewritten in the group elements basis as 
\begin{equation}
    \ket{GS(g\to\infty)}=\bigotimes_{\rm links}\frac{1}{\sqrt{|G|}}\sum_{g\in G} \ket{g}_l .
\end{equation}
Being a product state, it has no entanglement, and it has vanishing SRE both in representation and group-element basis because of the uniform superposition in the latter. Indeed, even for local dimension larger than 2, it remains an eigenstate of a generalized Pauli (clock) operator.

\begin{figure*}
    \begin{center}
        \includegraphics[width=\textwidth]{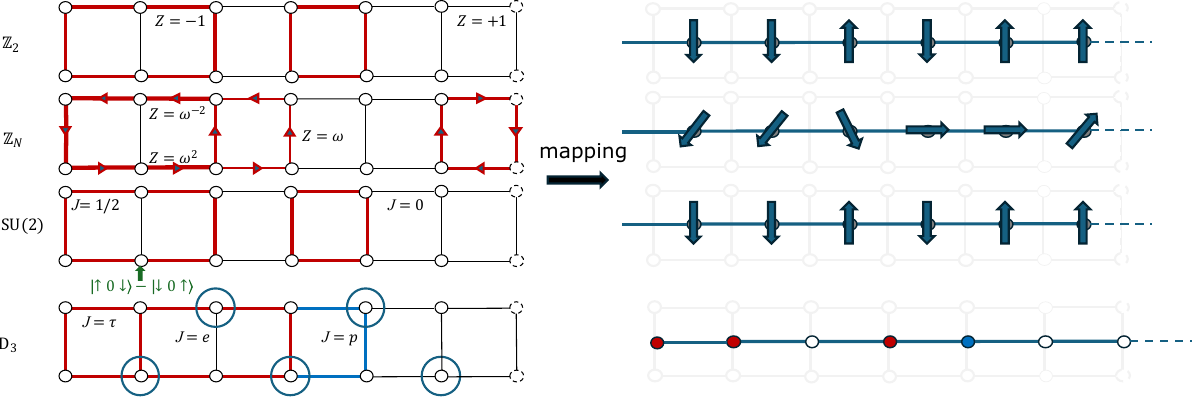}
        \caption{Examples of gauge-invariant configurations in the representation (electric) basis for the different models considered in this work. We highlight how these states appear after the mapping onto effective (1+1)D chains. The local basis for $\Ztwo$ and SU(2) is $\{\ket{\uparrow}, \ket{\downarrow}\}$, which correspond to the eigenstates of the Pauli-$Z$ operator. Their gauge-invariant states also share the same representation, while their Hamiltonians are markedly different.
        For $\ZN$, instead, we represent the eigenstates of the generalized phase ($Z$) operator with the corresponding angles, the $N$th-roots of unity, on the plane.
        For $\Dthree$, the circles identify the possible combinations of the three {\em irreps}---identity $e$, parity $p$, and fundamental $\tau$---into group-singlets. The internal structure of two-dimensional representations is not depicted.}
        \label{fig:lgtstates}
    \end{center}
\end{figure*}

The plaquette operator is the smallest Wilson loop $B_p = {\rm Tr}(U_1 U_2 U^\dagger_3 U^\dagger_4) + {\rm H.c.} $, consisting of the product of four gauge connections in a faithful representation of the group. These operators are diagonal in the group element basis ($\bra{h}U\ket{g} \propto \delta_{h,g}$) and act on the representation basis according to the Clebsch--Gordan coefficients (fusion rules) of the group.
Hence, they are not only responsible for four-body interactions but also can create a highly nontrivial entanglement structure in the GS depending on the specific symmetry group.
From the physical perspective, plaquette excitations correspond to magnetic vortices appearing in the system. The magnetic ground state, then, is associated with vanishing magnetic flux. This creates electric excitations, which, however, are not independent from one another: since they correspond to {\em irreps} that transform nontrivially under the gauge group ($J\neq e$), electric excitations around any given vertex are constrained by the gauge symmetry. They have to combine in such a way that the transformation on the vertex is trivial, i.e., they must form a singlet of the symmetry group. These generate the large entanglement structure appearing at weak coupling. The most celebrated example is the $\Ztwo$ Hamiltonian, which at $g\to 0$ corresponds to the toric code on a square lattice. The ground state is a uniform superposition of all possible Wilson loops,
\begin{equation}
    \ket{GS(g\to 0)}=\frac{1}{\sqrt{\mathcal{N}}}\sum_\Gamma \mathcal{W}_\Gamma \ket{GS(g\to\infty)}\ ,
\end{equation}
where $\mathcal{N}$ is the number of closed paths $\Gamma$ on the lattice and $\mathcal{W}_\Gamma=\bigotimes_{l\in\Gamma} X_l$ is corresponding Wilson loop.
A similar construction also holds for the other LGTs; the structure of the loops, however, becomes richer as it reflects the more intricate ways {\em irreps} can combine into singlets at each vertex. 
For instance, they can display bifurcations of electric fluxes, corresponding to vertices where an {\em irrep} ``splits'' into two or more nontrivial ones, following the group fusion rules.
Furthermore, representations with ${\rm dim}(J)>1$ can also have a nontrivial internal entanglement structure \footnote{Think of spin 1/2 singlets, for instance.} and contribute to increasing the ground state complexity in the magnetic phase.
Figure \ref{fig:lgtstates} reports examples of gauge-invariant states consisting of non-trivial electric loops for all models considered in this work. These states correspond to specific qubit or qudit configurations after the mapping onto the one-dimensional chains.
The circled vertices highlight how nontrivial {\em irreps} combine to form group singlets in non-Abelian theories.

The ladder geometry analyzed in this paper strongly constrains the shapes and dimensions of the Wilson loops appearing at weak coupling, but this general construction still holds.
Next, we will enter into more details of the three gauge groups considered in this work and their mapping onto one-dimensional chains.

\subsection{SU(2)}
For a SU(2) theory, making explicit the color-index $a$ in the non-Abelian electric field $E_l^a$, the pure-gauge Kogut--Susskind Hamiltonian of Eq.~\eqref{eq:HKSgeneral} becomes~\cite{kogut1975hamiltonian} 
\begin{equation}
    \label{eq:HKS}
    H = \frac{g^2}{2}\sum_{\rm links} (E_{l}^a)^2 - \frac{2}{g^2} \sum_{\rm plaquettes} B_p \\.
\end{equation}
Here, the SU(2) color-index $a$ is implicitly summed over. 
The plaquette operator is $B_p = {\rm Tr}(U_1 U_2 U^\dagger_3 U^\dagger_4 + {\rm H.c.}) $, where $U_l$ is the parallel transporter in the $J=1/2$ representation, acting on the $l-$th link of the plaquette. 
The gauge-field operators fulfill the canonical commutation relations
\begin{align}
    [L^a_l, U^{\alpha, \beta}_{l'}] & = -\delta_{l,l'}\frac{\sigma^a_{\alpha,\beta}}{2}U^{\beta,\gamma}_l  \ , \nonumber \\
    [R^a_l, U^{\alpha, \beta}_{l'}] & = \delta_{l,l'}U^{\alpha,\beta}_l\frac{\sigma^a_{\beta,\gamma}}{2} \ ,
\end{align}
where repeated indices are summed over. ${\bf L}_l$ and ${\bf R}_l$ are the left and right generators of the SU(2) gauge transformations and enter the electric field contribution as $|{\bf E}_l|^2 = |{\bf R}_l|^2=|{\bf L}_l|^2$.
    
The Hamiltonian in Eq.~\eqref{eq:HKS} is gauge invariant under local SU(2) transformations and fulfills the Gauss' laws $[H,G_{\bf v}]=0$ with generator of non-Abelian gauge transformations $G_{\bf v}=\sum_i {\bf L}_i +\sum_{o} {\bf R}_o$. $i$ and $o$ run over the ongoing and outgoing links, respectively, of the vertex ${\bf v}$.

In this work, we consider the case of a quasi-$(2+1)$D plaquette ladder, which has attracted recent interest in the context of digital quantum simulation and quantum annealers~\cite{zohar2015formulation,Shaw2020quantumalgorithms,klco2020su2,rahman2021su2, rahman2022self, grabowska2024fully, rahman2022realtime}.
By truncating the representations at $J=0$ and $J=1/2$ at each link, valid in the strong-coupling limit, it can be mapped to the Hamiltonian of an interacting spin-$1/2$ Hamiltonian~\cite{yao2023su}
\begin{align}
    \label{eq:SU2_chain}
    H_{\rm SU(2)} & = \sum_{i=0}^{N-1} \left[ h_{zz} Z_i Z_{i+1} + h_z Z_i + h_x (1-3  Z_{i-1}) X_i (1-3 Z_{i+1}) \right] \,.
\end{align}
The Pauli operator $Z_i$ counts the electric energy of the links around a given plaquette in the hardcore-gluon approximation, and $X_i$ corresponds to the plaquette operator that switches between the representations $J=0$ and $J=1/2$. 
The prefactors $(1-3 Z_{i-1})$ recover the correct Clebsch--Gordan coefficients of SU(2) when writing the group connection in the representation basis~\cite{zohar2015formulation,yao2023su}. 
The coefficients in Eq.~\eqref{eq:SU2_chain} are linked to the original LGT coupling $g$ through $h_{zz}=-3g^2/16$, $h_z=-2h_{zz}$, and $h_x=-1/8g^2$. 

The SU(2) plaquette ladder serves as a prototype of closely related, more complicated models. E.g., in the presence of dynamical fermions, a rich phase diagram emerges even in a (1+1)D chain as a function of the matter filling and of the matter--gauge-field coupling, including a meson BCS liquid phase, charge density waves, and tri-critical points compatible with a $\mathrm{SU}(2)_2$ Wess--Zumino--Novikov--Witten model~\cite{Silvi2017finitedensityphase}. In pure gauge SU(2) on an extended (2+1)D lattice, parametrizations of projected entangled-pair states show a transition between a gapped ``Higgs''-like and a gapless ``Coulomb''-like region~\cite{Zohar2016}, and mean-field calculations suggest a transition at $g=1$ in the loop configuration of the ground state~\cite{Raychowdhury2019}.

\subsection{\texorpdfstring{$\mathbb{Z}_N$}{ZN}}
\label{app:zn}
 \begin{table*}[t]
\centering
    \begin{tabular}{|c|c|c|c|c|c|}
        \hline
        $k,N$ & 2 & 3 & 4 & 5 &6 \\
        \hline
        0 &  & $\{\ket{1},\ket{2}\}$ & $\{\ket{1},\ket{3}\}$ & $\{\ket{1},\ket{4}\},\{\ket{2},\ket{3}\}$ & $\{\ket{1},\ket{5}\},\{\ket{2},\ket{4}\}$\\
        \hline
         1 & ${\color{red}\{\ket{0},\ket{1}\}}$ & $ {\color{red}\{\ket{0},\ket{2}\}}$ & $\{\ket{1},\ket{2}\},{\color{red}\{\ket{0},\ket{3}\}}$ & $\{\ket{0},\ket{4}\},{\color{red}\{\ket{1},\ket{3}\}}$ & $\{\ket{1},\ket{4}\},{\color{red}\{\ket{2},\ket{3}\}}$\\
        \hline
        2 &  & ${\color{red}\{\ket{0},\ket{1}\}}$ & ${\color{red}\{\ket{0},\ket{1},\ket{2},\ket{3}\}}$ & $\{\ket{1},\ket{2}\},\{\ket{0},\ket{3}\}$ & $\{\ket{0},\ket{4}\},\{\ket{1},\ket{3}\}$\\
        \hline
    \end{tabular}
\caption{Degenerate states for each pair $N,k$. Red highlighted are the pairs that dominantly contribute to the ground state at strong coupling.}
\label{tab: degeneracy}
\end{table*}

 $\mathbb{Z}_N$ LGTs on a plaquette ladder with periodic boundary conditions can be mapped onto one-dimensional clock models~\cite{Pradhan2024prd}.
 We start again from Eq.~\eqref{eq:HKS_general}. Being $\{ \ket{e_{k,l}}\}$ the electric basis on the link $l$, the action of the electric field operator gives $E_l \ket{e_{k,l}}=\omega^k\ket{e_{k,l}}$, where $\omega = \exp{(\frac{2\pi i}{N}})$. Labeling the upper, lower, and left rung legs of the $i-$th plaquette as 1, 2, and 0, respectively, the action of the plaquette operator reads $B_i = U_{i,2}U_{i+1,0}U_{i,1}^{\dagger}U_{i,0}^{\dagger}$, with $U_l \ket{e_{k,l}} = \ket{e_{k+1,l}}$. The commutation relations between the plaquette operators $B_i$ and the electric field $E_l$ read as
\begin{align}
    &B_iE_{i,2}= \omega^{-1} E_{i,2}B_i \quad B_iE_{i,1}= \omega E_{i,1}B_i \nonumber \\
    &E_{i,0}B_i=\omega^{-1}B_iE_{i,0} \quad E_{i,0}B_{i-1}=\omega B_{i-1}E_{i,0} \ .
    \label{eq: clock_comm_rules}
\end{align}
These commutation rules are conserved by the mapping on $\mathbb{Z}_N$ clock operators. We introduce the unitary matrices $X_i$ and $Z_i$, whose matrix elements are $(Z_i)_{mn} = \delta_{m,n}\omega^m$ and $(X_i)_{mn} = \delta_{m,n+1}$. Since $X_iZ_i=\omega^{-1}Z_iX_i$, one can verify that the last two relations in Eq.~\eqref{eq: clock_comm_rules} are satisfied by the following mappings
\begin{equation}
    B_i \rightarrow X_i \quad E_{i,0} \rightarrow Z_{i-1} Z_{i}^\dagger \ .
\end{equation}
By explicitly enforcing the Gauss constraints, it is possible to identify a mapping for the electric field operators on the upper and lower legs, such that
\begin{align}
    E_{i,1} \rightarrow Z_i^\dagger \quad E_{i,2} \rightarrow \omega^k Z_i \ ,
\end{align}
where $\omega^k$ identifies the background field, i.e., the superselection sector.
The transformed Hamiltonian can be written as
\begin{equation}\label{eq:H_ZN_app}
    H = - \sum_i \left[ \frac{g^2}{2} \left(Z_{i-1}^\dagger Z_i + (1 + \omega^k)Z_i\right) + \frac{1}{2g^2} X_i + \textrm{H.c.} \right] \ ,
\end{equation}
where $(1+\omega^k)Z_i$ depends on the superselection sector and represents a longitudinal field.

The structure of the ground state for the magnetic and electric phases can be found by looking at the $g\rightarrow0$ and $g\rightarrow +\infty$ limits.
In this dual representation, the weak-coupling limit becomes trivial, as the plaquette operators are reduced to a local transverse field. The magnetic ground state appears as a $\ZN$ paramagnet polarized in the $X$ direction. 

In the $g\rightarrow +\infty$ case (electric regime), the effective Hamiltonian consists of the first two terms in Eq.~\eqref{eq:H_ZN_app}. The term $\propto -Z_i^\dagger Z_{i+1}$ is minimized by product states of the form $\ket{a}^{\otimes L}$ in the computational basis, so we can expect its ground states to be degenerate with multiplicity $N$. The second term breaks this degeneracy and can be rewritten as 
\begin{equation}
    H_{Z}=-2\cos{\bigg( \frac{\pi k}{N}\bigg)}(\omega^{\frac{k}{2}}Z+\omega^{-\frac{k}{2}}Z^{\dagger}) \ ,
\end{equation}
which when acting on a basis state $\ket{a}$ has eigenvalue 
\begin{equation}
        E_a=-4\cos{\bigg( \frac{\pi k}{N}\bigg)}\cos{\bigg( \frac{2\pi}{N}\big(\frac{k}{2}+a\big)\bigg)} \ .
\end{equation}
When $k=\frac{N}{2}$, the contribution of the longitudinal term vanishes and the global $\Ztwo$ symmetry is not broken. Hence, the ground state is degenerate in the strong coupling regime.
Another possibility to obtain degenerate states is when the following equation is satisfied:
\begin{equation}
    \frac{1}{N}\big(\frac{k}{2}+j\big) = 1 - \frac{1}{N}\big(\frac{k}{2}+a\big) \rightarrow a+j = N - k \ .
\end{equation}
Notice that a degenerate state might not be the lowest energy state.
In Table \ref{tab: degeneracy}, we give list possible degeneracies for the cases considered in the main text, and highlight where they represent the dominant contribution to the ground state at large $g$.

\subsection{\texorpdfstring{$\Dthree$}{D3}}
\label{app:models_d3}
To analyze the nonstabilizerness in the case of finite non-Abelian LGTs, we study a $\Dthree$ LGT on the quasi-(2+1)D plaquette ladder. $\Dthree$, which represents the $6$ symmetries of an equilateral triangle, is the smallest non-Abelian group. Taking inspiration from Ref.~\cite{Munk_PRB2018}, the plaquette chain can be mapped on a (1+1)D system where each site represents one rung of the ladder, as depicted in Fig.~\ref{fig:figure1}(c). To do that, we apply a gauge transformation on each vertex, fixing the link on the left to the identity in the group element basis. Thanks to the gauge fixing, the plaquette term is simplified, going from a $4-$body term to a $2-$body term. Regarding the electric energy, instead, we will have two distinct terms. The first one is the usual electric term acting on the rungs, which reads
\begin{equation}
    H_E^{\textrm{rungs}} = -g^2\sum_i\sum_J \frac{\dim(J)}{|\G|}\varepsilon^J \hat{P}^J_i \ ,
\end{equation}
where $\hat{P}^J_i = \sum_h\chi^J(h)\theta^L_i(h) $ is the projection on the $J-$th irrep, $\chi^J(h)$ is the character of the group element $h$ in the $J-$th irrep, and $\varepsilon^J$ is the electric energy associated to each {\em irrep}. 
The prefactor $\frac{\dim(J)}{|\G|}\varepsilon^J$ is chosen such that $ \varepsilon^J$ are the eigenvalues of the electric Hamiltonian on a single rung.
Without loss of generality, we used $\theta^L$ when writing the projectors $P^J$, but one could equally well choose $\theta^R$.
The second term is due to the gauge fixing. Since the leg degrees of freedom are now fixed to the identity, we have to rewrite their electric field in terms of operators acting on the rungs only. This can be done by using explicitly the local symmetries, which generate a non-local term for every rung (because of the geometry of the ladder, the first rung is not affected by any terms). The full gauge-fixed Hamiltonian is
\begin{align}
    H_{\Dthree} = & -g^2\sum_i\sum_J\frac{\dim(J)}{|\G|}\varepsilon^J\hat{P}^J_i \nonumber\\
    & -2g^2\sum_i\sum_J\sum_h\frac{\dim(J)}{|\G|}\varepsilon^J\chi^J(h)\prod_{i'<i}\theta^L_{i'}(h) \nonumber\\
    & -\frac{1}{2g^2}\sum_i \left(\textrm{Tr}[U^\dagger_iU_{i+1}] + \textrm{H.c.}\right) \ .
\end{align}
By defining $\alpha^J = \frac{\dim(J)}{|\G|}\varepsilon^J$, and rewriting $\sum_h\chi^J(h)\prod_{i'<i}\theta_{i'}(h)$ as
\begin{equation}
    \prod_{i^\prime<i}\sum_h\chi^J(h)\theta^L_{i^\prime}(h) =\prod_{i^\prime<i}\hat{P}^J_{i^\prime} \ ,
\end{equation}
it is possible to write $H_{\Dthree}$ as in Eq.~\eqref{eq: D3Ham}.

\begin{figure}
    \centering
    \includegraphics[width=\columnwidth]{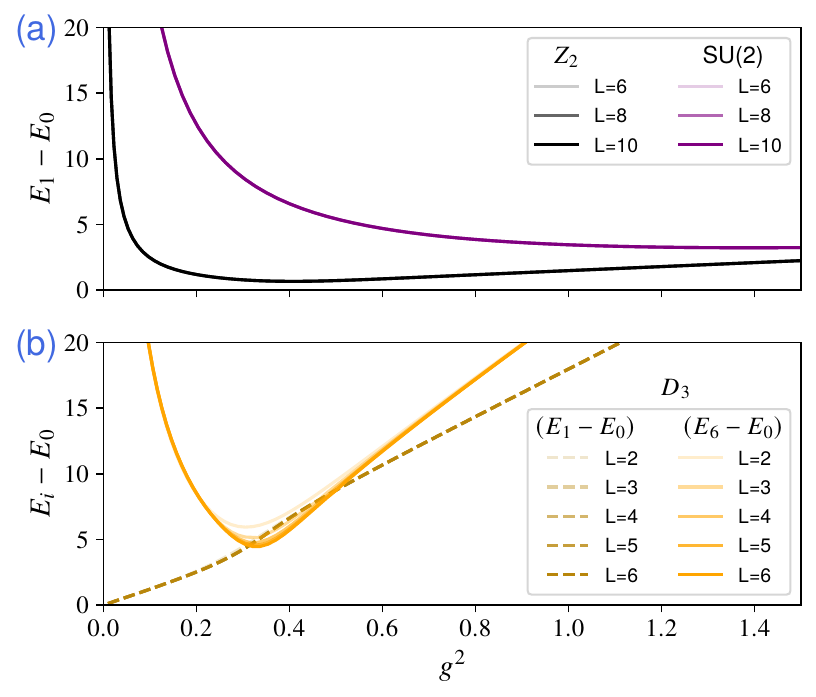}
    \caption{Energy gaps vs coupling $g^2$ for three different models considered in this work: $\Ztwo$, SU(2) (a), and $\Dthree$ (b). Neither displays significant finite-size scaling, indicating the lack of a phase transition but rather a smooth crossover between the electric and magnetic regimes.
    The panel (b) highlights the level crossing between different symmetry sectors of the global $\Dthree$  gauge transformation. Solid lines indicate the gap with the first excited states belonging to the same symmetry sector of the ground state, while dashed lines point to the manifold that transforms nontrivially under the global $\Dthree$ symmetry and becomes degenerate with the ground state at $g^2=0$.}
    \label{fig:gap_D3}
\end{figure}

For the results shown in the main text, we fixed $\varepsilon = (1,0,0)$ for the trivial, the parity, and the fundamental representation, respectively.
In Fig.~\ref{fig:gap_D3}(b), we depict the gap of the theory as a function of coupling strength. 
The system still retains a global $\Dthree$ gauge symmetry, so that each energy eigenstate transforms as one of the {\em irreps} of the group. This induces a $|\Dthree| = 6$-fold degeneracy in the ground state at $g^2=0$. At finite coupling, the ground state transforms trivially under the global symmetry, and the relevant energy gap is that between the ground state $E_0$ and the first excited state belonging to the same symmetry sector, $E_6$. The energy gap $E_6-E_0$ decreases with system size in the crossover regime but converges quickly to a finite minimum value, which occurs around a level crossing between different symmetry sectors. This behavior illustrates the absence of a phase transition in the thermodynamic limit, as discussed in the main text, but rather the magnetic and electric regimes remain smoothly connected.

Figure~\ref{fig:gap_D3}(a) shows the energy gap also for $\Ztwo$ and SU(2) models, where we do not observe a significant scaling with $L$ as well.

\section{Analytical computations of SRE for \texorpdfstring{$Z_N$}{ZN} LGTs at strong coupling} \label{app: anal}

In $\ZN$ LGTs with a background field, degeneracies in the spectrum at strong coupling may appear for specific combinations of $N$ and $k$, as explained in Appendix~\ref{app:zn}. See table~\ref{tab: degeneracy} for a summary.
In such cases, the electric ($g^2\gg 1$) ground state for $L$ qudits takes the form 
\begin{equation}
    \ket{\psi(g\to\infty)} = \frac{\ket{a}^{\otimes L}+\ket{b}^{\otimes L}}{\sqrt{2}}\,.
\end{equation}
For this state, we can compute $\sum_{P_i}|\braket{\psi|P_i|\psi}|^4$. 
We start by evaluating the expectation value of a single Pauli string $P_i =\otimes_{j=1}^L X_j^{r_j}Z_z^{s_j}$ is equal to 
\begin{align}
  \braket{\psi|P_i|\psi} = &\frac{1}{2} \bigg(\langle a |^{\otimes n} P_i \ket{a}^{\otimes n} +\langle b |^{\otimes n}  P_i \ket{b}^{\otimes n} \nonumber \\
  +& \langle a |^{\otimes n} P_i \ket{b}^{\otimes n}+ \langle b|^{\otimes n}P_i \ket{a}^{\otimes n}\bigg)  \\
=& \frac{1}{2} \left[ \prod_{j=1}^L\delta_{0,r_j}\left(\omega^{a\sum s_j} + \omega^{b\sum s_j}\right) \right.\nonumber \\
+ & \left. \prod_{j=1}^L \delta_{b,r_j\oplus a} \omega^{a\sum s_j}+ \prod_{j=1}^L \delta_{a,r_j\oplus b} \omega^{b\sum s_j} \right] \ .
\end{align}
Here, we used $Z^s\ket{a}=\omega^{as}\ket{a}$ and $\bra{b}X^r\ket{a}=\delta_{b,r\oplus a} $, where $\oplus$ is the summation modulo $N$ and $\omega = e^{\frac{2\pi i }{N}}$.
We can now sum over all possible Pauli strings
\begin{align}\label{eq:b4}   & \sum_{P_i}\left| \braket{\psi|P_i|\psi}\right|^4 = \frac{1}{2^4}\sum_{s_1,...,s_L=0}^{N}\left|\omega^{a\sum_{j=1}^{L}s_j}+\omega^{b\sum_{j=1}^{L}s_j}\right|^4\nonumber\\
    +&\frac{1}{2^4}\sum_{s_1,...,s_L=0}^{N}\sum_{r_1,...,r_L=1}^{N}\left|\omega^{a\sum_{j=1}^{L}s_j}\prod_{j=1}^{L}\delta_{b,r_j\oplus a}+\prod_{j=1}^{L}\delta_{a,r_j\oplus b}\omega^{b\sum_{j=1}^{L}s_j}\right|^4 \ ,
\end{align}
where we used the fact that different Kronecker deltas can not be satisfied simultaneously to split $|\braket{\psi|P_i|\psi}|^4$ into two distinct summations.
In the second line of Eq.~\eqref{eq:b4}, we can repeat the same trick, considering that the two deltas are simultaneously satisfied only if $2r=mN$, ($m$ integer), meaning that the angles $a$ and $b$ can differ at most by $\pi$. Despite the fact that they can be degenerate, such a pair can not constitute the ground state of the $\ZN$ theory at strong coupling, where the states aligned close to the background field are favored.
We can now simplify the expression by introducing an auxiliary variable $x=\sum_{j=1}^L s_j \in [0, L(N-1)]$ to write
\begin{align}
\sum_{P_i}\left| \braket{\psi|P_i|\psi}\right|^4
    =&\sum_{x = 0}^{L(N-1)} \frac{1}{2^4}c_x \left| \omega^{ax}+\omega^{bx} \right|^4+c_x \nonumber \\ 
    =&\sum_{x = 0}^{L(N-1)} c_x\bigg[\cos^4\bigg( \frac{\pi d x}{N}\bigg)+1\bigg]\,,
    \label{eq:analytic_nonstab}
\end{align}
where $c_x= \sum_{k=0}^{\frac{x}{d}}(-1)^k\binom{L}{k}\binom{x-kd+L-1}{L-1}$ is the multiplicity of the variable $x$ and $d=b-a$. 
For the states considered ($N = 2,...,6$ and $k=1$), using Eq.~\eqref{eq:analytic_nonstab} in Eq.~\eqref{eq: sre definition} gives a value for the SRE of $M_2=0.32$


\begin{thebibliography}{106}%
\makeatletter
\providecommand \@ifxundefined [1]{%
 \@ifx{#1\undefined}
}%
\providecommand \@ifnum [1]{%
 \ifnum #1\expandafter \@firstoftwo
 \else \expandafter \@secondoftwo
 \fi
}%
\providecommand \@ifx [1]{%
 \ifx #1\expandafter \@firstoftwo
 \else \expandafter \@secondoftwo
 \fi
}%
\providecommand \natexlab [1]{#1}%
\providecommand \enquote  [1]{``#1''}%
\providecommand \bibnamefont  [1]{#1}%
\providecommand \bibfnamefont [1]{#1}%
\providecommand \citenamefont [1]{#1}%
\providecommand \href@noop [0]{\@secondoftwo}%
\providecommand \href [0]{\begingroup \@sanitize@url \@href}%
\providecommand \@href[1]{\@@startlink{#1}\@@href}%
\providecommand \@@href[1]{\endgroup#1\@@endlink}%
\providecommand \@sanitize@url [0]{\catcode `\\12\catcode `\$12\catcode `\&12\catcode `\#12\catcode `\^12\catcode `\_12\catcode `\%12\relax}%
\providecommand \@@startlink[1]{}%
\providecommand \@@endlink[0]{}%
\providecommand \url  [0]{\begingroup\@sanitize@url \@url }%
\providecommand \@url [1]{\endgroup\@href {#1}{\urlprefix }}%
\providecommand \urlprefix  [0]{URL }%
\providecommand \Eprint [0]{\href }%
\providecommand \doibase [0]{https://doi.org/}%
\providecommand \selectlanguage [0]{\@gobble}%
\providecommand \bibinfo  [0]{\@secondoftwo}%
\providecommand \bibfield  [0]{\@secondoftwo}%
\providecommand \translation [1]{[#1]}%
\providecommand \BibitemOpen [0]{}%
\providecommand \bibitemStop [0]{}%
\providecommand \bibitemNoStop [0]{.\EOS\space}%
\providecommand \EOS [0]{\spacefactor3000\relax}%
\providecommand \BibitemShut  [1]{\csname bibitem#1\endcsname}%
\let\auto@bib@innerbib\@empty
\bibitem [{\citenamefont {Hauke}\ \emph {et~al.}(2012)\citenamefont {Hauke}, \citenamefont {Cucchietti}, \citenamefont {Tagliacozzo}, \citenamefont {Deutsch},\ and\ \citenamefont {Lewenstein}}]{Hauke_RepProgPhys2012}%
  \BibitemOpen
  \bibfield  {author} {\bibinfo {author} {\bibfnamefont {P.}~\bibnamefont {Hauke}}, \bibinfo {author} {\bibfnamefont {F.~M.}\ \bibnamefont {Cucchietti}}, \bibinfo {author} {\bibfnamefont {L.}~\bibnamefont {Tagliacozzo}}, \bibinfo {author} {\bibfnamefont {I.}~\bibnamefont {Deutsch}},\ and\ \bibinfo {author} {\bibfnamefont {M.}~\bibnamefont {Lewenstein}},\ }\bibfield  {title} {\bibinfo {title} {Can one trust quantum simulators?},\ }\href {https://doi.org/10.1088/0034-4885/75/8/082401} {\bibfield  {journal} {\bibinfo  {journal} {Reports on Progress in Physics}\ }\textbf {\bibinfo {volume} {75}},\ \bibinfo {pages} {082401} (\bibinfo {year} {2012})}\BibitemShut {NoStop}%
\bibitem [{\citenamefont {Georgescu}\ \emph {et~al.}(2014)\citenamefont {Georgescu}, \citenamefont {Ashhab},\ and\ \citenamefont {Nori}}]{Georgescu_RMP2014}%
  \BibitemOpen
  \bibfield  {author} {\bibinfo {author} {\bibfnamefont {I.~M.}\ \bibnamefont {Georgescu}}, \bibinfo {author} {\bibfnamefont {S.}~\bibnamefont {Ashhab}},\ and\ \bibinfo {author} {\bibfnamefont {F.}~\bibnamefont {Nori}},\ }\bibfield  {title} {\bibinfo {title} {Quantum simulation},\ }\href {https://doi.org/10.1103/RevModPhys.86.153} {\bibfield  {journal} {\bibinfo  {journal} {Rev. Mod. Phys.}\ }\textbf {\bibinfo {volume} {86}},\ \bibinfo {pages} {153} (\bibinfo {year} {2014})}\BibitemShut {NoStop}%
\bibitem [{\citenamefont {Daley}\ \emph {et~al.}(2022)\citenamefont {Daley}, \citenamefont {Bloch}, \citenamefont {Kokail}, \citenamefont {Flannigan}, \citenamefont {Pearson}, \citenamefont {Troyer},\ and\ \citenamefont {Zoller}}]{Daley2022}%
  \BibitemOpen
  \bibfield  {author} {\bibinfo {author} {\bibfnamefont {A.~J.}\ \bibnamefont {Daley}}, \bibinfo {author} {\bibfnamefont {I.}~\bibnamefont {Bloch}}, \bibinfo {author} {\bibfnamefont {C.}~\bibnamefont {Kokail}}, \bibinfo {author} {\bibfnamefont {S.}~\bibnamefont {Flannigan}}, \bibinfo {author} {\bibfnamefont {N.}~\bibnamefont {Pearson}}, \bibinfo {author} {\bibfnamefont {M.}~\bibnamefont {Troyer}},\ and\ \bibinfo {author} {\bibfnamefont {P.}~\bibnamefont {Zoller}},\ }\bibfield  {title} {\bibinfo {title} {Practical quantum advantage in quantum simulation},\ }\href {https://doi.org/10.1038/s41586-022-04940-6} {\bibfield  {journal} {\bibinfo  {journal} {Nature}\ }\textbf {\bibinfo {volume} {607}},\ \bibinfo {pages} {667} (\bibinfo {year} {2022})}\BibitemShut {NoStop}%
\bibitem [{\citenamefont {Dalmonte}\ and\ \citenamefont {Montangero}(2016)}]{dalmonte2016}%
  \BibitemOpen
  \bibfield  {author} {\bibinfo {author} {\bibfnamefont {M.}~\bibnamefont {Dalmonte}}\ and\ \bibinfo {author} {\bibfnamefont {S.}~\bibnamefont {Montangero}},\ }\bibfield  {title} {\bibinfo {title} {Lattice gauge theory simulations in the quantum information era},\ }\href {https://doi.org/10.1080/00107514.2016.1151199} {\bibfield  {journal} {\bibinfo  {journal} {Contemporary Physics}\ }\textbf {\bibinfo {volume} {57}},\ \bibinfo {pages} {388} (\bibinfo {year} {2016})}\BibitemShut {NoStop}%
\bibitem [{\citenamefont {Preskill}(2018)}]{Preskill_arxiv2018}%
  \BibitemOpen
  \bibfield  {author} {\bibinfo {author} {\bibfnamefont {J.}~\bibnamefont {Preskill}},\ }\href@noop {} {\bibinfo {title} {Simulating quantum field theory with a quantum computer}} (\bibinfo {year} {2018}),\ \Eprint {https://arxiv.org/abs/1811.10085} {arXiv:1811.10085} \BibitemShut {NoStop}%
\bibitem [{\citenamefont {Aidelsburger}\ \emph {et~al.}(2022)\citenamefont {Aidelsburger}, \citenamefont {Barbiero}, \citenamefont {Bermudez}, \citenamefont {Chanda}, \citenamefont {Dauphin}, \citenamefont {González-Cuadra}, \citenamefont {Grzybowski}, \citenamefont {Hands}, \citenamefont {Jendrzejewski}, \citenamefont {Jünemann}, \citenamefont {Juzeliūnas}, \citenamefont {Kasper}, \citenamefont {Piga}, \citenamefont {Ran}, \citenamefont {Rizzi}, \citenamefont {Sierra}, \citenamefont {Tagliacozzo}, \citenamefont {Tirrito}, \citenamefont {Zache}, \citenamefont {Zakrzewski}, \citenamefont {Zohar},\ and\ \citenamefont {Lewenstein}}]{Aidelsburger_LGT2021}%
  \BibitemOpen
  \bibfield  {author} {\bibinfo {author} {\bibfnamefont {M.}~\bibnamefont {Aidelsburger}}, \bibinfo {author} {\bibfnamefont {L.}~\bibnamefont {Barbiero}}, \bibinfo {author} {\bibfnamefont {A.}~\bibnamefont {Bermudez}}, \bibinfo {author} {\bibfnamefont {T.}~\bibnamefont {Chanda}}, \bibinfo {author} {\bibfnamefont {A.}~\bibnamefont {Dauphin}}, \bibinfo {author} {\bibfnamefont {D.}~\bibnamefont {González-Cuadra}}, \bibinfo {author} {\bibfnamefont {P.~R.}\ \bibnamefont {Grzybowski}}, \bibinfo {author} {\bibfnamefont {S.}~\bibnamefont {Hands}}, \bibinfo {author} {\bibfnamefont {F.}~\bibnamefont {Jendrzejewski}}, \bibinfo {author} {\bibfnamefont {J.}~\bibnamefont {Jünemann}}, \bibinfo {author} {\bibfnamefont {G.}~\bibnamefont {Juzeliūnas}}, \bibinfo {author} {\bibfnamefont {V.}~\bibnamefont {Kasper}}, \bibinfo {author} {\bibfnamefont {A.}~\bibnamefont {Piga}}, \bibinfo {author} {\bibfnamefont {S.-J.}\ \bibnamefont {Ran}}, \bibinfo {author} {\bibfnamefont {M.}~\bibnamefont {Rizzi}}, \bibinfo {author}
  {\bibfnamefont {G.}~\bibnamefont {Sierra}}, \bibinfo {author} {\bibfnamefont {L.}~\bibnamefont {Tagliacozzo}}, \bibinfo {author} {\bibfnamefont {E.}~\bibnamefont {Tirrito}}, \bibinfo {author} {\bibfnamefont {T.~V.}\ \bibnamefont {Zache}}, \bibinfo {author} {\bibfnamefont {J.}~\bibnamefont {Zakrzewski}}, \bibinfo {author} {\bibfnamefont {E.}~\bibnamefont {Zohar}},\ and\ \bibinfo {author} {\bibfnamefont {M.}~\bibnamefont {Lewenstein}},\ }\bibfield  {title} {\bibinfo {title} {Cold atoms meet lattice gauge theory},\ }\href {https://doi.org/10.1098/rsta.2021.0064} {\bibfield  {journal} {\bibinfo  {journal} {Philosophical Transactions of the Royal Society A: Mathematical, Physical and Engineering Sciences}\ }\textbf {\bibinfo {volume} {380}},\ \bibinfo {pages} {20210064} (\bibinfo {year} {2022})}\BibitemShut {NoStop}%
\bibitem [{\citenamefont {Zohar}(2022)}]{Zohar_PhilTransA2022}%
  \BibitemOpen
  \bibfield  {author} {\bibinfo {author} {\bibfnamefont {E.}~\bibnamefont {Zohar}},\ }\bibfield  {title} {\bibinfo {title} {Quantum simulation of lattice gauge theories in more than one space dimension-requirements, challenges and methods},\ }\href {https://doi.org/10.1098/rsta.2021.0069} {\bibfield  {journal} {\bibinfo  {journal} {Philosophical Transactions of the Royal Society A: Mathematical, Physical and Engineering Sciences}\ }\textbf {\bibinfo {volume} {380}},\ \bibinfo {pages} {20210069} (\bibinfo {year} {2022})}\BibitemShut {NoStop}%
\bibitem [{\citenamefont {Halimeh}\ \emph {et~al.}(2025)\citenamefont {Halimeh}, \citenamefont {Aidelsburger}, \citenamefont {Grusdt}, \citenamefont {Hauke},\ and\ \citenamefont {Yang}}]{halimeh2025cold}%
  \BibitemOpen
  \bibfield  {author} {\bibinfo {author} {\bibfnamefont {J.~C.}\ \bibnamefont {Halimeh}}, \bibinfo {author} {\bibfnamefont {M.}~\bibnamefont {Aidelsburger}}, \bibinfo {author} {\bibfnamefont {F.}~\bibnamefont {Grusdt}}, \bibinfo {author} {\bibfnamefont {P.}~\bibnamefont {Hauke}},\ and\ \bibinfo {author} {\bibfnamefont {B.}~\bibnamefont {Yang}},\ }\bibfield  {title} {\bibinfo {title} {Cold-atom quantum simulators of gauge theories},\ }\href {https://www.nature.com/articles/s41567-024-02721-8} {\bibfield  {journal} {\bibinfo  {journal} {Nature Physics}\ }\textbf {\bibinfo {volume} {21}},\ \bibinfo {pages} {25} (\bibinfo {year} {2025})}\BibitemShut {NoStop}%
\bibitem [{\citenamefont {Klco}\ \emph {et~al.}(2020)\citenamefont {Klco}, \citenamefont {Savage},\ and\ \citenamefont {Stryker}}]{klco2020su2}%
  \BibitemOpen
  \bibfield  {author} {\bibinfo {author} {\bibfnamefont {N.}~\bibnamefont {Klco}}, \bibinfo {author} {\bibfnamefont {M.~J.}\ \bibnamefont {Savage}},\ and\ \bibinfo {author} {\bibfnamefont {J.~R.}\ \bibnamefont {Stryker}},\ }\bibfield  {title} {\bibinfo {title} {{SU(2)} non-abelian gauge field theory in one dimension on digital quantum computers},\ }\href {https://doi.org/10.1103/PhysRevD.101.074512} {\bibfield  {journal} {\bibinfo  {journal} {Phys. Rev. D}\ }\textbf {\bibinfo {volume} {101}},\ \bibinfo {pages} {074512} (\bibinfo {year} {2020})}\BibitemShut {NoStop}%
\bibitem [{\citenamefont {Bauer}\ \emph {et~al.}(2023)\citenamefont {Bauer}, \citenamefont {Davoudi}, \citenamefont {Balantekin}, \citenamefont {Bhattacharya}, \citenamefont {Carena}, \citenamefont {de~Jong}, \citenamefont {Draper}, \citenamefont {El-Khadra}, \citenamefont {Gemelke}, \citenamefont {Hanada}, \citenamefont {Kharzeev}, \citenamefont {Lamm}, \citenamefont {Li}, \citenamefont {Liu}, \citenamefont {Lukin}, \citenamefont {Meurice}, \citenamefont {Monroe}, \citenamefont {Nachman}, \citenamefont {Pagano}, \citenamefont {Preskill}, \citenamefont {Rinaldi}, \citenamefont {Roggero}, \citenamefont {Santiago}, \citenamefont {Savage}, \citenamefont {Siddiqi}, \citenamefont {Siopsis}, \citenamefont {Van~Zanten}, \citenamefont {Wiebe}, \citenamefont {Yamauchi}, \citenamefont {Yeter-Aydeniz},\ and\ \citenamefont {Zorzetti}}]{Bauer_PRXQ2023}%
  \BibitemOpen
  \bibfield  {author} {\bibinfo {author} {\bibfnamefont {C.~W.}\ \bibnamefont {Bauer}}, \bibinfo {author} {\bibfnamefont {Z.}~\bibnamefont {Davoudi}}, \bibinfo {author} {\bibfnamefont {A.~B.}\ \bibnamefont {Balantekin}}, \bibinfo {author} {\bibfnamefont {T.}~\bibnamefont {Bhattacharya}}, \bibinfo {author} {\bibfnamefont {M.}~\bibnamefont {Carena}}, \bibinfo {author} {\bibfnamefont {W.~A.}\ \bibnamefont {de~Jong}}, \bibinfo {author} {\bibfnamefont {P.}~\bibnamefont {Draper}}, \bibinfo {author} {\bibfnamefont {A.}~\bibnamefont {El-Khadra}}, \bibinfo {author} {\bibfnamefont {N.}~\bibnamefont {Gemelke}}, \bibinfo {author} {\bibfnamefont {M.}~\bibnamefont {Hanada}}, \bibinfo {author} {\bibfnamefont {D.}~\bibnamefont {Kharzeev}}, \bibinfo {author} {\bibfnamefont {H.}~\bibnamefont {Lamm}}, \bibinfo {author} {\bibfnamefont {Y.-Y.}\ \bibnamefont {Li}}, \bibinfo {author} {\bibfnamefont {J.}~\bibnamefont {Liu}}, \bibinfo {author} {\bibfnamefont {M.}~\bibnamefont {Lukin}}, \bibinfo {author} {\bibfnamefont
  {Y.}~\bibnamefont {Meurice}}, \bibinfo {author} {\bibfnamefont {C.}~\bibnamefont {Monroe}}, \bibinfo {author} {\bibfnamefont {B.}~\bibnamefont {Nachman}}, \bibinfo {author} {\bibfnamefont {G.}~\bibnamefont {Pagano}}, \bibinfo {author} {\bibfnamefont {J.}~\bibnamefont {Preskill}}, \bibinfo {author} {\bibfnamefont {E.}~\bibnamefont {Rinaldi}}, \bibinfo {author} {\bibfnamefont {A.}~\bibnamefont {Roggero}}, \bibinfo {author} {\bibfnamefont {D.~I.}\ \bibnamefont {Santiago}}, \bibinfo {author} {\bibfnamefont {M.~J.}\ \bibnamefont {Savage}}, \bibinfo {author} {\bibfnamefont {I.}~\bibnamefont {Siddiqi}}, \bibinfo {author} {\bibfnamefont {G.}~\bibnamefont {Siopsis}}, \bibinfo {author} {\bibfnamefont {D.}~\bibnamefont {Van~Zanten}}, \bibinfo {author} {\bibfnamefont {N.}~\bibnamefont {Wiebe}}, \bibinfo {author} {\bibfnamefont {Y.}~\bibnamefont {Yamauchi}}, \bibinfo {author} {\bibfnamefont {K.}~\bibnamefont {Yeter-Aydeniz}},\ and\ \bibinfo {author} {\bibfnamefont {S.}~\bibnamefont {Zorzetti}},\ }\bibfield  {title}
  {\bibinfo {title} {Quantum simulation for high-energy physics},\ }\href {https://doi.org/10.1103/PRXQuantum.4.027001} {\bibfield  {journal} {\bibinfo  {journal} {PRX Quantum}\ }\textbf {\bibinfo {volume} {4}},\ \bibinfo {pages} {027001} (\bibinfo {year} {2023})}\BibitemShut {NoStop}%
\bibitem [{\citenamefont {Kitaev}(2002)}]{kitaev2002topological}%
  \BibitemOpen
  \bibfield  {author} {\bibinfo {author} {\bibfnamefont {A.}~\bibnamefont {Kitaev}},\ }\bibfield  {title} {\bibinfo {title} {Topological quantum codes and anyons},\ }in\ \href {https://userpages.cs.umbc.edu/lomonaco/ams/lecturenotes/Kitaev.pdf} {\emph {\bibinfo {booktitle} {Proceedings of Symposia in Applied Mathematics}}},\ Vol.~\bibinfo {volume} {58}\ (\bibinfo {year} {2002})\ pp.\ \bibinfo {pages} {267--272}\BibitemShut {NoStop}%
\bibitem [{\citenamefont {Kitaev}(2003)}]{KITAEV2003}%
  \BibitemOpen
  \bibfield  {author} {\bibinfo {author} {\bibfnamefont {A.}~\bibnamefont {Kitaev}},\ }\bibfield  {title} {\bibinfo {title} {Fault-tolerant quantum computation by anyons},\ }\href {https://doi.org/https://doi.org/10.1016/S0003-4916(02)00018-0} {\bibfield  {journal} {\bibinfo  {journal} {Annals of Physics}\ }\textbf {\bibinfo {volume} {303}},\ \bibinfo {pages} {2} (\bibinfo {year} {2003})}\BibitemShut {NoStop}%
\bibitem [{\citenamefont {Andersen}\ \emph {et~al.}(2023)\citenamefont {Andersen}, \citenamefont {Lensky}, \citenamefont {Kechedzhi}, \citenamefont {Drozdov}, \citenamefont {Bengtsson}, \citenamefont {Hong}, \citenamefont {Morvan}, \citenamefont {Mi}, \citenamefont {Opremcak}, \citenamefont {Acharya}, \citenamefont {Allen}, \citenamefont {Ansmann}, \citenamefont {Arute}, \citenamefont {Arya}, \citenamefont {Asfaw}, \citenamefont {Atalaya}, \citenamefont {Babbush}, \citenamefont {Bacon}, \citenamefont {Bardin}, \citenamefont {Bortoli}, \citenamefont {Bourassa}, \citenamefont {Bovaird}, \citenamefont {Brill}, \citenamefont {Broughton}, \citenamefont {Buckley}, \citenamefont {Buell}, \citenamefont {Burger}, \citenamefont {Burkett}, \citenamefont {Bushnell}, \citenamefont {Chen}, \citenamefont {Chiaro}, \citenamefont {Chik}, \citenamefont {Chou}, \citenamefont {Cogan}, \citenamefont {Collins}, \citenamefont {Conner}, \citenamefont {Courtney}, \citenamefont {Crook}, \citenamefont {Curtin}, \citenamefont {Debroy},
  \citenamefont {Del Toro~Barba}, \citenamefont {Demura}, \citenamefont {Dunsworth}, \citenamefont {Eppens}, \citenamefont {Erickson}, \citenamefont {Faoro}, \citenamefont {Farhi}, \citenamefont {Fatemi}, \citenamefont {Ferreira}, \citenamefont {Burgos}, \citenamefont {Forati}, \citenamefont {Fowler}, \citenamefont {Foxen}, \citenamefont {Giang}, \citenamefont {Gidney}, \citenamefont {Gilboa}, \citenamefont {Giustina}, \citenamefont {Gosula}, \citenamefont {Dau}, \citenamefont {Gross}, \citenamefont {Habegger}, \citenamefont {Hamilton}, \citenamefont {Hansen}, \citenamefont {Harrigan}, \citenamefont {Harrington}, \citenamefont {Heu}, \citenamefont {Hilton}, \citenamefont {Hoffmann}, \citenamefont {Huang}, \citenamefont {Huff}, \citenamefont {Huggins}, \citenamefont {Ioffe}, \citenamefont {Isakov}, \citenamefont {Iveland}, \citenamefont {Jeffrey}, \citenamefont {Jiang}, \citenamefont {Jones}, \citenamefont {Juhas}, \citenamefont {Kafri}, \citenamefont {Khattar}, \citenamefont {Khezri}, \citenamefont
  {Kieferov{\'a}}, \citenamefont {Kim}, \citenamefont {Kitaev}, \citenamefont {Klimov}, \citenamefont {Klots}, \citenamefont {Korotkov}, \citenamefont {Kostritsa}, \citenamefont {Kreikebaum}, \citenamefont {Landhuis}, \citenamefont {Laptev}, \citenamefont {Lau}, \citenamefont {Laws}, \citenamefont {Lee}, \citenamefont {Lee}, \citenamefont {Lester}, \citenamefont {Lill}, \citenamefont {Liu}, \citenamefont {Locharla}, \citenamefont {Lucero}, \citenamefont {Malone}, \citenamefont {Martin}, \citenamefont {McClean}, \citenamefont {McCourt}, \citenamefont {McEwen}, \citenamefont {Miao}, \citenamefont {Mieszala}, \citenamefont {Mohseni}, \citenamefont {Montazeri}, \citenamefont {Mount}, \citenamefont {Movassagh}, \citenamefont {Mruczkiewicz}, \citenamefont {Naaman}, \citenamefont {Neeley}, \citenamefont {Neill}, \citenamefont {Nersisyan}, \citenamefont {Newman}, \citenamefont {Ng}, \citenamefont {Nguyen}, \citenamefont {Nguyen}, \citenamefont {Niu}, \citenamefont {O'Brien}, \citenamefont {Omonije}, \citenamefont
  {Petukhov}, \citenamefont {Potter}, \citenamefont {Pryadko}, \citenamefont {Quintana}, \citenamefont {Rocque}, \citenamefont {Rubin}, \citenamefont {Saei}, \citenamefont {Sank}, \citenamefont {Sankaragomathi}, \citenamefont {Satzinger}, \citenamefont {Schurkus}, \citenamefont {Schuster}, \citenamefont {Shearn}, \citenamefont {Shorter}, \citenamefont {Shutty}, \citenamefont {Shvarts}, \citenamefont {Skruzny}, \citenamefont {Smith}, \citenamefont {Somma}, \citenamefont {Sterling}, \citenamefont {Strain}, \citenamefont {Szalay}, \citenamefont {Torres}, \citenamefont {Vidal}, \citenamefont {Villalonga}, \citenamefont {Heidweiller}, \citenamefont {White}, \citenamefont {Woo}, \citenamefont {Xing}, \citenamefont {Yao}, \citenamefont {Yeh}, \citenamefont {Yoo}, \citenamefont {Young}, \citenamefont {Zalcman}, \citenamefont {Zhang}, \citenamefont {Zhu}, \citenamefont {Zobrist}, \citenamefont {Neven}, \citenamefont {Boixo}, \citenamefont {Megrant}, \citenamefont {Kelly}, \citenamefont {Chen}, \citenamefont
  {Smelyanskiy}, \citenamefont {Kim}, \citenamefont {Aleiner}, \citenamefont {Roushan}, \citenamefont {AI},\ and\ \citenamefont {{Collaborators}}}]{Andersen_Nature2023}%
  \BibitemOpen
  \bibfield  {author} {\bibinfo {author} {\bibfnamefont {T.~I.}\ \bibnamefont {Andersen}}, \bibinfo {author} {\bibfnamefont {Y.~D.}\ \bibnamefont {Lensky}}, \bibinfo {author} {\bibfnamefont {K.}~\bibnamefont {Kechedzhi}}, \bibinfo {author} {\bibfnamefont {I.~K.}\ \bibnamefont {Drozdov}}, \bibinfo {author} {\bibfnamefont {A.}~\bibnamefont {Bengtsson}}, \bibinfo {author} {\bibfnamefont {S.}~\bibnamefont {Hong}}, \bibinfo {author} {\bibfnamefont {A.}~\bibnamefont {Morvan}}, \bibinfo {author} {\bibfnamefont {X.}~\bibnamefont {Mi}}, \bibinfo {author} {\bibfnamefont {A.}~\bibnamefont {Opremcak}}, \bibinfo {author} {\bibfnamefont {R.}~\bibnamefont {Acharya}}, \bibinfo {author} {\bibfnamefont {R.}~\bibnamefont {Allen}}, \bibinfo {author} {\bibfnamefont {M.}~\bibnamefont {Ansmann}}, \bibinfo {author} {\bibfnamefont {F.}~\bibnamefont {Arute}}, \bibinfo {author} {\bibfnamefont {K.}~\bibnamefont {Arya}}, \bibinfo {author} {\bibfnamefont {A.}~\bibnamefont {Asfaw}}, \bibinfo {author} {\bibfnamefont {J.}~\bibnamefont
  {Atalaya}}, \bibinfo {author} {\bibfnamefont {R.}~\bibnamefont {Babbush}}, \bibinfo {author} {\bibfnamefont {D.}~\bibnamefont {Bacon}}, \bibinfo {author} {\bibfnamefont {J.~C.}\ \bibnamefont {Bardin}}, \bibinfo {author} {\bibfnamefont {G.}~\bibnamefont {Bortoli}}, \bibinfo {author} {\bibfnamefont {A.}~\bibnamefont {Bourassa}}, \bibinfo {author} {\bibfnamefont {J.}~\bibnamefont {Bovaird}}, \bibinfo {author} {\bibfnamefont {L.}~\bibnamefont {Brill}}, \bibinfo {author} {\bibfnamefont {M.}~\bibnamefont {Broughton}}, \bibinfo {author} {\bibfnamefont {B.~B.}\ \bibnamefont {Buckley}}, \bibinfo {author} {\bibfnamefont {D.~A.}\ \bibnamefont {Buell}}, \bibinfo {author} {\bibfnamefont {T.}~\bibnamefont {Burger}}, \bibinfo {author} {\bibfnamefont {B.}~\bibnamefont {Burkett}}, \bibinfo {author} {\bibfnamefont {N.}~\bibnamefont {Bushnell}}, \bibinfo {author} {\bibfnamefont {Z.}~\bibnamefont {Chen}}, \bibinfo {author} {\bibfnamefont {B.}~\bibnamefont {Chiaro}}, \bibinfo {author} {\bibfnamefont {D.}~\bibnamefont {Chik}},
  \bibinfo {author} {\bibfnamefont {C.}~\bibnamefont {Chou}}, \bibinfo {author} {\bibfnamefont {J.}~\bibnamefont {Cogan}}, \bibinfo {author} {\bibfnamefont {R.}~\bibnamefont {Collins}}, \bibinfo {author} {\bibfnamefont {P.}~\bibnamefont {Conner}}, \bibinfo {author} {\bibfnamefont {W.}~\bibnamefont {Courtney}}, \bibinfo {author} {\bibfnamefont {A.~L.}\ \bibnamefont {Crook}}, \bibinfo {author} {\bibfnamefont {B.}~\bibnamefont {Curtin}}, \bibinfo {author} {\bibfnamefont {D.~M.}\ \bibnamefont {Debroy}}, \bibinfo {author} {\bibfnamefont {A.}~\bibnamefont {Del Toro~Barba}}, \bibinfo {author} {\bibfnamefont {S.}~\bibnamefont {Demura}}, \bibinfo {author} {\bibfnamefont {A.}~\bibnamefont {Dunsworth}}, \bibinfo {author} {\bibfnamefont {D.}~\bibnamefont {Eppens}}, \bibinfo {author} {\bibfnamefont {C.}~\bibnamefont {Erickson}}, \bibinfo {author} {\bibfnamefont {L.}~\bibnamefont {Faoro}}, \bibinfo {author} {\bibfnamefont {E.}~\bibnamefont {Farhi}}, \bibinfo {author} {\bibfnamefont {R.}~\bibnamefont {Fatemi}}, \bibinfo
  {author} {\bibfnamefont {V.~S.}\ \bibnamefont {Ferreira}}, \bibinfo {author} {\bibfnamefont {L.~F.}\ \bibnamefont {Burgos}}, \bibinfo {author} {\bibfnamefont {E.}~\bibnamefont {Forati}}, \bibinfo {author} {\bibfnamefont {A.~G.}\ \bibnamefont {Fowler}}, \bibinfo {author} {\bibfnamefont {B.}~\bibnamefont {Foxen}}, \bibinfo {author} {\bibfnamefont {W.}~\bibnamefont {Giang}}, \bibinfo {author} {\bibfnamefont {C.}~\bibnamefont {Gidney}}, \bibinfo {author} {\bibfnamefont {D.}~\bibnamefont {Gilboa}}, \bibinfo {author} {\bibfnamefont {M.}~\bibnamefont {Giustina}}, \bibinfo {author} {\bibfnamefont {R.}~\bibnamefont {Gosula}}, \bibinfo {author} {\bibfnamefont {A.~G.}\ \bibnamefont {Dau}}, \bibinfo {author} {\bibfnamefont {J.~A.}\ \bibnamefont {Gross}}, \bibinfo {author} {\bibfnamefont {S.}~\bibnamefont {Habegger}}, \bibinfo {author} {\bibfnamefont {M.~C.}\ \bibnamefont {Hamilton}}, \bibinfo {author} {\bibfnamefont {M.}~\bibnamefont {Hansen}}, \bibinfo {author} {\bibfnamefont {M.~P.}\ \bibnamefont {Harrigan}},
  \bibinfo {author} {\bibfnamefont {S.~D.}\ \bibnamefont {Harrington}}, \bibinfo {author} {\bibfnamefont {P.}~\bibnamefont {Heu}}, \bibinfo {author} {\bibfnamefont {J.}~\bibnamefont {Hilton}}, \bibinfo {author} {\bibfnamefont {M.~R.}\ \bibnamefont {Hoffmann}}, \bibinfo {author} {\bibfnamefont {T.}~\bibnamefont {Huang}}, \bibinfo {author} {\bibfnamefont {A.}~\bibnamefont {Huff}}, \bibinfo {author} {\bibfnamefont {W.~J.}\ \bibnamefont {Huggins}}, \bibinfo {author} {\bibfnamefont {L.~B.}\ \bibnamefont {Ioffe}}, \bibinfo {author} {\bibfnamefont {S.~V.}\ \bibnamefont {Isakov}}, \bibinfo {author} {\bibfnamefont {J.}~\bibnamefont {Iveland}}, \bibinfo {author} {\bibfnamefont {E.}~\bibnamefont {Jeffrey}}, \bibinfo {author} {\bibfnamefont {Z.}~\bibnamefont {Jiang}}, \bibinfo {author} {\bibfnamefont {C.}~\bibnamefont {Jones}}, \bibinfo {author} {\bibfnamefont {P.}~\bibnamefont {Juhas}}, \bibinfo {author} {\bibfnamefont {D.}~\bibnamefont {Kafri}}, \bibinfo {author} {\bibfnamefont {T.}~\bibnamefont {Khattar}}, \bibinfo
  {author} {\bibfnamefont {M.}~\bibnamefont {Khezri}}, \bibinfo {author} {\bibfnamefont {M.}~\bibnamefont {Kieferov{\'a}}}, \bibinfo {author} {\bibfnamefont {S.}~\bibnamefont {Kim}}, \bibinfo {author} {\bibfnamefont {A.}~\bibnamefont {Kitaev}}, \bibinfo {author} {\bibfnamefont {P.~V.}\ \bibnamefont {Klimov}}, \bibinfo {author} {\bibfnamefont {A.~R.}\ \bibnamefont {Klots}}, \bibinfo {author} {\bibfnamefont {A.~N.}\ \bibnamefont {Korotkov}}, \bibinfo {author} {\bibfnamefont {F.}~\bibnamefont {Kostritsa}}, \bibinfo {author} {\bibfnamefont {J.~M.}\ \bibnamefont {Kreikebaum}}, \bibinfo {author} {\bibfnamefont {D.}~\bibnamefont {Landhuis}}, \bibinfo {author} {\bibfnamefont {P.}~\bibnamefont {Laptev}}, \bibinfo {author} {\bibfnamefont {K.-M.}\ \bibnamefont {Lau}}, \bibinfo {author} {\bibfnamefont {L.}~\bibnamefont {Laws}}, \bibinfo {author} {\bibfnamefont {J.}~\bibnamefont {Lee}}, \bibinfo {author} {\bibfnamefont {K.~W.}\ \bibnamefont {Lee}}, \bibinfo {author} {\bibfnamefont {B.~J.}\ \bibnamefont {Lester}}, \bibinfo
  {author} {\bibfnamefont {A.~T.}\ \bibnamefont {Lill}}, \bibinfo {author} {\bibfnamefont {W.}~\bibnamefont {Liu}}, \bibinfo {author} {\bibfnamefont {A.}~\bibnamefont {Locharla}}, \bibinfo {author} {\bibfnamefont {E.}~\bibnamefont {Lucero}}, \bibinfo {author} {\bibfnamefont {F.~D.}\ \bibnamefont {Malone}}, \bibinfo {author} {\bibfnamefont {O.}~\bibnamefont {Martin}}, \bibinfo {author} {\bibfnamefont {J.~R.}\ \bibnamefont {McClean}}, \bibinfo {author} {\bibfnamefont {T.}~\bibnamefont {McCourt}}, \bibinfo {author} {\bibfnamefont {M.}~\bibnamefont {McEwen}}, \bibinfo {author} {\bibfnamefont {K.~C.}\ \bibnamefont {Miao}}, \bibinfo {author} {\bibfnamefont {A.}~\bibnamefont {Mieszala}}, \bibinfo {author} {\bibfnamefont {M.}~\bibnamefont {Mohseni}}, \bibinfo {author} {\bibfnamefont {S.}~\bibnamefont {Montazeri}}, \bibinfo {author} {\bibfnamefont {E.}~\bibnamefont {Mount}}, \bibinfo {author} {\bibfnamefont {R.}~\bibnamefont {Movassagh}}, \bibinfo {author} {\bibfnamefont {W.}~\bibnamefont {Mruczkiewicz}}, \bibinfo
  {author} {\bibfnamefont {O.}~\bibnamefont {Naaman}}, \bibinfo {author} {\bibfnamefont {M.}~\bibnamefont {Neeley}}, \bibinfo {author} {\bibfnamefont {C.}~\bibnamefont {Neill}}, \bibinfo {author} {\bibfnamefont {A.}~\bibnamefont {Nersisyan}}, \bibinfo {author} {\bibfnamefont {M.}~\bibnamefont {Newman}}, \bibinfo {author} {\bibfnamefont {J.~H.}\ \bibnamefont {Ng}}, \bibinfo {author} {\bibfnamefont {A.}~\bibnamefont {Nguyen}}, \bibinfo {author} {\bibfnamefont {M.}~\bibnamefont {Nguyen}}, \bibinfo {author} {\bibfnamefont {M.~Y.}\ \bibnamefont {Niu}}, \bibinfo {author} {\bibfnamefont {T.~E.}\ \bibnamefont {O'Brien}}, \bibinfo {author} {\bibfnamefont {S.}~\bibnamefont {Omonije}}, \bibinfo {author} {\bibfnamefont {A.}~\bibnamefont {Petukhov}}, \bibinfo {author} {\bibfnamefont {R.}~\bibnamefont {Potter}}, \bibinfo {author} {\bibfnamefont {L.~P.}\ \bibnamefont {Pryadko}}, \bibinfo {author} {\bibfnamefont {C.}~\bibnamefont {Quintana}}, \bibinfo {author} {\bibfnamefont {C.}~\bibnamefont {Rocque}}, \bibinfo {author}
  {\bibfnamefont {N.~C.}\ \bibnamefont {Rubin}}, \bibinfo {author} {\bibfnamefont {N.}~\bibnamefont {Saei}}, \bibinfo {author} {\bibfnamefont {D.}~\bibnamefont {Sank}}, \bibinfo {author} {\bibfnamefont {K.}~\bibnamefont {Sankaragomathi}}, \bibinfo {author} {\bibfnamefont {K.~J.}\ \bibnamefont {Satzinger}}, \bibinfo {author} {\bibfnamefont {H.~F.}\ \bibnamefont {Schurkus}}, \bibinfo {author} {\bibfnamefont {C.}~\bibnamefont {Schuster}}, \bibinfo {author} {\bibfnamefont {M.~J.}\ \bibnamefont {Shearn}}, \bibinfo {author} {\bibfnamefont {A.}~\bibnamefont {Shorter}}, \bibinfo {author} {\bibfnamefont {N.}~\bibnamefont {Shutty}}, \bibinfo {author} {\bibfnamefont {V.}~\bibnamefont {Shvarts}}, \bibinfo {author} {\bibfnamefont {J.}~\bibnamefont {Skruzny}}, \bibinfo {author} {\bibfnamefont {W.~C.}\ \bibnamefont {Smith}}, \bibinfo {author} {\bibfnamefont {R.}~\bibnamefont {Somma}}, \bibinfo {author} {\bibfnamefont {G.}~\bibnamefont {Sterling}}, \bibinfo {author} {\bibfnamefont {D.}~\bibnamefont {Strain}}, \bibinfo
  {author} {\bibfnamefont {M.}~\bibnamefont {Szalay}}, \bibinfo {author} {\bibfnamefont {A.}~\bibnamefont {Torres}}, \bibinfo {author} {\bibfnamefont {G.}~\bibnamefont {Vidal}}, \bibinfo {author} {\bibfnamefont {B.}~\bibnamefont {Villalonga}}, \bibinfo {author} {\bibfnamefont {C.~V.}\ \bibnamefont {Heidweiller}}, \bibinfo {author} {\bibfnamefont {T.}~\bibnamefont {White}}, \bibinfo {author} {\bibfnamefont {B.~W.~K.}\ \bibnamefont {Woo}}, \bibinfo {author} {\bibfnamefont {C.}~\bibnamefont {Xing}}, \bibinfo {author} {\bibfnamefont {Z.~J.}\ \bibnamefont {Yao}}, \bibinfo {author} {\bibfnamefont {P.}~\bibnamefont {Yeh}}, \bibinfo {author} {\bibfnamefont {J.}~\bibnamefont {Yoo}}, \bibinfo {author} {\bibfnamefont {G.}~\bibnamefont {Young}}, \bibinfo {author} {\bibfnamefont {A.}~\bibnamefont {Zalcman}}, \bibinfo {author} {\bibfnamefont {Y.}~\bibnamefont {Zhang}}, \bibinfo {author} {\bibfnamefont {N.}~\bibnamefont {Zhu}}, \bibinfo {author} {\bibfnamefont {N.}~\bibnamefont {Zobrist}}, \bibinfo {author} {\bibfnamefont
  {H.}~\bibnamefont {Neven}}, \bibinfo {author} {\bibfnamefont {S.}~\bibnamefont {Boixo}}, \bibinfo {author} {\bibfnamefont {A.}~\bibnamefont {Megrant}}, \bibinfo {author} {\bibfnamefont {J.}~\bibnamefont {Kelly}}, \bibinfo {author} {\bibfnamefont {Y.}~\bibnamefont {Chen}}, \bibinfo {author} {\bibfnamefont {V.}~\bibnamefont {Smelyanskiy}}, \bibinfo {author} {\bibfnamefont {E.-A.}\ \bibnamefont {Kim}}, \bibinfo {author} {\bibfnamefont {I.}~\bibnamefont {Aleiner}}, \bibinfo {author} {\bibfnamefont {P.}~\bibnamefont {Roushan}}, \bibinfo {author} {\bibfnamefont {G.~Q.}\ \bibnamefont {AI}},\ and\ \bibinfo {author} {\bibnamefont {{Collaborators}}},\ }\bibfield  {title} {\bibinfo {title} {Non-{A}belian braiding of graph vertices in a superconducting processor},\ }\href {https://doi.org/10.1038/s41586-023-05954-4} {\bibfield  {journal} {\bibinfo  {journal} {Nature}\ }\textbf {\bibinfo {volume} {618}},\ \bibinfo {pages} {264} (\bibinfo {year} {2023})}\BibitemShut {NoStop}%
\bibitem [{\citenamefont {Iqbal}\ \emph {et~al.}(2024)\citenamefont {Iqbal}, \citenamefont {Tantivasadakarn}, \citenamefont {Verresen}, \citenamefont {Campbell}, \citenamefont {Dreiling}, \citenamefont {Figgatt}, \citenamefont {Gaebler}, \citenamefont {Johansen}, \citenamefont {Mills}, \citenamefont {Moses}, \citenamefont {Pino}, \citenamefont {Ransford}, \citenamefont {Rowe}, \citenamefont {Siegfried}, \citenamefont {Stutz}, \citenamefont {Foss-Feig}, \citenamefont {Vishwanath},\ and\ \citenamefont {Dreyer}}]{Iqbal2024}%
  \BibitemOpen
  \bibfield  {author} {\bibinfo {author} {\bibfnamefont {M.}~\bibnamefont {Iqbal}}, \bibinfo {author} {\bibfnamefont {N.}~\bibnamefont {Tantivasadakarn}}, \bibinfo {author} {\bibfnamefont {R.}~\bibnamefont {Verresen}}, \bibinfo {author} {\bibfnamefont {S.~L.}\ \bibnamefont {Campbell}}, \bibinfo {author} {\bibfnamefont {J.~M.}\ \bibnamefont {Dreiling}}, \bibinfo {author} {\bibfnamefont {C.}~\bibnamefont {Figgatt}}, \bibinfo {author} {\bibfnamefont {J.~P.}\ \bibnamefont {Gaebler}}, \bibinfo {author} {\bibfnamefont {J.}~\bibnamefont {Johansen}}, \bibinfo {author} {\bibfnamefont {M.}~\bibnamefont {Mills}}, \bibinfo {author} {\bibfnamefont {S.~A.}\ \bibnamefont {Moses}}, \bibinfo {author} {\bibfnamefont {J.~M.}\ \bibnamefont {Pino}}, \bibinfo {author} {\bibfnamefont {A.}~\bibnamefont {Ransford}}, \bibinfo {author} {\bibfnamefont {M.}~\bibnamefont {Rowe}}, \bibinfo {author} {\bibfnamefont {P.}~\bibnamefont {Siegfried}}, \bibinfo {author} {\bibfnamefont {R.~P.}\ \bibnamefont {Stutz}}, \bibinfo {author}
  {\bibfnamefont {M.}~\bibnamefont {Foss-Feig}}, \bibinfo {author} {\bibfnamefont {A.}~\bibnamefont {Vishwanath}},\ and\ \bibinfo {author} {\bibfnamefont {H.}~\bibnamefont {Dreyer}},\ }\bibfield  {title} {\bibinfo {title} {Non-{A}belian topological order and anyons on a trapped-ion processor},\ }\href {https://doi.org/10.1038/s41586-023-06934-4} {\bibfield  {journal} {\bibinfo  {journal} {Nature}\ }\textbf {\bibinfo {volume} {626}},\ \bibinfo {pages} {505} (\bibinfo {year} {2024})}\BibitemShut {NoStop}%
\bibitem [{\citenamefont {Mezzacapo}\ \emph {et~al.}(2015)\citenamefont {Mezzacapo}, \citenamefont {Rico}, \citenamefont {Sab\'{\i}n}, \citenamefont {Egusquiza}, \citenamefont {Lamata},\ and\ \citenamefont {Solano}}]{Mezzacapo_PRL2015}%
  \BibitemOpen
  \bibfield  {author} {\bibinfo {author} {\bibfnamefont {A.}~\bibnamefont {Mezzacapo}}, \bibinfo {author} {\bibfnamefont {E.}~\bibnamefont {Rico}}, \bibinfo {author} {\bibfnamefont {C.}~\bibnamefont {Sab\'{\i}n}}, \bibinfo {author} {\bibfnamefont {I.~L.}\ \bibnamefont {Egusquiza}}, \bibinfo {author} {\bibfnamefont {L.}~\bibnamefont {Lamata}},\ and\ \bibinfo {author} {\bibfnamefont {E.}~\bibnamefont {Solano}},\ }\bibfield  {title} {\bibinfo {title} {Non-abelian su(2) lattice gauge theories in superconducting circuits},\ }\href {https://doi.org/10.1103/PhysRevLett.115.240502} {\bibfield  {journal} {\bibinfo  {journal} {Phys. Rev. Lett.}\ }\textbf {\bibinfo {volume} {115}},\ \bibinfo {pages} {240502} (\bibinfo {year} {2015})}\BibitemShut {NoStop}%
\bibitem [{\citenamefont {Atas}\ \emph {et~al.}(2021)\citenamefont {Atas}, \citenamefont {Zhang}, \citenamefont {Lewis}, \citenamefont {Jahanpour}, \citenamefont {Haase},\ and\ \citenamefont {Muschik}}]{Atas_NatComm2021}%
  \BibitemOpen
  \bibfield  {author} {\bibinfo {author} {\bibfnamefont {Y.~Y.}\ \bibnamefont {Atas}}, \bibinfo {author} {\bibfnamefont {J.}~\bibnamefont {Zhang}}, \bibinfo {author} {\bibfnamefont {R.}~\bibnamefont {Lewis}}, \bibinfo {author} {\bibfnamefont {A.}~\bibnamefont {Jahanpour}}, \bibinfo {author} {\bibfnamefont {J.~F.}\ \bibnamefont {Haase}},\ and\ \bibinfo {author} {\bibfnamefont {C.~A.}\ \bibnamefont {Muschik}},\ }\bibfield  {title} {\bibinfo {title} {{SU(2)} hadrons on a quantum computer via a variational approach},\ }\href {https://doi.org/10.1038/s41467-021-26825-4} {\bibfield  {journal} {\bibinfo  {journal} {Nature Communications}\ }\textbf {\bibinfo {volume} {12}},\ \bibinfo {pages} {6499} (\bibinfo {year} {2021})}\BibitemShut {NoStop}%
\bibitem [{\citenamefont {A~Rahman}\ \emph {et~al.}(2021)\citenamefont {A~Rahman}, \citenamefont {Lewis}, \citenamefont {Mendicelli},\ and\ \citenamefont {Powell}}]{rahman2021su2}%
  \BibitemOpen
  \bibfield  {author} {\bibinfo {author} {\bibfnamefont {S.}~\bibnamefont {A~Rahman}}, \bibinfo {author} {\bibfnamefont {R.}~\bibnamefont {Lewis}}, \bibinfo {author} {\bibfnamefont {E.}~\bibnamefont {Mendicelli}},\ and\ \bibinfo {author} {\bibfnamefont {S.}~\bibnamefont {Powell}},\ }\bibfield  {title} {\bibinfo {title} {{SU(2)} lattice gauge theory on a quantum annealer},\ }\href {https://doi.org/10.1103/PhysRevD.104.034501} {\bibfield  {journal} {\bibinfo  {journal} {Phys. Rev. D}\ }\textbf {\bibinfo {volume} {104}},\ \bibinfo {pages} {034501} (\bibinfo {year} {2021})}\BibitemShut {NoStop}%
\bibitem [{\citenamefont {Rahman}\ \emph {et~al.}(2022)\citenamefont {Rahman}, \citenamefont {Lewis}, \citenamefont {Mendicelli},\ and\ \citenamefont {Powell}}]{rahman2022realtime}%
  \BibitemOpen
  \bibfield  {author} {\bibinfo {author} {\bibfnamefont {S.~A.}\ \bibnamefont {Rahman}}, \bibinfo {author} {\bibfnamefont {R.}~\bibnamefont {Lewis}}, \bibinfo {author} {\bibfnamefont {E.}~\bibnamefont {Mendicelli}},\ and\ \bibinfo {author} {\bibfnamefont {S.}~\bibnamefont {Powell}},\ }\bibfield  {title} {\bibinfo {title} {Real time evolution and a traveling excitation in {SU(2)} pure gauge theory on a quantum computer},\ }\href {https://arxiv.org/abs/2210.11606} {\bibfield  {journal} {\bibinfo  {journal} {arXiv:2210.11606}\ } (\bibinfo {year} {2022})}\BibitemShut {NoStop}%
\bibitem [{\citenamefont {A~Rahman}\ \emph {et~al.}(2022)\citenamefont {A~Rahman}, \citenamefont {Lewis}, \citenamefont {Mendicelli},\ and\ \citenamefont {Powell}}]{rahman2022self}%
  \BibitemOpen
  \bibfield  {author} {\bibinfo {author} {\bibfnamefont {S.}~\bibnamefont {A~Rahman}}, \bibinfo {author} {\bibfnamefont {R.}~\bibnamefont {Lewis}}, \bibinfo {author} {\bibfnamefont {E.}~\bibnamefont {Mendicelli}},\ and\ \bibinfo {author} {\bibfnamefont {S.}~\bibnamefont {Powell}},\ }\bibfield  {title} {\bibinfo {title} {Self-mitigating trotter circuits for {SU(2)} lattice gauge theory on a quantum computer},\ }\href {https://doi.org/10.1103/PhysRevD.106.074502} {\bibfield  {journal} {\bibinfo  {journal} {Phys. Rev. D}\ }\textbf {\bibinfo {volume} {106}},\ \bibinfo {pages} {074502} (\bibinfo {year} {2022})}\BibitemShut {NoStop}%
\bibitem [{\citenamefont {Fromm}\ \emph {et~al.}(2023)\citenamefont {Fromm}, \citenamefont {Philipsen},\ and\ \citenamefont {Winterowd}}]{Fromm2023}%
  \BibitemOpen
  \bibfield  {author} {\bibinfo {author} {\bibfnamefont {M.}~\bibnamefont {Fromm}}, \bibinfo {author} {\bibfnamefont {O.}~\bibnamefont {Philipsen}},\ and\ \bibinfo {author} {\bibfnamefont {C.}~\bibnamefont {Winterowd}},\ }\bibfield  {title} {\bibinfo {title} {Dihedral lattice gauge theories on a quantum annealer},\ }\href {https://doi.org/10.1140/epjqt/s40507-023-00188-9} {\bibfield  {journal} {\bibinfo  {journal} {EPJ Quantum Technology}\ }\textbf {\bibinfo {volume} {10}},\ \bibinfo {pages} {31} (\bibinfo {year} {2023})}\BibitemShut {NoStop}%
\bibitem [{\citenamefont {Atas}\ \emph {et~al.}(2023)\citenamefont {Atas}, \citenamefont {Haase}, \citenamefont {Zhang}, \citenamefont {Wei}, \citenamefont {Pfaendler}, \citenamefont {Lewis},\ and\ \citenamefont {Muschik}}]{Atas_PRR2023}%
  \BibitemOpen
  \bibfield  {author} {\bibinfo {author} {\bibfnamefont {Y.~Y.}\ \bibnamefont {Atas}}, \bibinfo {author} {\bibfnamefont {J.~F.}\ \bibnamefont {Haase}}, \bibinfo {author} {\bibfnamefont {J.}~\bibnamefont {Zhang}}, \bibinfo {author} {\bibfnamefont {V.}~\bibnamefont {Wei}}, \bibinfo {author} {\bibfnamefont {S.~M.-L.}\ \bibnamefont {Pfaendler}}, \bibinfo {author} {\bibfnamefont {R.}~\bibnamefont {Lewis}},\ and\ \bibinfo {author} {\bibfnamefont {C.~A.}\ \bibnamefont {Muschik}},\ }\bibfield  {title} {\bibinfo {title} {Simulating one-dimensional quantum chromodynamics on a quantum computer: Real-time evolutions of tetra- and pentaquarks},\ }\href {https://doi.org/10.1103/PhysRevResearch.5.033184} {\bibfield  {journal} {\bibinfo  {journal} {Phys. Rev. Res.}\ }\textbf {\bibinfo {volume} {5}},\ \bibinfo {pages} {033184} (\bibinfo {year} {2023})}\BibitemShut {NoStop}%
\bibitem [{\citenamefont {Farrell}\ \emph {et~al.}(2023)\citenamefont {Farrell}, \citenamefont {Chernyshev}, \citenamefont {Powell}, \citenamefont {Zemlevskiy}, \citenamefont {Illa},\ and\ \citenamefont {Savage}}]{Farrell_PRD2023}%
  \BibitemOpen
  \bibfield  {author} {\bibinfo {author} {\bibfnamefont {R.~C.}\ \bibnamefont {Farrell}}, \bibinfo {author} {\bibfnamefont {I.~A.}\ \bibnamefont {Chernyshev}}, \bibinfo {author} {\bibfnamefont {S.~J.~M.}\ \bibnamefont {Powell}}, \bibinfo {author} {\bibfnamefont {N.~A.}\ \bibnamefont {Zemlevskiy}}, \bibinfo {author} {\bibfnamefont {M.}~\bibnamefont {Illa}},\ and\ \bibinfo {author} {\bibfnamefont {M.~J.}\ \bibnamefont {Savage}},\ }\bibfield  {title} {\bibinfo {title} {Preparations for quantum simulations of quantum chromodynamics in $1+1$ dimensions. i. axial gauge},\ }\href {https://doi.org/10.1103/PhysRevD.107.054512} {\bibfield  {journal} {\bibinfo  {journal} {Phys. Rev. D}\ }\textbf {\bibinfo {volume} {107}},\ \bibinfo {pages} {054512} (\bibinfo {year} {2023})}\BibitemShut {NoStop}%
\bibitem [{\citenamefont {Turro}\ \emph {et~al.}(2024)\citenamefont {Turro}, \citenamefont {Ciavarella},\ and\ \citenamefont {Yao}}]{Turro_PRD2024}%
  \BibitemOpen
  \bibfield  {author} {\bibinfo {author} {\bibfnamefont {F.}~\bibnamefont {Turro}}, \bibinfo {author} {\bibfnamefont {A.}~\bibnamefont {Ciavarella}},\ and\ \bibinfo {author} {\bibfnamefont {X.}~\bibnamefont {Yao}},\ }\bibfield  {title} {\bibinfo {title} {Classical and quantum computing of shear viscosity for $(2+1)d$ su(2) gauge theory},\ }\href {https://doi.org/10.1103/PhysRevD.109.114511} {\bibfield  {journal} {\bibinfo  {journal} {Phys. Rev. D}\ }\textbf {\bibinfo {volume} {109}},\ \bibinfo {pages} {114511} (\bibinfo {year} {2024})}\BibitemShut {NoStop}%
\bibitem [{\citenamefont {Than}\ \emph {et~al.}(2024)\citenamefont {Than}, \citenamefont {Atas}, \citenamefont {Chakraborty}, \citenamefont {Zhang}, \citenamefont {Diaz}, \citenamefont {Wen}, \citenamefont {Liu}, \citenamefont {Lewis}, \citenamefont {Green}, \citenamefont {Muschik},\ and\ \citenamefont {Linke}}]{Than_2024}%
  \BibitemOpen
  \bibfield  {author} {\bibinfo {author} {\bibfnamefont {A.~T.}\ \bibnamefont {Than}}, \bibinfo {author} {\bibfnamefont {Y.~Y.}\ \bibnamefont {Atas}}, \bibinfo {author} {\bibfnamefont {A.}~\bibnamefont {Chakraborty}}, \bibinfo {author} {\bibfnamefont {J.}~\bibnamefont {Zhang}}, \bibinfo {author} {\bibfnamefont {M.~T.}\ \bibnamefont {Diaz}}, \bibinfo {author} {\bibfnamefont {K.}~\bibnamefont {Wen}}, \bibinfo {author} {\bibfnamefont {X.}~\bibnamefont {Liu}}, \bibinfo {author} {\bibfnamefont {R.}~\bibnamefont {Lewis}}, \bibinfo {author} {\bibfnamefont {A.~M.}\ \bibnamefont {Green}}, \bibinfo {author} {\bibfnamefont {C.~A.}\ \bibnamefont {Muschik}},\ and\ \bibinfo {author} {\bibfnamefont {N.~M.}\ \bibnamefont {Linke}},\ }\bibfield  {title} {\bibinfo {title} {The phase diagram of quantum chromodynamics in one dimension on a quantum computer},\ }\href {https://arxiv.org/abs/2501.00579} {\bibfield  {journal} {\bibinfo  {journal} {arXiv:2501.00579}\ } (\bibinfo {year} {2024})}\BibitemShut {NoStop}%
\bibitem [{\citenamefont {Ciavarella}\ and\ \citenamefont {Bauer}(2024)}]{Ciavarella_PRL2024}%
  \BibitemOpen
  \bibfield  {author} {\bibinfo {author} {\bibfnamefont {A.~N.}\ \bibnamefont {Ciavarella}}\ and\ \bibinfo {author} {\bibfnamefont {C.~W.}\ \bibnamefont {Bauer}},\ }\bibfield  {title} {\bibinfo {title} {Quantum simulation of su(3) lattice yang-mills theory at leading order in large-${N}_{c}$ expansion},\ }\href {https://doi.org/10.1103/PhysRevLett.133.111901} {\bibfield  {journal} {\bibinfo  {journal} {Phys. Rev. Lett.}\ }\textbf {\bibinfo {volume} {133}},\ \bibinfo {pages} {111901} (\bibinfo {year} {2024})}\BibitemShut {NoStop}%
\bibitem [{\citenamefont {Liu}\ and\ \citenamefont {Winter}(2022)}]{liu2022many}%
  \BibitemOpen
  \bibfield  {author} {\bibinfo {author} {\bibfnamefont {Z.-W.}\ \bibnamefont {Liu}}\ and\ \bibinfo {author} {\bibfnamefont {A.}~\bibnamefont {Winter}},\ }\bibfield  {title} {\bibinfo {title} {Many-body quantum magic},\ }\href {https://doi.org/10.1103/PRXQuantum.3.020333} {\bibfield  {journal} {\bibinfo  {journal} {PRX Quantum}\ }\textbf {\bibinfo {volume} {3}},\ \bibinfo {pages} {020333} (\bibinfo {year} {2022})}\BibitemShut {NoStop}%
\bibitem [{\citenamefont {Leone}\ \emph {et~al.}(2022)\citenamefont {Leone}, \citenamefont {Oliviero},\ and\ \citenamefont {Hamma}}]{leone2022stabilizer}%
  \BibitemOpen
  \bibfield  {author} {\bibinfo {author} {\bibfnamefont {L.}~\bibnamefont {Leone}}, \bibinfo {author} {\bibfnamefont {S.~F.~E.}\ \bibnamefont {Oliviero}},\ and\ \bibinfo {author} {\bibfnamefont {A.}~\bibnamefont {Hamma}},\ }\bibfield  {title} {\bibinfo {title} {Stabilizer r\'enyi entropy},\ }\href {https://doi.org/10.1103/PhysRevLett.128.050402} {\bibfield  {journal} {\bibinfo  {journal} {Phys. Rev. Lett.}\ }\textbf {\bibinfo {volume} {128}},\ \bibinfo {pages} {050402} (\bibinfo {year} {2022})}\BibitemShut {NoStop}%
\bibitem [{\citenamefont {Walter}\ \emph {et~al.}(2016)\citenamefont {Walter}, \citenamefont {Gross},\ and\ \citenamefont {Eisert}}]{walter2016multipartite}%
  \BibitemOpen
  \bibfield  {author} {\bibinfo {author} {\bibfnamefont {M.}~\bibnamefont {Walter}}, \bibinfo {author} {\bibfnamefont {D.}~\bibnamefont {Gross}},\ and\ \bibinfo {author} {\bibfnamefont {J.}~\bibnamefont {Eisert}},\ }\bibfield  {title} {\bibinfo {title} {Multipartite entanglement},\ }\href {https://onlinelibrary.wiley.com/doi/abs/10.1002/9783527805785.ch14} {\bibfield  {journal} {\bibinfo  {journal} {Quantum Information: From Foundations to Quantum Technology Applications}\ ,\ \bibinfo {pages} {293}} (\bibinfo {year} {2016})}\BibitemShut {NoStop}%
\bibitem [{\citenamefont {Bengtsson}\ and\ \citenamefont {Zyczkowski}(2016)}]{bengtsson2016brief}%
  \BibitemOpen
  \bibfield  {author} {\bibinfo {author} {\bibfnamefont {I.}~\bibnamefont {Bengtsson}}\ and\ \bibinfo {author} {\bibfnamefont {K.}~\bibnamefont {Zyczkowski}},\ }\bibfield  {title} {\bibinfo {title} {A brief introduction to multipartite entanglement},\ }\href {https://arxiv.org/abs/1612.07747} {\bibfield  {journal} {\bibinfo  {journal} {arXiv:1612.07747}\ } (\bibinfo {year} {2016})}\BibitemShut {NoStop}%
\bibitem [{\citenamefont {Froland}\ \emph {et~al.}(2025)\citenamefont {Froland}, \citenamefont {Zache}, \citenamefont {Ott},\ and\ \citenamefont {Mueller}}]{froland2025entanglement}%
  \BibitemOpen
  \bibfield  {author} {\bibinfo {author} {\bibfnamefont {H.}~\bibnamefont {Froland}}, \bibinfo {author} {\bibfnamefont {T.~V.}\ \bibnamefont {Zache}}, \bibinfo {author} {\bibfnamefont {R.}~\bibnamefont {Ott}},\ and\ \bibinfo {author} {\bibfnamefont {N.}~\bibnamefont {Mueller}},\ }\bibfield  {title} {\bibinfo {title} {Entanglement structure of non-gaussian states and how to measure it},\ }\href {https://doi.org/10.1103/pnp2-g1g5} {\bibfield  {journal} {\bibinfo  {journal} {Phys. Rev. Lett.}\ }\textbf {\bibinfo {volume} {135}},\ \bibinfo {pages} {040201} (\bibinfo {year} {2025})}\BibitemShut {NoStop}%
\bibitem [{\citenamefont {Hebenstreit}\ \emph {et~al.}(2019)\citenamefont {Hebenstreit}, \citenamefont {Jozsa}, \citenamefont {Kraus}, \citenamefont {Strelchuk},\ and\ \citenamefont {Yoganathan}}]{hebenstreit2019all}%
  \BibitemOpen
  \bibfield  {author} {\bibinfo {author} {\bibfnamefont {M.}~\bibnamefont {Hebenstreit}}, \bibinfo {author} {\bibfnamefont {R.}~\bibnamefont {Jozsa}}, \bibinfo {author} {\bibfnamefont {B.}~\bibnamefont {Kraus}}, \bibinfo {author} {\bibfnamefont {S.}~\bibnamefont {Strelchuk}},\ and\ \bibinfo {author} {\bibfnamefont {M.}~\bibnamefont {Yoganathan}},\ }\bibfield  {title} {\bibinfo {title} {All pure fermionic non-gaussian states are magic states for matchgate computations},\ }\href {https://doi.org/10.1103/PhysRevLett.123.080503} {\bibfield  {journal} {\bibinfo  {journal} {Physical review letters}\ }\textbf {\bibinfo {volume} {123}},\ \bibinfo {pages} {080503} (\bibinfo {year} {2019})}\BibitemShut {NoStop}%
\bibitem [{\citenamefont {Collura}\ \emph {et~al.}(2024)\citenamefont {Collura}, \citenamefont {De~Nardis}, \citenamefont {Alba},\ and\ \citenamefont {Lami}}]{collura2024quantum}%
  \BibitemOpen
  \bibfield  {author} {\bibinfo {author} {\bibfnamefont {M.}~\bibnamefont {Collura}}, \bibinfo {author} {\bibfnamefont {J.}~\bibnamefont {De~Nardis}}, \bibinfo {author} {\bibfnamefont {V.}~\bibnamefont {Alba}},\ and\ \bibinfo {author} {\bibfnamefont {G.}~\bibnamefont {Lami}},\ }\bibfield  {title} {\bibinfo {title} {The quantum magic of fermionic gaussian states},\ }\href {https://arxiv.org/abs/2412.05367} {\bibfield  {journal} {\bibinfo  {journal} {arXiv:2412.05367}\ } (\bibinfo {year} {2024})}\BibitemShut {NoStop}%
\bibitem [{\citenamefont {Sierant}\ \emph {et~al.}(2025)\citenamefont {Sierant}, \citenamefont {Stornati},\ and\ \citenamefont {Turkeshi}}]{sierant2025fermionic}%
  \BibitemOpen
  \bibfield  {author} {\bibinfo {author} {\bibfnamefont {P.}~\bibnamefont {Sierant}}, \bibinfo {author} {\bibfnamefont {P.}~\bibnamefont {Stornati}},\ and\ \bibinfo {author} {\bibfnamefont {X.}~\bibnamefont {Turkeshi}},\ }\bibfield  {title} {\bibinfo {title} {Fermionic magic resources of quantum many-body systems},\ }\href {https://arxiv.org/abs/2506.00116} {\bibfield  {journal} {\bibinfo  {journal} {arXiv:2506.00116}\ } (\bibinfo {year} {2025})}\BibitemShut {NoStop}%
\bibitem [{\citenamefont {Jozsa}\ and\ \citenamefont {Linden}(2003)}]{jozsa2003role}%
  \BibitemOpen
  \bibfield  {author} {\bibinfo {author} {\bibfnamefont {R.}~\bibnamefont {Jozsa}}\ and\ \bibinfo {author} {\bibfnamefont {N.}~\bibnamefont {Linden}},\ }\bibfield  {title} {\bibinfo {title} {On the role of entanglement in quantum-computational speed-up},\ }\href {https://doi.org/https://doi.org/10.1098/rspa.2002.1097} {\bibfield  {journal} {\bibinfo  {journal} {Proceedings of the Royal Society of London. Series A: Mathematical, Physical and Engineering Sciences}\ }\textbf {\bibinfo {volume} {459}},\ \bibinfo {pages} {2011} (\bibinfo {year} {2003})}\BibitemShut {NoStop}%
\bibitem [{\citenamefont {Gottesman}(1998)}]{gottesman1998heisenberg}%
  \BibitemOpen
  \bibfield  {author} {\bibinfo {author} {\bibfnamefont {D.}~\bibnamefont {Gottesman}},\ }\bibfield  {title} {\bibinfo {title} {The heisenberg representation of quantum computers},\ }\href {https://arxiv.org/abs/quant-ph/9807006} {\bibfield  {journal} {\bibinfo  {journal} {arXiv preprint quant-ph/9807006}\ } (\bibinfo {year} {1998})}\BibitemShut {NoStop}%
\bibitem [{\citenamefont {Aaronson}\ and\ \citenamefont {Gottesman}(2004)}]{aaronson2004improved}%
  \BibitemOpen
  \bibfield  {author} {\bibinfo {author} {\bibfnamefont {S.}~\bibnamefont {Aaronson}}\ and\ \bibinfo {author} {\bibfnamefont {D.}~\bibnamefont {Gottesman}},\ }\bibfield  {title} {\bibinfo {title} {Improved simulation of stabilizer circuits},\ }\href {https://doi.org/10.1103/PhysRevA.70.052328} {\bibfield  {journal} {\bibinfo  {journal} {Phys. Rev. A}\ }\textbf {\bibinfo {volume} {70}},\ \bibinfo {pages} {052328} (\bibinfo {year} {2004})}\BibitemShut {NoStop}%
\bibitem [{\citenamefont {Nest}(2008)}]{nest2008classical}%
  \BibitemOpen
  \bibfield  {author} {\bibinfo {author} {\bibfnamefont {M.}~\bibnamefont {Nest}},\ }\bibfield  {title} {\bibinfo {title} {Classical simulation of quantum computation, the gottesman-knill theorem, and slightly beyond},\ }\href {https://arxiv.org/abs/0811.0898} {\bibfield  {journal} {\bibinfo  {journal} {arXiv:0811.0898}\ } (\bibinfo {year} {2008})}\BibitemShut {NoStop}%
\bibitem [{\citenamefont {Bravyi}\ and\ \citenamefont {Kitaev}(2005)}]{bravyi2005universal}%
  \BibitemOpen
  \bibfield  {author} {\bibinfo {author} {\bibfnamefont {S.}~\bibnamefont {Bravyi}}\ and\ \bibinfo {author} {\bibfnamefont {A.}~\bibnamefont {Kitaev}},\ }\bibfield  {title} {\bibinfo {title} {Universal quantum computation with ideal clifford gates and noisy ancillas},\ }\href {https://doi.org/10.1103/PhysRevA.71.022316} {\bibfield  {journal} {\bibinfo  {journal} {Phys. Rev. A}\ }\textbf {\bibinfo {volume} {71}},\ \bibinfo {pages} {022316} (\bibinfo {year} {2005})}\BibitemShut {NoStop}%
\bibitem [{\citenamefont {Fowler}\ \emph {et~al.}(2012)\citenamefont {Fowler}, \citenamefont {Mariantoni}, \citenamefont {Martinis},\ and\ \citenamefont {Cleland}}]{fowler2012surface}%
  \BibitemOpen
  \bibfield  {author} {\bibinfo {author} {\bibfnamefont {A.~G.}\ \bibnamefont {Fowler}}, \bibinfo {author} {\bibfnamefont {M.}~\bibnamefont {Mariantoni}}, \bibinfo {author} {\bibfnamefont {J.~M.}\ \bibnamefont {Martinis}},\ and\ \bibinfo {author} {\bibfnamefont {A.~N.}\ \bibnamefont {Cleland}},\ }\bibfield  {title} {\bibinfo {title} {Surface codes: Towards practical large-scale quantum computation},\ }\href {https://doi.org/10.1103/PhysRevA.86.032324} {\bibfield  {journal} {\bibinfo  {journal} {Phys. Rev. A}\ }\textbf {\bibinfo {volume} {86}},\ \bibinfo {pages} {032324} (\bibinfo {year} {2012})}\BibitemShut {NoStop}%
\bibitem [{\citenamefont {Falc\~ao}\ \emph {et~al.}(2025)\citenamefont {Falc\~ao}, \citenamefont {Tarabunga}, \citenamefont {Frau}, \citenamefont {Tirrito}, \citenamefont {Zakrzewski},\ and\ \citenamefont {Dalmonte}}]{Falcao_PRB2025}%
  \BibitemOpen
  \bibfield  {author} {\bibinfo {author} {\bibfnamefont {P.~R.~N.}\ \bibnamefont {Falc\~ao}}, \bibinfo {author} {\bibfnamefont {P.~S.}\ \bibnamefont {Tarabunga}}, \bibinfo {author} {\bibfnamefont {M.}~\bibnamefont {Frau}}, \bibinfo {author} {\bibfnamefont {E.}~\bibnamefont {Tirrito}}, \bibinfo {author} {\bibfnamefont {J.}~\bibnamefont {Zakrzewski}},\ and\ \bibinfo {author} {\bibfnamefont {M.}~\bibnamefont {Dalmonte}},\ }\bibfield  {title} {\bibinfo {title} {Nonstabilizerness in u(1) lattice gauge theory},\ }\href {https://doi.org/10.1103/PhysRevB.111.L081102} {\bibfield  {journal} {\bibinfo  {journal} {Phys. Rev. B}\ }\textbf {\bibinfo {volume} {111}},\ \bibinfo {pages} {L081102} (\bibinfo {year} {2025})}\BibitemShut {NoStop}%
\bibitem [{\citenamefont {Valiant}(2001)}]{valiant2001quantum}%
  \BibitemOpen
  \bibfield  {author} {\bibinfo {author} {\bibfnamefont {L.~G.}\ \bibnamefont {Valiant}},\ }\bibfield  {title} {\bibinfo {title} {Quantum computers that can be simulated classically in polynomial time},\ }in\ \href {https://doi.org/10.1145/380752.380785} {\emph {\bibinfo {booktitle} {Proceedings of the Thirty-Third Annual ACM Symposium on Theory of Computing}}},\ \bibinfo {series and number} {STOC '01}\ (\bibinfo  {publisher} {Association for Computing Machinery},\ \bibinfo {address} {New York, NY, USA},\ \bibinfo {year} {2001})\ p.\ \bibinfo {pages} {114–123}\BibitemShut {NoStop}%
\bibitem [{\citenamefont {Terhal}\ and\ \citenamefont {DiVincenzo}(2002)}]{terhal2002classical}%
  \BibitemOpen
  \bibfield  {author} {\bibinfo {author} {\bibfnamefont {B.~M.}\ \bibnamefont {Terhal}}\ and\ \bibinfo {author} {\bibfnamefont {D.~P.}\ \bibnamefont {DiVincenzo}},\ }\bibfield  {title} {\bibinfo {title} {Classical simulation of noninteracting-fermion quantum circuits},\ }\href {https://doi.org/10.1103/PhysRevA.65.032325} {\bibfield  {journal} {\bibinfo  {journal} {Phys. Rev. A}\ }\textbf {\bibinfo {volume} {65}},\ \bibinfo {pages} {032325} (\bibinfo {year} {2002})}\BibitemShut {NoStop}%
\bibitem [{\citenamefont {Jozsa}\ and\ \citenamefont {Miyake}(2008)}]{jozsa2008matchgates}%
  \BibitemOpen
  \bibfield  {author} {\bibinfo {author} {\bibfnamefont {R.}~\bibnamefont {Jozsa}}\ and\ \bibinfo {author} {\bibfnamefont {A.}~\bibnamefont {Miyake}},\ }\bibfield  {title} {\bibinfo {title} {Matchgates and classical simulation of quantum circuits},\ }\href {https://royalsocietypublishing.org/doi/abs/10.1098/rspa.2008.0189} {\bibfield  {journal} {\bibinfo  {journal} {Proceedings of the Royal Society A: Mathematical, Physical and Engineering Sciences}\ }\textbf {\bibinfo {volume} {464}},\ \bibinfo {pages} {3089} (\bibinfo {year} {2008})}\BibitemShut {NoStop}%
\bibitem [{\citenamefont {Brod}(2016)}]{Brod_PRA2016}%
  \BibitemOpen
  \bibfield  {author} {\bibinfo {author} {\bibfnamefont {D.~J.}\ \bibnamefont {Brod}},\ }\bibfield  {title} {\bibinfo {title} {Efficient classical simulation of matchgate circuits with generalized inputs and measurements},\ }\href {https://doi.org/10.1103/PhysRevA.93.062332} {\bibfield  {journal} {\bibinfo  {journal} {Phys. Rev. A}\ }\textbf {\bibinfo {volume} {93}},\ \bibinfo {pages} {062332} (\bibinfo {year} {2016})}\BibitemShut {NoStop}%
\bibitem [{\citenamefont {Singha~Roy}\ \emph {et~al.}(2022)\citenamefont {Singha~Roy}, \citenamefont {Carl},\ and\ \citenamefont {Hauke}}]{roy2022genuine}%
  \BibitemOpen
  \bibfield  {author} {\bibinfo {author} {\bibfnamefont {S.}~\bibnamefont {Singha~Roy}}, \bibinfo {author} {\bibfnamefont {L.}~\bibnamefont {Carl}},\ and\ \bibinfo {author} {\bibfnamefont {P.}~\bibnamefont {Hauke}},\ }\bibfield  {title} {\bibinfo {title} {Genuine multipartite entanglement in a one-dimensional bose-hubbard model with frustrated hopping},\ }\href {https://doi.org/10.1103/PhysRevB.106.195158} {\bibfield  {journal} {\bibinfo  {journal} {Phys. Rev. B}\ }\textbf {\bibinfo {volume} {106}},\ \bibinfo {pages} {195158} (\bibinfo {year} {2022})}\BibitemShut {NoStop}%
\bibitem [{\citenamefont {Lakkaraju}\ \emph {et~al.}(2024)\citenamefont {Lakkaraju}, \citenamefont {Haldar},\ and\ \citenamefont {Sen(De)}}]{lakkaraju2024predicting}%
  \BibitemOpen
  \bibfield  {author} {\bibinfo {author} {\bibfnamefont {L.~G.~C.}\ \bibnamefont {Lakkaraju}}, \bibinfo {author} {\bibfnamefont {S.~K.}\ \bibnamefont {Haldar}},\ and\ \bibinfo {author} {\bibfnamefont {A.}~\bibnamefont {Sen(De)}},\ }\bibfield  {title} {\bibinfo {title} {Predicting a topological quantum phase transition from dynamics via multisite entanglement},\ }\href {https://doi.org/10.1103/PhysRevA.109.022436} {\bibfield  {journal} {\bibinfo  {journal} {Phys. Rev. A}\ }\textbf {\bibinfo {volume} {109}},\ \bibinfo {pages} {022436} (\bibinfo {year} {2024})}\BibitemShut {NoStop}%
\bibitem [{\citenamefont {Sapui}\ \emph {et~al.}(2025)\citenamefont {Sapui}, \citenamefont {Agarwal}, \citenamefont {Konar}, \citenamefont {Lakkaraju},\ and\ \citenamefont {De}}]{sapui2025genuine}%
  \BibitemOpen
  \bibfield  {author} {\bibinfo {author} {\bibfnamefont {T.}~\bibnamefont {Sapui}}, \bibinfo {author} {\bibfnamefont {K.~D.}\ \bibnamefont {Agarwal}}, \bibinfo {author} {\bibfnamefont {T.~K.}\ \bibnamefont {Konar}}, \bibinfo {author} {\bibfnamefont {L.~G.~C.}\ \bibnamefont {Lakkaraju}},\ and\ \bibinfo {author} {\bibfnamefont {A.~S.}\ \bibnamefont {De}},\ }\bibfield  {title} {\bibinfo {title} {Genuine multipartite entanglement as a probe of many-body localization in disordered spin chains with dzyaloshinskii-moriya interactions},\ }\href {https://arxiv.org/abs/2507.22795} {\bibfield  {journal} {\bibinfo  {journal} {arXiv:2507.22795}\ } (\bibinfo {year} {2025})}\BibitemShut {NoStop}%
\bibitem [{\citenamefont {Jasser}\ \emph {et~al.}(2025)\citenamefont {Jasser}, \citenamefont {Odavic},\ and\ \citenamefont {Hamma}}]{jasser2025stabilizer}%
  \BibitemOpen
  \bibfield  {author} {\bibinfo {author} {\bibfnamefont {B.}~\bibnamefont {Jasser}}, \bibinfo {author} {\bibfnamefont {J.}~\bibnamefont {Odavic}},\ and\ \bibinfo {author} {\bibfnamefont {A.}~\bibnamefont {Hamma}},\ }\bibfield  {title} {\bibinfo {title} {Stabilizer entropy and entanglement complexity in the {S}achdev-{Y}e-{K}itaev model},\ }\href {https://arxiv.org/abs/2502.03093} {\bibfield  {journal} {\bibinfo  {journal} {arXiv:2502.03093}\ } (\bibinfo {year} {2025})}\BibitemShut {NoStop}%
\bibitem [{\citenamefont {Santra}\ \emph {et~al.}(2025{\natexlab{a}})\citenamefont {Santra}, \citenamefont {Windey}, \citenamefont {Bandyopadhyay}, \citenamefont {Legramandi},\ and\ \citenamefont {Hauke}}]{santra2025complexity}%
  \BibitemOpen
  \bibfield  {author} {\bibinfo {author} {\bibfnamefont {G.~C.}\ \bibnamefont {Santra}}, \bibinfo {author} {\bibfnamefont {A.}~\bibnamefont {Windey}}, \bibinfo {author} {\bibfnamefont {S.}~\bibnamefont {Bandyopadhyay}}, \bibinfo {author} {\bibfnamefont {A.}~\bibnamefont {Legramandi}},\ and\ \bibinfo {author} {\bibfnamefont {P.}~\bibnamefont {Hauke}},\ }\bibfield  {title} {\bibinfo {title} {Complexity transitions in chaotic quantum systems: Nonstabilizerness, entanglement, and fractal dimension in {SYK} and random matrix models},\ }\href {https://arxiv.org/abs/2505.09707} {\bibfield  {journal} {\bibinfo  {journal} {arXiv:2505.09707}\ } (\bibinfo {year} {2025}{\natexlab{a}})}\BibitemShut {NoStop}%
\bibitem [{\citenamefont {Br\"okemeier}\ \emph {et~al.}(2025)\citenamefont {Br\"okemeier}, \citenamefont {Hengstenberg}, \citenamefont {Keeble}, \citenamefont {Robin}, \citenamefont {Rocco},\ and\ \citenamefont {Savage}}]{brokemeier2025quantum}%
  \BibitemOpen
  \bibfield  {author} {\bibinfo {author} {\bibfnamefont {F.}~\bibnamefont {Br\"okemeier}}, \bibinfo {author} {\bibfnamefont {S.~M.}\ \bibnamefont {Hengstenberg}}, \bibinfo {author} {\bibfnamefont {J.~W.~T.}\ \bibnamefont {Keeble}}, \bibinfo {author} {\bibfnamefont {C.~E.~P.}\ \bibnamefont {Robin}}, \bibinfo {author} {\bibfnamefont {F.}~\bibnamefont {Rocco}},\ and\ \bibinfo {author} {\bibfnamefont {M.~J.}\ \bibnamefont {Savage}},\ }\bibfield  {title} {\bibinfo {title} {Quantum magic and multipartite entanglement in the structure of nuclei},\ }\href {https://doi.org/10.1103/PhysRevC.111.034317} {\bibfield  {journal} {\bibinfo  {journal} {Phys. Rev. C}\ }\textbf {\bibinfo {volume} {111}},\ \bibinfo {pages} {034317} (\bibinfo {year} {2025})}\BibitemShut {NoStop}%
\bibitem [{\citenamefont {Chernyshev}\ \emph {et~al.}(2025)\citenamefont {Chernyshev}, \citenamefont {Robin},\ and\ \citenamefont {Savage}}]{cherynshev2025quantum}%
  \BibitemOpen
  \bibfield  {author} {\bibinfo {author} {\bibfnamefont {I.}~\bibnamefont {Chernyshev}}, \bibinfo {author} {\bibfnamefont {C.~E.~P.}\ \bibnamefont {Robin}},\ and\ \bibinfo {author} {\bibfnamefont {M.~J.}\ \bibnamefont {Savage}},\ }\bibfield  {title} {\bibinfo {title} {Quantum magic and computational complexity in the neutrino sector},\ }\href {https://doi.org/10.1103/PhysRevResearch.7.023228} {\bibfield  {journal} {\bibinfo  {journal} {Phys. Rev. Res.}\ }\textbf {\bibinfo {volume} {7}},\ \bibinfo {pages} {023228} (\bibinfo {year} {2025})}\BibitemShut {NoStop}%
\bibitem [{\citenamefont {Viscardi}\ \emph {et~al.}(2025)\citenamefont {Viscardi}, \citenamefont {Dalmonte}, \citenamefont {Hamma},\ and\ \citenamefont {Tirrito}}]{viscardi2025interplay}%
  \BibitemOpen
  \bibfield  {author} {\bibinfo {author} {\bibfnamefont {M.}~\bibnamefont {Viscardi}}, \bibinfo {author} {\bibfnamefont {M.}~\bibnamefont {Dalmonte}}, \bibinfo {author} {\bibfnamefont {A.}~\bibnamefont {Hamma}},\ and\ \bibinfo {author} {\bibfnamefont {E.}~\bibnamefont {Tirrito}},\ }\bibfield  {title} {\bibinfo {title} {Interplay of entanglement structures and stabilizer entropy in spin models},\ }\href {https://arxiv.org/abs/2503.08620} {\bibfield  {journal} {\bibinfo  {journal} {arXiv:2503.08620}\ } (\bibinfo {year} {2025})}\BibitemShut {NoStop}%
\bibitem [{\citenamefont {Fux}\ \emph {et~al.}(2024)\citenamefont {Fux}, \citenamefont {Tirrito}, \citenamefont {Dalmonte},\ and\ \citenamefont {Fazio}}]{fux2024entanglement}%
  \BibitemOpen
  \bibfield  {author} {\bibinfo {author} {\bibfnamefont {G.~E.}\ \bibnamefont {Fux}}, \bibinfo {author} {\bibfnamefont {E.}~\bibnamefont {Tirrito}}, \bibinfo {author} {\bibfnamefont {M.}~\bibnamefont {Dalmonte}},\ and\ \bibinfo {author} {\bibfnamefont {R.}~\bibnamefont {Fazio}},\ }\bibfield  {title} {\bibinfo {title} {Entanglement -- nonstabilizerness separation in hybrid quantum circuits},\ }\href {https://doi.org/10.1103/PhysRevResearch.6.L042030} {\bibfield  {journal} {\bibinfo  {journal} {Phys. Rev. Res.}\ }\textbf {\bibinfo {volume} {6}},\ \bibinfo {pages} {L042030} (\bibinfo {year} {2024})}\BibitemShut {NoStop}%
\bibitem [{\citenamefont {Kumar}\ \emph {et~al.}(2024)\citenamefont {Kumar}, \citenamefont {Konar}, \citenamefont {Lakkaraju},\ and\ \citenamefont {Sen(De)}}]{KUMAR2024129668}%
  \BibitemOpen
  \bibfield  {author} {\bibinfo {author} {\bibfnamefont {P.}~\bibnamefont {Kumar}}, \bibinfo {author} {\bibfnamefont {T.~K.}\ \bibnamefont {Konar}}, \bibinfo {author} {\bibfnamefont {L.~G.~C.}\ \bibnamefont {Lakkaraju}},\ and\ \bibinfo {author} {\bibfnamefont {A.}~\bibnamefont {Sen(De)}},\ }\bibfield  {title} {\bibinfo {title} {Quantum resources in {H}arrow-{H}assidim-{L}loyd algorithm},\ }\href {https://doi.org/https://doi.org/10.1016/j.physleta.2024.129668} {\bibfield  {journal} {\bibinfo  {journal} {Physics Letters A}\ }\textbf {\bibinfo {volume} {517}},\ \bibinfo {pages} {129668} (\bibinfo {year} {2024})}\BibitemShut {NoStop}%
\bibitem [{\citenamefont {Santra}\ \emph {et~al.}(2025{\natexlab{b}})\citenamefont {Santra}, \citenamefont {Roy}, \citenamefont {Egger},\ and\ \citenamefont {Hauke}}]{santra2025genuine}%
  \BibitemOpen
  \bibfield  {author} {\bibinfo {author} {\bibfnamefont {G.~C.}\ \bibnamefont {Santra}}, \bibinfo {author} {\bibfnamefont {S.~S.}\ \bibnamefont {Roy}}, \bibinfo {author} {\bibfnamefont {D.~J.}\ \bibnamefont {Egger}},\ and\ \bibinfo {author} {\bibfnamefont {P.}~\bibnamefont {Hauke}},\ }\bibfield  {title} {\bibinfo {title} {Genuine multipartite entanglement in quantum optimization},\ }\href {https://doi.org/10.1103/PhysRevA.111.022434} {\bibfield  {journal} {\bibinfo  {journal} {Phys. Rev. A}\ }\textbf {\bibinfo {volume} {111}},\ \bibinfo {pages} {022434} (\bibinfo {year} {2025}{\natexlab{b}})}\BibitemShut {NoStop}%
\bibitem [{\citenamefont {Turkeshi}\ \emph {et~al.}(2025{\natexlab{a}})\citenamefont {Turkeshi}, \citenamefont {Tirrito},\ and\ \citenamefont {Sierant}}]{turkeshi2025magic}%
  \BibitemOpen
  \bibfield  {author} {\bibinfo {author} {\bibfnamefont {X.}~\bibnamefont {Turkeshi}}, \bibinfo {author} {\bibfnamefont {E.}~\bibnamefont {Tirrito}},\ and\ \bibinfo {author} {\bibfnamefont {P.}~\bibnamefont {Sierant}},\ }\bibfield  {title} {\bibinfo {title} {Magic spreading in random quantum circuits},\ }\href {https://www.nature.com/articles/s41467-025-57704-x} {\bibfield  {journal} {\bibinfo  {journal} {Nature Communications}\ }\textbf {\bibinfo {volume} {16}},\ \bibinfo {pages} {2575} (\bibinfo {year} {2025}{\natexlab{a}})}\BibitemShut {NoStop}%
\bibitem [{\citenamefont {Capecci}\ \emph {et~al.}(2025)\citenamefont {Capecci}, \citenamefont {Santra}, \citenamefont {Bottarelli}, \citenamefont {Tirrito},\ and\ \citenamefont {Hauke}}]{capecci2025role}%
  \BibitemOpen
  \bibfield  {author} {\bibinfo {author} {\bibfnamefont {C.}~\bibnamefont {Capecci}}, \bibinfo {author} {\bibfnamefont {G.~C.}\ \bibnamefont {Santra}}, \bibinfo {author} {\bibfnamefont {A.}~\bibnamefont {Bottarelli}}, \bibinfo {author} {\bibfnamefont {E.}~\bibnamefont {Tirrito}},\ and\ \bibinfo {author} {\bibfnamefont {P.}~\bibnamefont {Hauke}},\ }\bibfield  {title} {\bibinfo {title} {Role of nonstabilizerness in quantum optimization},\ }\href {https://arxiv.org/abs/2505.17185} {\bibfield  {journal} {\bibinfo  {journal} {arXiv:2505.17185}\ } (\bibinfo {year} {2025})}\BibitemShut {NoStop}%
\bibitem [{\citenamefont {Barbiero}\ \emph {et~al.}(2019)\citenamefont {Barbiero}, \citenamefont {Schweizer}, \citenamefont {Aidelsburger}, \citenamefont {Demler}, \citenamefont {Goldman},\ and\ \citenamefont {Grusdt}}]{Barbiero2019}%
  \BibitemOpen
  \bibfield  {author} {\bibinfo {author} {\bibfnamefont {L.}~\bibnamefont {Barbiero}}, \bibinfo {author} {\bibfnamefont {C.}~\bibnamefont {Schweizer}}, \bibinfo {author} {\bibfnamefont {M.}~\bibnamefont {Aidelsburger}}, \bibinfo {author} {\bibfnamefont {E.}~\bibnamefont {Demler}}, \bibinfo {author} {\bibfnamefont {N.}~\bibnamefont {Goldman}},\ and\ \bibinfo {author} {\bibfnamefont {F.}~\bibnamefont {Grusdt}},\ }\bibfield  {title} {\bibinfo {title} {Coupling ultracold matter to dynamical gauge fields in optical lattices: From flux attachment to $\mathbb{Z}_2$ lattice gauge theories},\ }\href {https://doi.org/10.1126/sciadv.aav7444} {\bibfield  {journal} {\bibinfo  {journal} {Science Advances}\ }\textbf {\bibinfo {volume} {5}},\ \bibinfo {pages} {eaav7444} (\bibinfo {year} {2019})}\BibitemShut {NoStop}%
\bibitem [{\citenamefont {Gonz\'alez-Cuadra}\ \emph {et~al.}(2020)\citenamefont {Gonz\'alez-Cuadra}, \citenamefont {Tagliacozzo}, \citenamefont {Lewenstein},\ and\ \citenamefont {Bermudez}}]{Gonzalezcuadra_PRX2020}%
  \BibitemOpen
  \bibfield  {author} {\bibinfo {author} {\bibfnamefont {D.}~\bibnamefont {Gonz\'alez-Cuadra}}, \bibinfo {author} {\bibfnamefont {L.}~\bibnamefont {Tagliacozzo}}, \bibinfo {author} {\bibfnamefont {M.}~\bibnamefont {Lewenstein}},\ and\ \bibinfo {author} {\bibfnamefont {A.}~\bibnamefont {Bermudez}},\ }\bibfield  {title} {\bibinfo {title} {Robust topological order in fermionic $\mathbb{Z}_{2}$ gauge theories: From aharonov-bohm instability to soliton-induced deconfinement},\ }\href {https://doi.org/10.1103/PhysRevX.10.041007} {\bibfield  {journal} {\bibinfo  {journal} {Phys. Rev. X}\ }\textbf {\bibinfo {volume} {10}},\ \bibinfo {pages} {041007} (\bibinfo {year} {2020})}\BibitemShut {NoStop}%
\bibitem [{\citenamefont {Grabowska}\ \emph {et~al.}(2024)\citenamefont {Grabowska}, \citenamefont {Kane},\ and\ \citenamefont {Bauer}}]{grabowska2024fully}%
  \BibitemOpen
  \bibfield  {author} {\bibinfo {author} {\bibfnamefont {D.~M.}\ \bibnamefont {Grabowska}}, \bibinfo {author} {\bibfnamefont {C.~F.}\ \bibnamefont {Kane}},\ and\ \bibinfo {author} {\bibfnamefont {C.~W.}\ \bibnamefont {Bauer}},\ }\bibfield  {title} {\bibinfo {title} {A fully gauge-fixed {SU(2)} hamiltonian for quantum simulations},\ }\href {https://arxiv.org/abs/2409.10610} {\bibfield  {journal} {\bibinfo  {journal} {arXiv:2409.10610}\ } (\bibinfo {year} {2024})}\BibitemShut {NoStop}%
\bibitem [{\citenamefont {Yao}(2023)}]{yao2023su}%
  \BibitemOpen
  \bibfield  {author} {\bibinfo {author} {\bibfnamefont {X.}~\bibnamefont {Yao}},\ }\bibfield  {title} {\bibinfo {title} {{SU(2)} gauge theory in $2+1$ dimensions on a plaquette chain obeys the eigenstate thermalization hypothesis},\ }\href {https://doi.org/10.1103/PhysRevD.108.L031504} {\bibfield  {journal} {\bibinfo  {journal} {Phys. Rev. D}\ }\textbf {\bibinfo {volume} {108}},\ \bibinfo {pages} {L031504} (\bibinfo {year} {2023})}\BibitemShut {NoStop}%
\bibitem [{\citenamefont {Ballini}\ \emph {et~al.}(2025)\citenamefont {Ballini}, \citenamefont {Mildenberger}, \citenamefont {Wauters},\ and\ \citenamefont {Hauke}}]{Ballini_2024}%
  \BibitemOpen
  \bibfield  {author} {\bibinfo {author} {\bibfnamefont {E.}~\bibnamefont {Ballini}}, \bibinfo {author} {\bibfnamefont {J.}~\bibnamefont {Mildenberger}}, \bibinfo {author} {\bibfnamefont {M.~M.}\ \bibnamefont {Wauters}},\ and\ \bibinfo {author} {\bibfnamefont {P.}~\bibnamefont {Hauke}},\ }\bibfield  {title} {\bibinfo {title} {Symmetry verification for noisy quantum simulations of non-abelian lattice gauge theories},\ }\href {https://quantum-journal.org/papers/q-2025-07-22-1802/} {\bibfield  {journal} {\bibinfo  {journal} {Quantum}\ }\textbf {\bibinfo {volume} {9}},\ \bibinfo {pages} {1802} (\bibinfo {year} {2025})}\BibitemShut {NoStop}%
\bibitem [{\citenamefont {Wang}\ \emph {et~al.}(2025)\citenamefont {Wang}, \citenamefont {Sun},\ and\ \citenamefont {Zheng}}]{Wang_PRR2025}%
  \BibitemOpen
  \bibfield  {author} {\bibinfo {author} {\bibfnamefont {J.}~\bibnamefont {Wang}}, \bibinfo {author} {\bibfnamefont {X.}~\bibnamefont {Sun}},\ and\ \bibinfo {author} {\bibfnamefont {W.}~\bibnamefont {Zheng}},\ }\bibfield  {title} {\bibinfo {title} {Quantum simulation with gauge fixing: From ising lattice gauge theory to dynamical flux model},\ }\href {https://doi.org/10.1103/PhysRevResearch.7.013311} {\bibfield  {journal} {\bibinfo  {journal} {Phys. Rev. Res.}\ }\textbf {\bibinfo {volume} {7}},\ \bibinfo {pages} {013311} (\bibinfo {year} {2025})}\BibitemShut {NoStop}%
\bibitem [{\citenamefont {Fontana}\ \emph {et~al.}(2025)\citenamefont {Fontana}, \citenamefont {Miranda-Riaza},\ and\ \citenamefont {Celi}}]{Fontana_PRX2025}%
  \BibitemOpen
  \bibfield  {author} {\bibinfo {author} {\bibfnamefont {P.}~\bibnamefont {Fontana}}, \bibinfo {author} {\bibfnamefont {M.}~\bibnamefont {Miranda-Riaza}},\ and\ \bibinfo {author} {\bibfnamefont {A.}~\bibnamefont {Celi}},\ }\bibfield  {title} {\bibinfo {title} {Efficient finite-resource formulation of non-abelian lattice gauge theories beyond one dimension},\ }\href {https://doi.org/10.1103/k9p6-c649} {\bibfield  {journal} {\bibinfo  {journal} {Phys. Rev. X}\ }\textbf {\bibinfo {volume} {15}},\ \bibinfo {pages} {031065} (\bibinfo {year} {2025})}\BibitemShut {NoStop}%
\bibitem [{\citenamefont {Kogut}\ and\ \citenamefont {Susskind}(1975)}]{kogut1975hamiltonian}%
  \BibitemOpen
  \bibfield  {author} {\bibinfo {author} {\bibfnamefont {J.}~\bibnamefont {Kogut}}\ and\ \bibinfo {author} {\bibfnamefont {L.}~\bibnamefont {Susskind}},\ }\bibfield  {title} {\bibinfo {title} {Hamiltonian formulation of wilson's lattice gauge theories},\ }\href {https://doi.org/10.1103/PhysRevD.11.395} {\bibfield  {journal} {\bibinfo  {journal} {Phys. Rev. D}\ }\textbf {\bibinfo {volume} {11}},\ \bibinfo {pages} {395} (\bibinfo {year} {1975})}\BibitemShut {NoStop}%
\bibitem [{\citenamefont {Zohar}\ and\ \citenamefont {Burrello}(2015)}]{zohar2015formulation}%
  \BibitemOpen
  \bibfield  {author} {\bibinfo {author} {\bibfnamefont {E.}~\bibnamefont {Zohar}}\ and\ \bibinfo {author} {\bibfnamefont {M.}~\bibnamefont {Burrello}},\ }\bibfield  {title} {\bibinfo {title} {Formulation of lattice gauge theories for quantum simulations},\ }\href {https://doi.org/10.1103/PhysRevD.91.054506} {\bibfield  {journal} {\bibinfo  {journal} {Phys. Rev. D}\ }\textbf {\bibinfo {volume} {91}},\ \bibinfo {pages} {054506} (\bibinfo {year} {2015})}\BibitemShut {NoStop}%
\bibitem [{\citenamefont {Mariani}\ \emph {et~al.}(2023)\citenamefont {Mariani}, \citenamefont {Pradhan},\ and\ \citenamefont {Ercolessi}}]{Mariani_PRD2023}%
  \BibitemOpen
  \bibfield  {author} {\bibinfo {author} {\bibfnamefont {A.}~\bibnamefont {Mariani}}, \bibinfo {author} {\bibfnamefont {S.}~\bibnamefont {Pradhan}},\ and\ \bibinfo {author} {\bibfnamefont {E.}~\bibnamefont {Ercolessi}},\ }\bibfield  {title} {\bibinfo {title} {Hamiltonians and gauge-invariant hilbert space for lattice yang-mills-like theories with finite gauge group},\ }\href {https://doi.org/10.1103/PhysRevD.107.114513} {\bibfield  {journal} {\bibinfo  {journal} {Phys. Rev. D}\ }\textbf {\bibinfo {volume} {107}},\ \bibinfo {pages} {114513} (\bibinfo {year} {2023})}\BibitemShut {NoStop}%
\bibitem [{\citenamefont {Lamm}\ \emph {et~al.}(2019)\citenamefont {Lamm}, \citenamefont {Lawrence},\ and\ \citenamefont {Yamauchi}}]{lamm2019prd}%
  \BibitemOpen
  \bibfield  {author} {\bibinfo {author} {\bibfnamefont {H.}~\bibnamefont {Lamm}}, \bibinfo {author} {\bibfnamefont {S.}~\bibnamefont {Lawrence}},\ and\ \bibinfo {author} {\bibfnamefont {Y.}~\bibnamefont {Yamauchi}} (\bibinfo {collaboration} {NuQS Collaboration}),\ }\bibfield  {title} {\bibinfo {title} {General methods for digital quantum simulation of gauge theories},\ }\href {https://doi.org/10.1103/PhysRevD.100.034518} {\bibfield  {journal} {\bibinfo  {journal} {Phys. Rev. D}\ }\textbf {\bibinfo {volume} {100}},\ \bibinfo {pages} {034518} (\bibinfo {year} {2019})}\BibitemShut {NoStop}%
\bibitem [{Note1()}]{Note1}%
  \BibitemOpen
  \bibinfo {note} {In extended (2+1)D geometries, the system can undergo a phase transition at a finite coupling, while in the ladder geometry it occurs only at $g_c^2=0$.}\BibitemShut {Stop}%
\bibitem [{\citenamefont {Fradkin}\ and\ \citenamefont {Susskind}(1978)}]{fradkin1978}%
  \BibitemOpen
  \bibfield  {author} {\bibinfo {author} {\bibfnamefont {E.}~\bibnamefont {Fradkin}}\ and\ \bibinfo {author} {\bibfnamefont {L.}~\bibnamefont {Susskind}},\ }\bibfield  {title} {\bibinfo {title} {Order and disorder in gauge systems and magnets},\ }\href {https://doi.org/10.1103/PhysRevD.17.2637} {\bibfield  {journal} {\bibinfo  {journal} {Phys. Rev. D}\ }\textbf {\bibinfo {volume} {17}},\ \bibinfo {pages} {2637} (\bibinfo {year} {1978})}\BibitemShut {NoStop}%
\bibitem [{\citenamefont {Nayak}\ \emph {et~al.}(2008)\citenamefont {Nayak}, \citenamefont {Simon}, \citenamefont {Stern}, \citenamefont {Freedman},\ and\ \citenamefont {Das~Sarma}}]{Nayak_RMP2008}%
  \BibitemOpen
  \bibfield  {author} {\bibinfo {author} {\bibfnamefont {C.}~\bibnamefont {Nayak}}, \bibinfo {author} {\bibfnamefont {S.~H.}\ \bibnamefont {Simon}}, \bibinfo {author} {\bibfnamefont {A.}~\bibnamefont {Stern}}, \bibinfo {author} {\bibfnamefont {M.}~\bibnamefont {Freedman}},\ and\ \bibinfo {author} {\bibfnamefont {S.}~\bibnamefont {Das~Sarma}},\ }\bibfield  {title} {\bibinfo {title} {Non-abelian anyons and topological quantum computation},\ }\href {https://doi.org/10.1103/RevModPhys.80.1083} {\bibfield  {journal} {\bibinfo  {journal} {Rev. Mod. Phys.}\ }\textbf {\bibinfo {volume} {80}},\ \bibinfo {pages} {1083} (\bibinfo {year} {2008})}\BibitemShut {NoStop}%
\bibitem [{\citenamefont {Zohar}\ \emph {et~al.}(2016)\citenamefont {Zohar}, \citenamefont {Wahl}, \citenamefont {Burrello},\ and\ \citenamefont {Cirac}}]{Zohar2016}%
  \BibitemOpen
  \bibfield  {author} {\bibinfo {author} {\bibfnamefont {E.}~\bibnamefont {Zohar}}, \bibinfo {author} {\bibfnamefont {T.~B.}\ \bibnamefont {Wahl}}, \bibinfo {author} {\bibfnamefont {M.}~\bibnamefont {Burrello}},\ and\ \bibinfo {author} {\bibfnamefont {J.~I.}\ \bibnamefont {Cirac}},\ }\bibfield  {title} {\bibinfo {title} {Projected entangled pair states with non-abelian gauge symmetries: An {SU(2)} study},\ }\href {https://doi.org/https://doi.org/10.1016/j.aop.2016.08.008} {\bibfield  {journal} {\bibinfo  {journal} {Annals of Physics}\ }\textbf {\bibinfo {volume} {374}},\ \bibinfo {pages} {84} (\bibinfo {year} {2016})}\BibitemShut {NoStop}%
\bibitem [{\citenamefont {Munk}\ \emph {et~al.}(2018)\citenamefont {Munk}, \citenamefont {Rasmussen},\ and\ \citenamefont {Burrello}}]{Munk_PRB2018}%
  \BibitemOpen
  \bibfield  {author} {\bibinfo {author} {\bibfnamefont {M.~I.~K.}\ \bibnamefont {Munk}}, \bibinfo {author} {\bibfnamefont {A.}~\bibnamefont {Rasmussen}},\ and\ \bibinfo {author} {\bibfnamefont {M.}~\bibnamefont {Burrello}},\ }\bibfield  {title} {\bibinfo {title} {Dyonic zero-energy modes},\ }\href {https://doi.org/10.1103/PhysRevB.98.245135} {\bibfield  {journal} {\bibinfo  {journal} {Phys. Rev. B}\ }\textbf {\bibinfo {volume} {98}},\ \bibinfo {pages} {245135} (\bibinfo {year} {2018})}\BibitemShut {NoStop}%
\bibitem [{\citenamefont {Semeghini}\ \emph {et~al.}(2021)\citenamefont {Semeghini}, \citenamefont {Levine}, \citenamefont {Keesling}, \citenamefont {Ebadi}, \citenamefont {Wang}, \citenamefont {Bluvstein}, \citenamefont {Verresen}, \citenamefont {Pichler}, \citenamefont {Kalinowski}, \citenamefont {Samajdar}, \citenamefont {Omran}, \citenamefont {Sachdev}, \citenamefont {Vishwanath}, \citenamefont {Greiner}, \citenamefont {Vuletić},\ and\ \citenamefont {Lukin}}]{Semeghini_Sci2021}%
  \BibitemOpen
  \bibfield  {author} {\bibinfo {author} {\bibfnamefont {G.}~\bibnamefont {Semeghini}}, \bibinfo {author} {\bibfnamefont {H.}~\bibnamefont {Levine}}, \bibinfo {author} {\bibfnamefont {A.}~\bibnamefont {Keesling}}, \bibinfo {author} {\bibfnamefont {S.}~\bibnamefont {Ebadi}}, \bibinfo {author} {\bibfnamefont {T.~T.}\ \bibnamefont {Wang}}, \bibinfo {author} {\bibfnamefont {D.}~\bibnamefont {Bluvstein}}, \bibinfo {author} {\bibfnamefont {R.}~\bibnamefont {Verresen}}, \bibinfo {author} {\bibfnamefont {H.}~\bibnamefont {Pichler}}, \bibinfo {author} {\bibfnamefont {M.}~\bibnamefont {Kalinowski}}, \bibinfo {author} {\bibfnamefont {R.}~\bibnamefont {Samajdar}}, \bibinfo {author} {\bibfnamefont {A.}~\bibnamefont {Omran}}, \bibinfo {author} {\bibfnamefont {S.}~\bibnamefont {Sachdev}}, \bibinfo {author} {\bibfnamefont {A.}~\bibnamefont {Vishwanath}}, \bibinfo {author} {\bibfnamefont {M.}~\bibnamefont {Greiner}}, \bibinfo {author} {\bibfnamefont {V.}~\bibnamefont {Vuletić}},\ and\ \bibinfo {author} {\bibfnamefont
  {M.~D.}\ \bibnamefont {Lukin}},\ }\bibfield  {title} {\bibinfo {title} {Probing topological spin liquids on a programmable quantum simulator},\ }\href {https://doi.org/10.1126/science.abi8794} {\bibfield  {journal} {\bibinfo  {journal} {Science}\ }\textbf {\bibinfo {volume} {374}},\ \bibinfo {pages} {1242} (\bibinfo {year} {2021})}\BibitemShut {NoStop}%
\bibitem [{\citenamefont {Satzinger}\ \emph {et~al.}(2021)\citenamefont {Satzinger}, \citenamefont {Liu}, \citenamefont {Smith}, \citenamefont {Knapp}, \citenamefont {Newman}, \citenamefont {Jones}, \citenamefont {Chen}, \citenamefont {Quintana}, \citenamefont {Mi}, \citenamefont {Dunsworth}, \citenamefont {Gidney}, \citenamefont {Aleiner}, \citenamefont {Arute}, \citenamefont {Arya}, \citenamefont {Atalaya}, \citenamefont {Babbush}, \citenamefont {Bardin}, \citenamefont {Barends}, \citenamefont {Basso}, \citenamefont {Bengtsson}, \citenamefont {Bilmes}, \citenamefont {Broughton}, \citenamefont {Buckley}, \citenamefont {Buell}, \citenamefont {Burkett}, \citenamefont {Bushnell}, \citenamefont {Chiaro}, \citenamefont {Collins}, \citenamefont {Courtney}, \citenamefont {Demura}, \citenamefont {Derk}, \citenamefont {Eppens}, \citenamefont {Erickson}, \citenamefont {Faoro}, \citenamefont {Farhi}, \citenamefont {Fowler}, \citenamefont {Foxen}, \citenamefont {Giustina}, \citenamefont {Greene}, \citenamefont {Gross},
  \citenamefont {Harrigan}, \citenamefont {Harrington}, \citenamefont {Hilton}, \citenamefont {Hong}, \citenamefont {Huang}, \citenamefont {Huggins}, \citenamefont {Ioffe}, \citenamefont {Isakov}, \citenamefont {Jeffrey}, \citenamefont {Jiang}, \citenamefont {Kafri}, \citenamefont {Kechedzhi}, \citenamefont {Khattar}, \citenamefont {Kim}, \citenamefont {Klimov}, \citenamefont {Korotkov}, \citenamefont {Kostritsa}, \citenamefont {Landhuis}, \citenamefont {Laptev}, \citenamefont {Locharla}, \citenamefont {Lucero}, \citenamefont {Martin}, \citenamefont {McClean}, \citenamefont {McEwen}, \citenamefont {Miao}, \citenamefont {Mohseni}, \citenamefont {Montazeri}, \citenamefont {Mruczkiewicz}, \citenamefont {Mutus}, \citenamefont {Naaman}, \citenamefont {Neeley}, \citenamefont {Neill}, \citenamefont {Niu}, \citenamefont {O’Brien}, \citenamefont {Opremcak}, \citenamefont {Pató}, \citenamefont {Petukhov}, \citenamefont {Rubin}, \citenamefont {Sank}, \citenamefont {Shvarts}, \citenamefont {Strain}, \citenamefont
  {Szalay}, \citenamefont {Villalonga}, \citenamefont {White}, \citenamefont {Yao}, \citenamefont {Yeh}, \citenamefont {Yoo}, \citenamefont {Zalcman}, \citenamefont {Neven}, \citenamefont {Boixo}, \citenamefont {Megrant}, \citenamefont {Chen}, \citenamefont {Kelly}, \citenamefont {Smelyanskiy}, \citenamefont {Kitaev}, \citenamefont {Knap}, \citenamefont {Pollmann},\ and\ \citenamefont {Roushan}}]{Satzinger_Science2021}%
  \BibitemOpen
  \bibfield  {author} {\bibinfo {author} {\bibfnamefont {K.~J.}\ \bibnamefont {Satzinger}}, \bibinfo {author} {\bibfnamefont {Y.-J.}\ \bibnamefont {Liu}}, \bibinfo {author} {\bibfnamefont {A.}~\bibnamefont {Smith}}, \bibinfo {author} {\bibfnamefont {C.}~\bibnamefont {Knapp}}, \bibinfo {author} {\bibfnamefont {M.}~\bibnamefont {Newman}}, \bibinfo {author} {\bibfnamefont {C.}~\bibnamefont {Jones}}, \bibinfo {author} {\bibfnamefont {Z.}~\bibnamefont {Chen}}, \bibinfo {author} {\bibfnamefont {C.}~\bibnamefont {Quintana}}, \bibinfo {author} {\bibfnamefont {X.}~\bibnamefont {Mi}}, \bibinfo {author} {\bibfnamefont {A.}~\bibnamefont {Dunsworth}}, \bibinfo {author} {\bibfnamefont {C.}~\bibnamefont {Gidney}}, \bibinfo {author} {\bibfnamefont {I.}~\bibnamefont {Aleiner}}, \bibinfo {author} {\bibfnamefont {F.}~\bibnamefont {Arute}}, \bibinfo {author} {\bibfnamefont {K.}~\bibnamefont {Arya}}, \bibinfo {author} {\bibfnamefont {J.}~\bibnamefont {Atalaya}}, \bibinfo {author} {\bibfnamefont {R.}~\bibnamefont {Babbush}},
  \bibinfo {author} {\bibfnamefont {J.~C.}\ \bibnamefont {Bardin}}, \bibinfo {author} {\bibfnamefont {R.}~\bibnamefont {Barends}}, \bibinfo {author} {\bibfnamefont {J.}~\bibnamefont {Basso}}, \bibinfo {author} {\bibfnamefont {A.}~\bibnamefont {Bengtsson}}, \bibinfo {author} {\bibfnamefont {A.}~\bibnamefont {Bilmes}}, \bibinfo {author} {\bibfnamefont {M.}~\bibnamefont {Broughton}}, \bibinfo {author} {\bibfnamefont {B.~B.}\ \bibnamefont {Buckley}}, \bibinfo {author} {\bibfnamefont {D.~A.}\ \bibnamefont {Buell}}, \bibinfo {author} {\bibfnamefont {B.}~\bibnamefont {Burkett}}, \bibinfo {author} {\bibfnamefont {N.}~\bibnamefont {Bushnell}}, \bibinfo {author} {\bibfnamefont {B.}~\bibnamefont {Chiaro}}, \bibinfo {author} {\bibfnamefont {R.}~\bibnamefont {Collins}}, \bibinfo {author} {\bibfnamefont {W.}~\bibnamefont {Courtney}}, \bibinfo {author} {\bibfnamefont {S.}~\bibnamefont {Demura}}, \bibinfo {author} {\bibfnamefont {A.~R.}\ \bibnamefont {Derk}}, \bibinfo {author} {\bibfnamefont {D.}~\bibnamefont {Eppens}},
  \bibinfo {author} {\bibfnamefont {C.}~\bibnamefont {Erickson}}, \bibinfo {author} {\bibfnamefont {L.}~\bibnamefont {Faoro}}, \bibinfo {author} {\bibfnamefont {E.}~\bibnamefont {Farhi}}, \bibinfo {author} {\bibfnamefont {A.~G.}\ \bibnamefont {Fowler}}, \bibinfo {author} {\bibfnamefont {B.}~\bibnamefont {Foxen}}, \bibinfo {author} {\bibfnamefont {M.}~\bibnamefont {Giustina}}, \bibinfo {author} {\bibfnamefont {A.}~\bibnamefont {Greene}}, \bibinfo {author} {\bibfnamefont {J.~A.}\ \bibnamefont {Gross}}, \bibinfo {author} {\bibfnamefont {M.~P.}\ \bibnamefont {Harrigan}}, \bibinfo {author} {\bibfnamefont {S.~D.}\ \bibnamefont {Harrington}}, \bibinfo {author} {\bibfnamefont {J.}~\bibnamefont {Hilton}}, \bibinfo {author} {\bibfnamefont {S.}~\bibnamefont {Hong}}, \bibinfo {author} {\bibfnamefont {T.}~\bibnamefont {Huang}}, \bibinfo {author} {\bibfnamefont {W.~J.}\ \bibnamefont {Huggins}}, \bibinfo {author} {\bibfnamefont {L.~B.}\ \bibnamefont {Ioffe}}, \bibinfo {author} {\bibfnamefont {S.~V.}\ \bibnamefont {Isakov}},
  \bibinfo {author} {\bibfnamefont {E.}~\bibnamefont {Jeffrey}}, \bibinfo {author} {\bibfnamefont {Z.}~\bibnamefont {Jiang}}, \bibinfo {author} {\bibfnamefont {D.}~\bibnamefont {Kafri}}, \bibinfo {author} {\bibfnamefont {K.}~\bibnamefont {Kechedzhi}}, \bibinfo {author} {\bibfnamefont {T.}~\bibnamefont {Khattar}}, \bibinfo {author} {\bibfnamefont {S.}~\bibnamefont {Kim}}, \bibinfo {author} {\bibfnamefont {P.~V.}\ \bibnamefont {Klimov}}, \bibinfo {author} {\bibfnamefont {A.~N.}\ \bibnamefont {Korotkov}}, \bibinfo {author} {\bibfnamefont {F.}~\bibnamefont {Kostritsa}}, \bibinfo {author} {\bibfnamefont {D.}~\bibnamefont {Landhuis}}, \bibinfo {author} {\bibfnamefont {P.}~\bibnamefont {Laptev}}, \bibinfo {author} {\bibfnamefont {A.}~\bibnamefont {Locharla}}, \bibinfo {author} {\bibfnamefont {E.}~\bibnamefont {Lucero}}, \bibinfo {author} {\bibfnamefont {O.}~\bibnamefont {Martin}}, \bibinfo {author} {\bibfnamefont {J.~R.}\ \bibnamefont {McClean}}, \bibinfo {author} {\bibfnamefont {M.}~\bibnamefont {McEwen}}, \bibinfo
  {author} {\bibfnamefont {K.~C.}\ \bibnamefont {Miao}}, \bibinfo {author} {\bibfnamefont {M.}~\bibnamefont {Mohseni}}, \bibinfo {author} {\bibfnamefont {S.}~\bibnamefont {Montazeri}}, \bibinfo {author} {\bibfnamefont {W.}~\bibnamefont {Mruczkiewicz}}, \bibinfo {author} {\bibfnamefont {J.}~\bibnamefont {Mutus}}, \bibinfo {author} {\bibfnamefont {O.}~\bibnamefont {Naaman}}, \bibinfo {author} {\bibfnamefont {M.}~\bibnamefont {Neeley}}, \bibinfo {author} {\bibfnamefont {C.}~\bibnamefont {Neill}}, \bibinfo {author} {\bibfnamefont {M.~Y.}\ \bibnamefont {Niu}}, \bibinfo {author} {\bibfnamefont {T.~E.}\ \bibnamefont {O’Brien}}, \bibinfo {author} {\bibfnamefont {A.}~\bibnamefont {Opremcak}}, \bibinfo {author} {\bibfnamefont {B.}~\bibnamefont {Pató}}, \bibinfo {author} {\bibfnamefont {A.}~\bibnamefont {Petukhov}}, \bibinfo {author} {\bibfnamefont {N.~C.}\ \bibnamefont {Rubin}}, \bibinfo {author} {\bibfnamefont {D.}~\bibnamefont {Sank}}, \bibinfo {author} {\bibfnamefont {V.}~\bibnamefont {Shvarts}}, \bibinfo
  {author} {\bibfnamefont {D.}~\bibnamefont {Strain}}, \bibinfo {author} {\bibfnamefont {M.}~\bibnamefont {Szalay}}, \bibinfo {author} {\bibfnamefont {B.}~\bibnamefont {Villalonga}}, \bibinfo {author} {\bibfnamefont {T.~C.}\ \bibnamefont {White}}, \bibinfo {author} {\bibfnamefont {Z.}~\bibnamefont {Yao}}, \bibinfo {author} {\bibfnamefont {P.}~\bibnamefont {Yeh}}, \bibinfo {author} {\bibfnamefont {J.}~\bibnamefont {Yoo}}, \bibinfo {author} {\bibfnamefont {A.}~\bibnamefont {Zalcman}}, \bibinfo {author} {\bibfnamefont {H.}~\bibnamefont {Neven}}, \bibinfo {author} {\bibfnamefont {S.}~\bibnamefont {Boixo}}, \bibinfo {author} {\bibfnamefont {A.}~\bibnamefont {Megrant}}, \bibinfo {author} {\bibfnamefont {Y.}~\bibnamefont {Chen}}, \bibinfo {author} {\bibfnamefont {J.}~\bibnamefont {Kelly}}, \bibinfo {author} {\bibfnamefont {V.}~\bibnamefont {Smelyanskiy}}, \bibinfo {author} {\bibfnamefont {A.}~\bibnamefont {Kitaev}}, \bibinfo {author} {\bibfnamefont {M.}~\bibnamefont {Knap}}, \bibinfo {author} {\bibfnamefont
  {F.}~\bibnamefont {Pollmann}},\ and\ \bibinfo {author} {\bibfnamefont {P.}~\bibnamefont {Roushan}},\ }\bibfield  {title} {\bibinfo {title} {Realizing topologically ordered states on a quantum processor},\ }\href {https://doi.org/10.1126/science.abi8378} {\bibfield  {journal} {\bibinfo  {journal} {Science}\ }\textbf {\bibinfo {volume} {374}},\ \bibinfo {pages} {1237} (\bibinfo {year} {2021})}\BibitemShut {NoStop}%
\bibitem [{\citenamefont {Lumia}\ \emph {et~al.}(2022)\citenamefont {Lumia}, \citenamefont {Torta}, \citenamefont {Mbeng}, \citenamefont {Santoro}, \citenamefont {Ercolessi}, \citenamefont {Burrello},\ and\ \citenamefont {Wauters}}]{Lumia_PRXQ2022}%
  \BibitemOpen
  \bibfield  {author} {\bibinfo {author} {\bibfnamefont {L.}~\bibnamefont {Lumia}}, \bibinfo {author} {\bibfnamefont {P.}~\bibnamefont {Torta}}, \bibinfo {author} {\bibfnamefont {G.~B.}\ \bibnamefont {Mbeng}}, \bibinfo {author} {\bibfnamefont {G.~E.}\ \bibnamefont {Santoro}}, \bibinfo {author} {\bibfnamefont {E.}~\bibnamefont {Ercolessi}}, \bibinfo {author} {\bibfnamefont {M.}~\bibnamefont {Burrello}},\ and\ \bibinfo {author} {\bibfnamefont {M.~M.}\ \bibnamefont {Wauters}},\ }\bibfield  {title} {\bibinfo {title} {Two-dimensional $\mathbb{Z}_{2}$ lattice gauge theory on a near-term quantum simulator: Variational quantum optimization, confinement, and topological order},\ }\href {https://doi.org/10.1103/PRXQuantum.3.020320} {\bibfield  {journal} {\bibinfo  {journal} {PRX Quantum}\ }\textbf {\bibinfo {volume} {3}},\ \bibinfo {pages} {020320} (\bibinfo {year} {2022})}\BibitemShut {NoStop}%
\bibitem [{\citenamefont {Cataldi}\ \emph {et~al.}(2024)\citenamefont {Cataldi}, \citenamefont {Magnifico}, \citenamefont {Silvi},\ and\ \citenamefont {Montangero}}]{Cataldi_PRR2024}%
  \BibitemOpen
  \bibfield  {author} {\bibinfo {author} {\bibfnamefont {G.}~\bibnamefont {Cataldi}}, \bibinfo {author} {\bibfnamefont {G.}~\bibnamefont {Magnifico}}, \bibinfo {author} {\bibfnamefont {P.}~\bibnamefont {Silvi}},\ and\ \bibinfo {author} {\bibfnamefont {S.}~\bibnamefont {Montangero}},\ }\bibfield  {title} {\bibinfo {title} {Simulating $(2+1)\mathrm{D}$ {SU(2)} yang-mills lattice gauge theory at finite density with tensor networks},\ }\href {https://doi.org/10.1103/PhysRevResearch.6.033057} {\bibfield  {journal} {\bibinfo  {journal} {Phys. Rev. Res.}\ }\textbf {\bibinfo {volume} {6}},\ \bibinfo {pages} {033057} (\bibinfo {year} {2024})}\BibitemShut {NoStop}%
\bibitem [{\citenamefont {Pradhan}\ \emph {et~al.}(2024)\citenamefont {Pradhan}, \citenamefont {Maroncelli},\ and\ \citenamefont {Ercolessi}}]{Pradhan2024prd}%
  \BibitemOpen
  \bibfield  {author} {\bibinfo {author} {\bibfnamefont {S.}~\bibnamefont {Pradhan}}, \bibinfo {author} {\bibfnamefont {A.}~\bibnamefont {Maroncelli}},\ and\ \bibinfo {author} {\bibfnamefont {E.}~\bibnamefont {Ercolessi}},\ }\bibfield  {title} {\bibinfo {title} {Discrete abelian lattice gauge theories on a ladder and their dualities with quantum clock models},\ }\href {https://doi.org/10.1103/PhysRevB.109.064410} {\bibfield  {journal} {\bibinfo  {journal} {Phys. Rev. B}\ }\textbf {\bibinfo {volume} {109}},\ \bibinfo {pages} {064410} (\bibinfo {year} {2024})}\BibitemShut {NoStop}%
\bibitem [{\citenamefont {Silvi}\ \emph {et~al.}(2017)\citenamefont {Silvi}, \citenamefont {Rico}, \citenamefont {Dalmonte}, \citenamefont {Tschirsich},\ and\ \citenamefont {Montangero}}]{Silvi2017finitedensityphase}%
  \BibitemOpen
  \bibfield  {author} {\bibinfo {author} {\bibfnamefont {P.}~\bibnamefont {Silvi}}, \bibinfo {author} {\bibfnamefont {E.}~\bibnamefont {Rico}}, \bibinfo {author} {\bibfnamefont {M.}~\bibnamefont {Dalmonte}}, \bibinfo {author} {\bibfnamefont {F.}~\bibnamefont {Tschirsich}},\ and\ \bibinfo {author} {\bibfnamefont {S.}~\bibnamefont {Montangero}},\ }\bibfield  {title} {\bibinfo {title} {Finite-density phase diagram of a {$(1+1)-d$} non-abelian lattice gauge theory with tensor networks},\ }\href {https://doi.org/10.22331/q-2017-04-25-9} {\bibfield  {journal} {\bibinfo  {journal} {{Quantum}}\ }\textbf {\bibinfo {volume} {1}},\ \bibinfo {pages} {9} (\bibinfo {year} {2017})}\BibitemShut {NoStop}%
\bibitem [{\citenamefont {Di~Meglio}\ \emph {et~al.}(2024)\citenamefont {Di~Meglio}, \citenamefont {Jansen}, \citenamefont {Tavernelli}, \citenamefont {Alexandrou}, \citenamefont {Arunachalam}, \citenamefont {Bauer}, \citenamefont {Borras}, \citenamefont {Carrazza}, \citenamefont {Crippa}, \citenamefont {Croft}, \citenamefont {de~Putter}, \citenamefont {Delgado}, \citenamefont {Dunjko}, \citenamefont {Egger}, \citenamefont {Fern\'andez-Combarro}, \citenamefont {Fuchs}, \citenamefont {Funcke}, \citenamefont {Gonz\'alez-Cuadra}, \citenamefont {Grossi}, \citenamefont {Halimeh}, \citenamefont {Holmes}, \citenamefont {K\"uhn}, \citenamefont {Lacroix}, \citenamefont {Lewis}, \citenamefont {Lucchesi}, \citenamefont {Martinez}, \citenamefont {Meloni}, \citenamefont {Mezzacapo}, \citenamefont {Montangero}, \citenamefont {Nagano}, \citenamefont {Pascuzzi}, \citenamefont {Radescu}, \citenamefont {Ortega}, \citenamefont {Roggero}, \citenamefont {Schuhmacher}, \citenamefont {Seixas}, \citenamefont {Silvi}, \citenamefont
  {Spentzouris}, \citenamefont {Tacchino}, \citenamefont {Temme}, \citenamefont {Terashi}, \citenamefont {Tura}, \citenamefont {T\"uys\"uz}, \citenamefont {Vallecorsa}, \citenamefont {Wiese}, \citenamefont {Yoo},\ and\ \citenamefont {Zhang}}]{DiMeglio_PRXQ2024}%
  \BibitemOpen
  \bibfield  {author} {\bibinfo {author} {\bibfnamefont {A.}~\bibnamefont {Di~Meglio}}, \bibinfo {author} {\bibfnamefont {K.}~\bibnamefont {Jansen}}, \bibinfo {author} {\bibfnamefont {I.}~\bibnamefont {Tavernelli}}, \bibinfo {author} {\bibfnamefont {C.}~\bibnamefont {Alexandrou}}, \bibinfo {author} {\bibfnamefont {S.}~\bibnamefont {Arunachalam}}, \bibinfo {author} {\bibfnamefont {C.~W.}\ \bibnamefont {Bauer}}, \bibinfo {author} {\bibfnamefont {K.}~\bibnamefont {Borras}}, \bibinfo {author} {\bibfnamefont {S.}~\bibnamefont {Carrazza}}, \bibinfo {author} {\bibfnamefont {A.}~\bibnamefont {Crippa}}, \bibinfo {author} {\bibfnamefont {V.}~\bibnamefont {Croft}}, \bibinfo {author} {\bibfnamefont {R.}~\bibnamefont {de~Putter}}, \bibinfo {author} {\bibfnamefont {A.}~\bibnamefont {Delgado}}, \bibinfo {author} {\bibfnamefont {V.}~\bibnamefont {Dunjko}}, \bibinfo {author} {\bibfnamefont {D.~J.}\ \bibnamefont {Egger}}, \bibinfo {author} {\bibfnamefont {E.}~\bibnamefont {Fern\'andez-Combarro}}, \bibinfo {author}
  {\bibfnamefont {E.}~\bibnamefont {Fuchs}}, \bibinfo {author} {\bibfnamefont {L.}~\bibnamefont {Funcke}}, \bibinfo {author} {\bibfnamefont {D.}~\bibnamefont {Gonz\'alez-Cuadra}}, \bibinfo {author} {\bibfnamefont {M.}~\bibnamefont {Grossi}}, \bibinfo {author} {\bibfnamefont {J.~C.}\ \bibnamefont {Halimeh}}, \bibinfo {author} {\bibfnamefont {Z.}~\bibnamefont {Holmes}}, \bibinfo {author} {\bibfnamefont {S.}~\bibnamefont {K\"uhn}}, \bibinfo {author} {\bibfnamefont {D.}~\bibnamefont {Lacroix}}, \bibinfo {author} {\bibfnamefont {R.}~\bibnamefont {Lewis}}, \bibinfo {author} {\bibfnamefont {D.}~\bibnamefont {Lucchesi}}, \bibinfo {author} {\bibfnamefont {M.~L.}\ \bibnamefont {Martinez}}, \bibinfo {author} {\bibfnamefont {F.}~\bibnamefont {Meloni}}, \bibinfo {author} {\bibfnamefont {A.}~\bibnamefont {Mezzacapo}}, \bibinfo {author} {\bibfnamefont {S.}~\bibnamefont {Montangero}}, \bibinfo {author} {\bibfnamefont {L.}~\bibnamefont {Nagano}}, \bibinfo {author} {\bibfnamefont {V.~R.}\ \bibnamefont {Pascuzzi}}, \bibinfo
  {author} {\bibfnamefont {V.}~\bibnamefont {Radescu}}, \bibinfo {author} {\bibfnamefont {E.~R.}\ \bibnamefont {Ortega}}, \bibinfo {author} {\bibfnamefont {A.}~\bibnamefont {Roggero}}, \bibinfo {author} {\bibfnamefont {J.}~\bibnamefont {Schuhmacher}}, \bibinfo {author} {\bibfnamefont {J.}~\bibnamefont {Seixas}}, \bibinfo {author} {\bibfnamefont {P.}~\bibnamefont {Silvi}}, \bibinfo {author} {\bibfnamefont {P.}~\bibnamefont {Spentzouris}}, \bibinfo {author} {\bibfnamefont {F.}~\bibnamefont {Tacchino}}, \bibinfo {author} {\bibfnamefont {K.}~\bibnamefont {Temme}}, \bibinfo {author} {\bibfnamefont {K.}~\bibnamefont {Terashi}}, \bibinfo {author} {\bibfnamefont {J.}~\bibnamefont {Tura}}, \bibinfo {author} {\bibfnamefont {C.}~\bibnamefont {T\"uys\"uz}}, \bibinfo {author} {\bibfnamefont {S.}~\bibnamefont {Vallecorsa}}, \bibinfo {author} {\bibfnamefont {U.-J.}\ \bibnamefont {Wiese}}, \bibinfo {author} {\bibfnamefont {S.}~\bibnamefont {Yoo}},\ and\ \bibinfo {author} {\bibfnamefont {J.}~\bibnamefont {Zhang}},\ }\bibfield
   {title} {\bibinfo {title} {Quantum computing for high-energy physics: State of the art and challenges},\ }\href {https://doi.org/10.1103/PRXQuantum.5.037001} {\bibfield  {journal} {\bibinfo  {journal} {PRX Quantum}\ }\textbf {\bibinfo {volume} {5}},\ \bibinfo {pages} {037001} (\bibinfo {year} {2024})}\BibitemShut {NoStop}%
\bibitem [{\citenamefont {Gonz\'alez-Cuadra}\ \emph {et~al.}(2022)\citenamefont {Gonz\'alez-Cuadra}, \citenamefont {Zache}, \citenamefont {Carrasco}, \citenamefont {Kraus},\ and\ \citenamefont {Zoller}}]{Gonzalezcuadra_PRL2022}%
  \BibitemOpen
  \bibfield  {author} {\bibinfo {author} {\bibfnamefont {D.}~\bibnamefont {Gonz\'alez-Cuadra}}, \bibinfo {author} {\bibfnamefont {T.~V.}\ \bibnamefont {Zache}}, \bibinfo {author} {\bibfnamefont {J.}~\bibnamefont {Carrasco}}, \bibinfo {author} {\bibfnamefont {B.}~\bibnamefont {Kraus}},\ and\ \bibinfo {author} {\bibfnamefont {P.}~\bibnamefont {Zoller}},\ }\bibfield  {title} {\bibinfo {title} {Hardware efficient quantum simulation of non-abelian gauge theories with qudits on rydberg platforms},\ }\href {https://doi.org/10.1103/PhysRevLett.129.160501} {\bibfield  {journal} {\bibinfo  {journal} {Phys. Rev. Lett.}\ }\textbf {\bibinfo {volume} {129}},\ \bibinfo {pages} {160501} (\bibinfo {year} {2022})}\BibitemShut {NoStop}%
\bibitem [{\citenamefont {Gustafson}\ \emph {et~al.}(2022)\citenamefont {Gustafson}, \citenamefont {Lamm}, \citenamefont {Lovelace},\ and\ \citenamefont {Musk}}]{Gustafson_PRD2022}%
  \BibitemOpen
  \bibfield  {author} {\bibinfo {author} {\bibfnamefont {E.~J.}\ \bibnamefont {Gustafson}}, \bibinfo {author} {\bibfnamefont {H.}~\bibnamefont {Lamm}}, \bibinfo {author} {\bibfnamefont {F.}~\bibnamefont {Lovelace}},\ and\ \bibinfo {author} {\bibfnamefont {D.}~\bibnamefont {Musk}},\ }\bibfield  {title} {\bibinfo {title} {Primitive quantum gates for an $su(2)$ discrete subgroup: Binary tetrahedral},\ }\href {https://doi.org/10.1103/PhysRevD.106.114501} {\bibfield  {journal} {\bibinfo  {journal} {Phys. Rev. D}\ }\textbf {\bibinfo {volume} {106}},\ \bibinfo {pages} {114501} (\bibinfo {year} {2022})}\BibitemShut {NoStop}%
\bibitem [{\citenamefont {Gaz}\ \emph {et~al.}(2025)\citenamefont {Gaz}, \citenamefont {Popov}, \citenamefont {Pardo}, \citenamefont {Lewenstein}, \citenamefont {Hauke},\ and\ \citenamefont {Zohar}}]{Gaz_PRR2025}%
  \BibitemOpen
  \bibfield  {author} {\bibinfo {author} {\bibfnamefont {E.}~\bibnamefont {Gaz}}, \bibinfo {author} {\bibfnamefont {P.~P.}\ \bibnamefont {Popov}}, \bibinfo {author} {\bibfnamefont {G.}~\bibnamefont {Pardo}}, \bibinfo {author} {\bibfnamefont {M.}~\bibnamefont {Lewenstein}}, \bibinfo {author} {\bibfnamefont {P.}~\bibnamefont {Hauke}},\ and\ \bibinfo {author} {\bibfnamefont {E.}~\bibnamefont {Zohar}},\ }\bibfield  {title} {\bibinfo {title} {Quantum simulation of non-abelian lattice gauge theories: A variational approach to ${\mathbb{d}}_{8}$ with dynamical matter},\ }\href {https://doi.org/10.1103/l8b6-h5s5} {\bibfield  {journal} {\bibinfo  {journal} {Phys. Rev. Res.}\ }\textbf {\bibinfo {volume} {7}},\ \bibinfo {pages} {033012} (\bibinfo {year} {2025})}\BibitemShut {NoStop}%
\bibitem [{\citenamefont {Notarnicola}\ \emph {et~al.}(2015)\citenamefont {Notarnicola}, \citenamefont {Ercolessi}, \citenamefont {Facchi}, \citenamefont {Marmo}, \citenamefont {Pascazio},\ and\ \citenamefont {Pepe}}]{Notarnicola_JPA2015}%
  \BibitemOpen
  \bibfield  {author} {\bibinfo {author} {\bibfnamefont {S.}~\bibnamefont {Notarnicola}}, \bibinfo {author} {\bibfnamefont {E.}~\bibnamefont {Ercolessi}}, \bibinfo {author} {\bibfnamefont {P.}~\bibnamefont {Facchi}}, \bibinfo {author} {\bibfnamefont {G.}~\bibnamefont {Marmo}}, \bibinfo {author} {\bibfnamefont {S.}~\bibnamefont {Pascazio}},\ and\ \bibinfo {author} {\bibfnamefont {F.~V.}\ \bibnamefont {Pepe}},\ }\bibfield  {title} {\bibinfo {title} {Discrete abelian gauge theories for quantum simulations of qed},\ }\href {https://doi.org/10.1088/1751-8113/48/30/30FT01} {\bibfield  {journal} {\bibinfo  {journal} {Journal of Physics A: Mathematical and Theoretical}\ }\textbf {\bibinfo {volume} {48}},\ \bibinfo {pages} {30FT01} (\bibinfo {year} {2015})}\BibitemShut {NoStop}%
\bibitem [{\citenamefont {Magnifico}\ \emph {et~al.}(2019)\citenamefont {Magnifico}, \citenamefont {Vodola}, \citenamefont {Ercolessi}, \citenamefont {Kumar}, \citenamefont {M\"uller},\ and\ \citenamefont {Bermudez}}]{Magnifico_PRB2019}%
  \BibitemOpen
  \bibfield  {author} {\bibinfo {author} {\bibfnamefont {G.}~\bibnamefont {Magnifico}}, \bibinfo {author} {\bibfnamefont {D.}~\bibnamefont {Vodola}}, \bibinfo {author} {\bibfnamefont {E.}~\bibnamefont {Ercolessi}}, \bibinfo {author} {\bibfnamefont {S.~P.}\ \bibnamefont {Kumar}}, \bibinfo {author} {\bibfnamefont {M.}~\bibnamefont {M\"uller}},\ and\ \bibinfo {author} {\bibfnamefont {A.}~\bibnamefont {Bermudez}},\ }\bibfield  {title} {\bibinfo {title} {${\mathbb{z}}_{N}$ gauge theories coupled to topological fermions: ${\mathrm{qed}}_{2}$ with a quantum mechanical $\ensuremath{\theta}$ angle},\ }\href {https://doi.org/10.1103/PhysRevB.100.115152} {\bibfield  {journal} {\bibinfo  {journal} {Phys. Rev. B}\ }\textbf {\bibinfo {volume} {100}},\ \bibinfo {pages} {115152} (\bibinfo {year} {2019})}\BibitemShut {NoStop}%
\bibitem [{\citenamefont {Magnifico}\ \emph {et~al.}(2020)\citenamefont {Magnifico}, \citenamefont {Dalmonte}, \citenamefont {Facchi}, \citenamefont {Pascazio}, \citenamefont {Pepe},\ and\ \citenamefont {Ercolessi}}]{Magnifico_Quantum2020}%
  \BibitemOpen
  \bibfield  {author} {\bibinfo {author} {\bibfnamefont {G.}~\bibnamefont {Magnifico}}, \bibinfo {author} {\bibfnamefont {M.}~\bibnamefont {Dalmonte}}, \bibinfo {author} {\bibfnamefont {P.}~\bibnamefont {Facchi}}, \bibinfo {author} {\bibfnamefont {S.}~\bibnamefont {Pascazio}}, \bibinfo {author} {\bibfnamefont {F.~V.}\ \bibnamefont {Pepe}},\ and\ \bibinfo {author} {\bibfnamefont {E.}~\bibnamefont {Ercolessi}},\ }\bibfield  {title} {\bibinfo {title} {Real {T}ime {D}ynamics and {C}onfinement in the {$\mathbb{Z}_{n}$} {S}chwinger-{W}eyl lattice model for 1+1 {QED}},\ }\href {https://doi.org/10.22331/q-2020-06-15-281} {\bibfield  {journal} {\bibinfo  {journal} {{Quantum}}\ }\textbf {\bibinfo {volume} {4}},\ \bibinfo {pages} {281} (\bibinfo {year} {2020})}\BibitemShut {NoStop}%
\bibitem [{\citenamefont {Meth}\ \emph {et~al.}(2025)\citenamefont {Meth}, \citenamefont {Zhang}, \citenamefont {Haase}, \citenamefont {Edmunds}, \citenamefont {Postler}, \citenamefont {Jena}, \citenamefont {Steiner}, \citenamefont {Dellantonio}, \citenamefont {Blatt}, \citenamefont {Zoller}, \citenamefont {Monz}, \citenamefont {Schindler}, \citenamefont {Muschik},\ and\ \citenamefont {Ringbauer}}]{meth2023simulating}%
  \BibitemOpen
  \bibfield  {author} {\bibinfo {author} {\bibfnamefont {M.}~\bibnamefont {Meth}}, \bibinfo {author} {\bibfnamefont {J.}~\bibnamefont {Zhang}}, \bibinfo {author} {\bibfnamefont {J.~F.}\ \bibnamefont {Haase}}, \bibinfo {author} {\bibfnamefont {C.}~\bibnamefont {Edmunds}}, \bibinfo {author} {\bibfnamefont {L.}~\bibnamefont {Postler}}, \bibinfo {author} {\bibfnamefont {A.~J.}\ \bibnamefont {Jena}}, \bibinfo {author} {\bibfnamefont {A.}~\bibnamefont {Steiner}}, \bibinfo {author} {\bibfnamefont {L.}~\bibnamefont {Dellantonio}}, \bibinfo {author} {\bibfnamefont {R.}~\bibnamefont {Blatt}}, \bibinfo {author} {\bibfnamefont {P.}~\bibnamefont {Zoller}}, \bibinfo {author} {\bibfnamefont {T.}~\bibnamefont {Monz}}, \bibinfo {author} {\bibfnamefont {P.}~\bibnamefont {Schindler}}, \bibinfo {author} {\bibfnamefont {C.}~\bibnamefont {Muschik}},\ and\ \bibinfo {author} {\bibfnamefont {M.}~\bibnamefont {Ringbauer}},\ }\bibfield  {title} {\bibinfo {title} {Simulating two-dimensional lattice gauge theories on a qudit quantum
  computer},\ }\href {https://doi.org/10.1038/s41567-025-02797-w} {\bibfield  {journal} {\bibinfo  {journal} {Nature Physics}\ }\textbf {\bibinfo {volume} {21}},\ \bibinfo {pages} {570} (\bibinfo {year} {2025})}\BibitemShut {NoStop}%
\bibitem [{\citenamefont {Calliari}\ \emph {et~al.}(2025)\citenamefont {Calliari}, \citenamefont {Ott}, \citenamefont {Pichler},\ and\ \citenamefont {Zache}}]{calliari_2025}%
  \BibitemOpen
  \bibfield  {author} {\bibinfo {author} {\bibfnamefont {G.}~\bibnamefont {Calliari}}, \bibinfo {author} {\bibfnamefont {R.}~\bibnamefont {Ott}}, \bibinfo {author} {\bibfnamefont {H.}~\bibnamefont {Pichler}},\ and\ \bibinfo {author} {\bibfnamefont {T.~V.}\ \bibnamefont {Zache}},\ }\bibfield  {title} {\bibinfo {title} {Field digitization scaling in a {$\mathbb{Z}_N \subset U(1)$} symmetric model},\ }\href {https://arxiv.org/abs/2507.22984} {\bibfield  {journal} {\bibinfo  {journal} {arXiv:2507.22984}\ } (\bibinfo {year} {2025})}\BibitemShut {NoStop}%
\bibitem [{\citenamefont {Shimony}(1995)}]{shimony1995degree}%
  \BibitemOpen
  \bibfield  {author} {\bibinfo {author} {\bibfnamefont {A.}~\bibnamefont {Shimony}},\ }\bibfield  {title} {\bibinfo {title} {Degree of entanglement},\ }\href {https://nyaspubs.onlinelibrary.wiley.com/doi/epdf/10.1111/j.1749-6632.1995.tb39008.x} {\bibfield  {journal} {\bibinfo  {journal} {Annals of the New York Academy of Sciences}\ }\textbf {\bibinfo {volume} {755}},\ \bibinfo {pages} {675} (\bibinfo {year} {1995})}\BibitemShut {NoStop}%
\bibitem [{\citenamefont {Barnum}\ and\ \citenamefont {Linden}(2001)}]{barnum2001monotones}%
  \BibitemOpen
  \bibfield  {author} {\bibinfo {author} {\bibfnamefont {H.}~\bibnamefont {Barnum}}\ and\ \bibinfo {author} {\bibfnamefont {N.}~\bibnamefont {Linden}},\ }\bibfield  {title} {\bibinfo {title} {Monotones and invariants for multi-particle quantum states},\ }\href {https://iopscience.iop.org/article/10.1088/0305-4470/34/35/305} {\bibfield  {journal} {\bibinfo  {journal} {Journal of Physics A: Mathematical and General}\ }\textbf {\bibinfo {volume} {34}},\ \bibinfo {pages} {6787} (\bibinfo {year} {2001})}\BibitemShut {NoStop}%
\bibitem [{\citenamefont {Wei}\ and\ \citenamefont {Goldbart}(2003)}]{wei2003geometric}%
  \BibitemOpen
  \bibfield  {author} {\bibinfo {author} {\bibfnamefont {T.-C.}\ \bibnamefont {Wei}}\ and\ \bibinfo {author} {\bibfnamefont {P.~M.}\ \bibnamefont {Goldbart}},\ }\bibfield  {title} {\bibinfo {title} {Geometric measure of entanglement and applications to bipartite and multipartite quantum states},\ }\href {https://doi.org/10.1103/PhysRevA.68.042307} {\bibfield  {journal} {\bibinfo  {journal} {Phys. Rev. A}\ }\textbf {\bibinfo {volume} {68}},\ \bibinfo {pages} {042307} (\bibinfo {year} {2003})}\BibitemShut {NoStop}%
\bibitem [{\citenamefont {Sen(De)}\ and\ \citenamefont {Sen}(2010)}]{sende2010pra}%
  \BibitemOpen
  \bibfield  {author} {\bibinfo {author} {\bibfnamefont {A.}~\bibnamefont {Sen(De)}}\ and\ \bibinfo {author} {\bibfnamefont {U.}~\bibnamefont {Sen}},\ }\bibfield  {title} {\bibinfo {title} {Channel capacities versus entanglement measures in multiparty quantum states},\ }\href {https://doi.org/10.1103/PhysRevA.81.012308} {\bibfield  {journal} {\bibinfo  {journal} {Phys. Rev. A}\ }\textbf {\bibinfo {volume} {81}},\ \bibinfo {pages} {012308} (\bibinfo {year} {2010})}\BibitemShut {NoStop}%
\bibitem [{\citenamefont {Biswas}\ \emph {et~al.}(2014)\citenamefont {Biswas}, \citenamefont {Prabhu}, \citenamefont {Sen(De)},\ and\ \citenamefont {Sen}}]{biswas2014pra}%
  \BibitemOpen
  \bibfield  {author} {\bibinfo {author} {\bibfnamefont {A.}~\bibnamefont {Biswas}}, \bibinfo {author} {\bibfnamefont {R.}~\bibnamefont {Prabhu}}, \bibinfo {author} {\bibfnamefont {A.}~\bibnamefont {Sen(De)}},\ and\ \bibinfo {author} {\bibfnamefont {U.}~\bibnamefont {Sen}},\ }\bibfield  {title} {\bibinfo {title} {Genuine-multipartite-entanglement trends in gapless-to-gapped transitions of quantum spin systems},\ }\href {https://doi.org/10.1103/PhysRevA.90.032301} {\bibfield  {journal} {\bibinfo  {journal} {Phys. Rev. A}\ }\textbf {\bibinfo {volume} {90}},\ \bibinfo {pages} {032301} (\bibinfo {year} {2014})}\BibitemShut {NoStop}%
\bibitem [{\citenamefont {Das}\ \emph {et~al.}(2016)\citenamefont {Das}, \citenamefont {Roy}, \citenamefont {Bagchi}, \citenamefont {Misra}, \citenamefont {Sen(De)},\ and\ \citenamefont {Sen}}]{das2016generalized}%
  \BibitemOpen
  \bibfield  {author} {\bibinfo {author} {\bibfnamefont {T.}~\bibnamefont {Das}}, \bibinfo {author} {\bibfnamefont {S.~S.}\ \bibnamefont {Roy}}, \bibinfo {author} {\bibfnamefont {S.}~\bibnamefont {Bagchi}}, \bibinfo {author} {\bibfnamefont {A.}~\bibnamefont {Misra}}, \bibinfo {author} {\bibfnamefont {A.}~\bibnamefont {Sen(De)}},\ and\ \bibinfo {author} {\bibfnamefont {U.}~\bibnamefont {Sen}},\ }\bibfield  {title} {\bibinfo {title} {Generalized geometric measure of entanglement for multiparty mixed states},\ }\href {https://doi.org/10.1103/PhysRevA.94.022336} {\bibfield  {journal} {\bibinfo  {journal} {Phys. Rev. A}\ }\textbf {\bibinfo {volume} {94}},\ \bibinfo {pages} {022336} (\bibinfo {year} {2016})}\BibitemShut {NoStop}%
\bibitem [{\citenamefont {Haug}\ and\ \citenamefont {Piroli}(2023)}]{haug2023stabilizer}%
  \BibitemOpen
  \bibfield  {author} {\bibinfo {author} {\bibfnamefont {T.}~\bibnamefont {Haug}}\ and\ \bibinfo {author} {\bibfnamefont {L.}~\bibnamefont {Piroli}},\ }\bibfield  {title} {\bibinfo {title} {Stabilizer entropies and nonstabilizerness monotones},\ }\href {https://doi.org/10.22331/q-2023-08-28-1092} {\bibfield  {journal} {\bibinfo  {journal} {Quantum}\ }\textbf {\bibinfo {volume} {7}},\ \bibinfo {pages} {1092} (\bibinfo {year} {2023})}\BibitemShut {NoStop}%
\bibitem [{\citenamefont {Gross}\ \emph {et~al.}(2021)\citenamefont {Gross}, \citenamefont {Nezami},\ and\ \citenamefont {Walter}}]{gross2021schurweylduality}%
  \BibitemOpen
  \bibfield  {author} {\bibinfo {author} {\bibfnamefont {D.}~\bibnamefont {Gross}}, \bibinfo {author} {\bibfnamefont {S.}~\bibnamefont {Nezami}},\ and\ \bibinfo {author} {\bibfnamefont {M.}~\bibnamefont {Walter}},\ }\bibfield  {title} {\bibinfo {title} {Schur–weyl duality for the clifford group with applications: Property testing, a robust hudson theorem, and de finetti representations},\ }\href {https://doi.org/10.1007/s00220-021-04118-7} {\bibfield  {journal} {\bibinfo  {journal} {Commun. Math. Phys.}\ }\textbf {\bibinfo {volume} {385}},\ \bibinfo {pages} {1325} (\bibinfo {year} {2021})}\BibitemShut {NoStop}%
\bibitem [{\citenamefont {Niroula}\ \emph {et~al.}(2024)\citenamefont {Niroula}, \citenamefont {Wang}, \citenamefont {Johri}, \citenamefont {Zhu}, \citenamefont {Monroe}, \citenamefont {Noel},\ and\ \citenamefont {Gullans}}]{Niroula2024}%
  \BibitemOpen
  \bibfield  {author} {\bibinfo {author} {\bibfnamefont {C.~D.}\ \bibnamefont {Niroula}, \bibfnamefont {Pradeep~andWhite}}, \bibinfo {author} {\bibfnamefont {Q.}~\bibnamefont {Wang}}, \bibinfo {author} {\bibfnamefont {S.}~\bibnamefont {Johri}}, \bibinfo {author} {\bibfnamefont {D.}~\bibnamefont {Zhu}}, \bibinfo {author} {\bibfnamefont {C.}~\bibnamefont {Monroe}}, \bibinfo {author} {\bibfnamefont {C.}~\bibnamefont {Noel}},\ and\ \bibinfo {author} {\bibfnamefont {M.~J.}\ \bibnamefont {Gullans}},\ }\bibfield  {title} {\bibinfo {title} {Phase transition in magic with random quantum circuits},\ }\href {https://doi.org/10.1038/s41567-024-02637-3} {\bibfield  {journal} {\bibinfo  {journal} {Nature Physics}\ }\textbf {\bibinfo {volume} {20}},\ \bibinfo {pages} {1786–1792} (\bibinfo {year} {2024})}\BibitemShut {NoStop}%
\bibitem [{\citenamefont {Bera}\ and\ \citenamefont {Singha~Roy}(2020)}]{bera2020growth}%
  \BibitemOpen
  \bibfield  {author} {\bibinfo {author} {\bibfnamefont {A.}~\bibnamefont {Bera}}\ and\ \bibinfo {author} {\bibfnamefont {S.}~\bibnamefont {Singha~Roy}},\ }\bibfield  {title} {\bibinfo {title} {Growth of genuine multipartite entanglement in random unitary circuits},\ }\href {https://doi.org/10.1103/PhysRevA.102.062431} {\bibfield  {journal} {\bibinfo  {journal} {Phys. Rev. A}\ }\textbf {\bibinfo {volume} {102}},\ \bibinfo {pages} {062431} (\bibinfo {year} {2020})}\BibitemShut {NoStop}%
\bibitem [{\citenamefont {Turkeshi}\ \emph {et~al.}(2025{\natexlab{b}})\citenamefont {Turkeshi}, \citenamefont {Dymarsky},\ and\ \citenamefont {Sierant}}]{turkeshi2025pauli}%
  \BibitemOpen
  \bibfield  {author} {\bibinfo {author} {\bibfnamefont {X.}~\bibnamefont {Turkeshi}}, \bibinfo {author} {\bibfnamefont {A.}~\bibnamefont {Dymarsky}},\ and\ \bibinfo {author} {\bibfnamefont {P.}~\bibnamefont {Sierant}},\ }\bibfield  {title} {\bibinfo {title} {Pauli spectrum and nonstabilizerness of typical quantum many-body states},\ }\href {https://doi.org/10.1103/PhysRevB.111.054301} {\bibfield  {journal} {\bibinfo  {journal} {Phys. Rev. B}\ }\textbf {\bibinfo {volume} {111}},\ \bibinfo {pages} {054301} (\bibinfo {year} {2025}{\natexlab{b}})}\BibitemShut {NoStop}%
\bibitem [{Note2()}]{Note2}%
  \BibitemOpen
  \bibinfo {note} {In an extended (2+1)D system, the crossover becomes a continuous phase transition, at least for $\protect \mathbb {Z}_N$ LGTs that can be mapped to Potts models. For SU(2), there are indications of a phase transition~\cite {Cataldi_PRR2024}, but numerical limitations prevent a precise characterization of the critical behavior.}\BibitemShut {Stop}%
\bibitem [{Note3()}]{Note3}%
  \BibitemOpen
  \bibinfo {note} {This would not be the case in the full Hilbert space, where the ground-state at $g=0$ is an entangled quantum-spin liquid.}\BibitemShut {Stop}%
\bibitem [{\citenamefont {Smacchia}\ \emph {et~al.}(2011)\citenamefont {Smacchia}, \citenamefont {Amico}, \citenamefont {Facchi}, \citenamefont {Fazio}, \citenamefont {Florio}, \citenamefont {Pascazio},\ and\ \citenamefont {Vedral}}]{Smacchia_PRA2011}%
  \BibitemOpen
  \bibfield  {author} {\bibinfo {author} {\bibfnamefont {P.}~\bibnamefont {Smacchia}}, \bibinfo {author} {\bibfnamefont {L.}~\bibnamefont {Amico}}, \bibinfo {author} {\bibfnamefont {P.}~\bibnamefont {Facchi}}, \bibinfo {author} {\bibfnamefont {R.}~\bibnamefont {Fazio}}, \bibinfo {author} {\bibfnamefont {G.}~\bibnamefont {Florio}}, \bibinfo {author} {\bibfnamefont {S.}~\bibnamefont {Pascazio}},\ and\ \bibinfo {author} {\bibfnamefont {V.}~\bibnamefont {Vedral}},\ }\bibfield  {title} {\bibinfo {title} {Statistical mechanics of the cluster ising model},\ }\href {https://doi.org/10.1103/PhysRevA.84.022304} {\bibfield  {journal} {\bibinfo  {journal} {Phys. Rev. A}\ }\textbf {\bibinfo {volume} {84}},\ \bibinfo {pages} {022304} (\bibinfo {year} {2011})}\BibitemShut {NoStop}%
\bibitem [{\citenamefont {Potter}\ and\ \citenamefont {Vasseur}(2016)}]{TheChosenOne}%
  \BibitemOpen
  \bibfield  {author} {\bibinfo {author} {\bibfnamefont {A.~C.}\ \bibnamefont {Potter}}\ and\ \bibinfo {author} {\bibfnamefont {R.}~\bibnamefont {Vasseur}},\ }\bibfield  {title} {\bibinfo {title} {Symmetry constraints on many-body localization},\ }\href {https://doi.org/10.1103/PhysRevB.94.224206} {\bibfield  {journal} {\bibinfo  {journal} {Phys. Rev. B}\ }\textbf {\bibinfo {volume} {94}},\ \bibinfo {pages} {224206} (\bibinfo {year} {2016})}\BibitemShut {NoStop}%
\bibitem [{Note4()}]{Note4}%
  \BibitemOpen
  \bibinfo {note} {Think of spin 1/2 singlets, for instance.}\BibitemShut {Stop}%
\bibitem [{\citenamefont {Shaw}\ \emph {et~al.}(2020)\citenamefont {Shaw}, \citenamefont {Lougovski}, \citenamefont {Stryker},\ and\ \citenamefont {Wiebe}}]{Shaw2020quantumalgorithms}%
  \BibitemOpen
  \bibfield  {author} {\bibinfo {author} {\bibfnamefont {A.~F.}\ \bibnamefont {Shaw}}, \bibinfo {author} {\bibfnamefont {P.}~\bibnamefont {Lougovski}}, \bibinfo {author} {\bibfnamefont {J.~R.}\ \bibnamefont {Stryker}},\ and\ \bibinfo {author} {\bibfnamefont {N.}~\bibnamefont {Wiebe}},\ }\bibfield  {title} {\bibinfo {title} {Quantum {A}lgorithms for {S}imulating the {L}attice {S}chwinger {M}odel},\ }\href {https://doi.org/10.22331/q-2020-08-10-306} {\bibfield  {journal} {\bibinfo  {journal} {{Quantum}}\ }\textbf {\bibinfo {volume} {4}},\ \bibinfo {pages} {306} (\bibinfo {year} {2020})}\BibitemShut {NoStop}%
\bibitem [{\citenamefont {Raychowdhury}(2019)}]{Raychowdhury2019}%
  \BibitemOpen
  \bibfield  {author} {\bibinfo {author} {\bibfnamefont {I.}~\bibnamefont {Raychowdhury}},\ }\bibfield  {title} {\bibinfo {title} {Low energy spectrum of su(2) lattice gauge theory},\ }\href {https://doi.org/10.1140/epjc/s10052-019-6753-0} {\bibfield  {journal} {\bibinfo  {journal} {The European Physical Journal C}\ }\textbf {\bibinfo {volume} {79}},\ \bibinfo {pages} {235} (\bibinfo {year} {2019})}\BibitemShut {NoStop}%
\end{thebibliography}
\end{document}